\documentclass[11pt,a4paper]{article}
\pdfoutput=1
\usepackage[pdftex]{graphics}
\usepackage{jheppub}
\usepackage{amsmath,amssymb,amsfonts,cancel}
\usepackage{enumitem}
\usepackage{multirow}
\usepackage{array,booktabs}
\usepackage{tikz}
\usepackage{slashed}
\usepackage{verbatim}
\usepackage{float}

\newcommand{\eq}{\begin{equation}}
\newcommand{\eqe}{\end{equation}}
\newcommand{\eqa}{\begin{eqnarray}}
\newcommand{\eqae}{\end{eqnarray}}

\newcommand{\bn}{\begin{enumerate}}
\newcommand{\en}{\end{enumerate}}
\newcommand{\bl}{\begin{align}}
\newcommand{\el}{\end{align}}
\def\ie{\begin{equation}\begin{aligned}}
\def\fe{\end{aligned}\end{equation}}

\def\jmath{{j}}
\def\bl#1\el{\begin{align} #1 \end{align}}
\def\bg#1\eg{\begin{gather} #1 \end{gather}}

\def\bld#1\eld{\begin{aligned} #1 \end{aligned}}
\def\bgd#1\egd{\begin{gathered} #1 \end{gathered}}

\def\bC{{\mathbb{C}}}

\usetikzlibrary{snakes}
\usetikzlibrary{shapes.misc}
\usetikzlibrary{decorations.markings}

\tikzset{line/.style={line width=0.25mm},
curve/.style={line,smooth,tension=1},
->-/.style={decoration={
  markings,
  mark=at position #1 with {\arrow[>=stealth]{>}}},postaction={decorate}},
-<-/.style={decoration={
  markings,
  mark=at position #1 with {\arrow[>=stealth]{<}}},postaction={decorate}},
}

\title{Bulk locality from the celestial amplitude}

\author[1,2]{Chi-Ming Chang} 
\author[3,4]{Yu-tin Huang}
\author[3]{Zi-Xun Huang}
\author[3]{Wei Li}

\affiliation[1]{Yau Mathematical Sciences Center (YMSC), Tsinghua University, Beijing, 100084, China}
\affiliation[2]{Beijing Institute of Mathematical Sciences and Applications (BIMSA), Beijing, 101408, China}
\affiliation[3]{Department of Physics and Center for Theoretical Physics, National Taiwan University, Taipei, Taiwan 106}
\affiliation[4]{Physics Division, National Center for Theoretical Sciences, Taipei 10617, Taiwan}

\emailAdd{cmchang@tsinghua.edu.cn}
\emailAdd{yutinyt@gmail.com}
\emailAdd{r09222060@ntu.edu.tw}
\emailAdd{r07222072@ntu.edu.tw}

\abstract{In this paper, we study the implications of bulk locality on the celestial amplitude. In the context of the four-point amplitude, the fact that the bulk S-matrix factorizes locally in poles of Mandelstam variables is reflected in the imaginary part of the celestial amplitude. In particular, on the real axis in the complex plane of the boost weight, the imaginary part of the celestial amplitude can be given as a positive expansion on the Poincar\'e partial waves, which are nothing but the projection of flat-space spinning polynomials onto the celestial sphere. Furthermore, we derive the celestial dispersion relation, which relates the imaginary part to the residue of the celestial amplitude for negative even integer boost weight. The latter is precisely the projection of low energy EFT coefficients onto the celestial sphere. We demonstrate these properties explicitly on the open and closed string celestial amplitudes. Finally, we give an explicit expansion of the Poincar\'e partial waves in terms of 2D conformal partial waves.}

\begin{document}

\maketitle

\newpage
\section{Introduction}

In recent years, there has been steady progress on understanding the 2D holographic description of 4D flat-space scattering amplitudes \cite{Pasterski:2016qvg,Cardona:2017keg,Strominger:2017zoo,Pasterski:2017kqt,Pasterski:2017ylz,Lam:2017ofc,Banerjee:2017jeg,Schreiber:2017jsr,Banerjee:2018gce,Stieberger:2018edy,Donnay:2018neh,Stieberger:2018onx,Banerjee:2019aoy,Fan:2019emx,Pate:2019mfs,Puhm:2019zbl,Law:2019glh,Pate:2019lpp,Donnay:2020guq,Law:2020xcf,Arkani-Hamed:2020gyp} introduced by the pioneering work~\cite{Pasterski:2016qvg,Pasterski:2017kqt}, where one replaces the asymptotic state of the scattering amplitudes with boost eigenstates. In particular, the action of  the Lorentz group ${\rm SL}(2, \bC)$ on the kinematic data is recast into the M\"{o}bius transform on the celestial sphere, and the scattering amplitude is reinterpreted as a correlation function for some two-dimensional conformal field theory (CFT), termed the celestial amplitude. For amplitudes of massless external particles, this change of basis is implemented by a Mellin transform and the quantum numbers of the primary operator $(h,\bar{h})$ is related to the helicity ($\ell$) of the external particle as $\ell=h-\bar{h}$, while the dimension   $\Delta=h+\bar{h}$ is in-principle unconstrained.\footnote{Completeness relation of the conformal partial waves requires $\Delta=1+i\mathbb{R}$.}  

As the celestial amplitudes (denoted as $\tilde{\mathcal{A}}_n$ for $n$ states) are defined on boost eigenstates, which superpose all energies, the usual Wilsonian decoupling of UV/IR physics no long applies and is only well-defined for theories equipped with a UV completion. This combined with the lack of local observables in quantum gravity, makes the celestial amplitude the prime arena to study general properties of consistent quantum gravity theories~ \cite{Arkani-Hamed:2020gyp}.  Motivated by this, it will be desirable to derive the general set of consistency conditions for $\tilde{\mathcal{A}}$. 

The analytic properties of flat-space amplitude is an intensely studied subject and is relatively well-understood within the realm of perturbation theory. Thus, the ``projection" of these properties onto the celestial sphere should serve as the primary constraint. For massless amplitudes, one of the simplest universal behaviors are the soft limits. However, due to the superposition of all energies, for $\tilde{\mathcal{A}}_n$ the fate of soft theorems were unclear. The first progress toward elucidating the image of flat-space soft theorems were taken in~\cite{Fan:2019emx, Pate:2019mfs, Nandan:2019jas}, where the limit was realized in the limit where the conformal dimensions are taken to 1. More precisely, these ``conformal soft limits" of the celestial amplitudes lead to various conformal Ward identities associated with the holomorphic currents that generate the Kac-Moody symmetry in gauge theory \cite{Pate:2019mfs}, or the BMS supertranslation and the Virasoro symmetries in gravity \cite{Puhm:2019zbl, Donnay:2020guq}. The conformal soft theorems then constrain the leading operator product expansion (OPE) coefficients in the celestial CFT \cite{Pate:2019lpp}.

On the other hand, the actions of the Poincar\'e symmetry on the celestial sphere are explicitly worked out in \cite{Stieberger:2018onx}, and their constraints on the four and lower point celestial amplitudes are investigated in detailed in \cite{Law:2019glh}. As a consequence a new set of expansion basis for $\tilde{\mathcal{A}}_4$, the Poincar\'e (relativistic) partial waves, was proposed in \cite{Law:2020xcf} analogous to the conformal partial wave expansion of four-point functions in CFT.

The above progresses focuses on the various symmetry properties of the celestial amplitudes and the constraints followed from them. An obvious gap is the role of flat-space factorization, which encodes \textit{locality}, through where singularity might occur, and \text{unitary} that governs the residue or discontinuity across the singularity. Indeed it is often the case that these considerations alone are sufficient to completely determine the flat-space amplitude. In view of this one would like to ask:

$\quad\quad\quad\quad\quad\emph{How does the bulk-locality reflect on celestial amplitudes?}$

\noindent Limit analysis was done for three-dimensional scalar exchange~\cite{Lam:2017ofc}, as well as four-dimensional massive external legs~\cite{Cardona:2017keg}. In this paper, we aim to address this question in general for the four-point massless celestial amplitudes.

Poincar\'e invariance fixes the form of $\tilde{\mathcal{A}}_4$ up to a function that depends only on the real conformal cross ratio $z=\bar z={z_{12}z_{34}\over z_{13}z_{24}}$, the total conformal dimension (boost weight) $\beta=\Delta_1+\Delta_2+\Delta_3+\Delta_4-4$, and the helicities $\ell_i$, for $i=1,\cdots,4$ of the external particles \cite{Stieberger:2018onx,Law:2019glh}. 
The resulting function $\Psi(\beta,\ell_i,z)$ is related to the scattering amplitude in the plane wave basis by a Mellin transform
\begin{equation}
\Psi(\beta,\ell_i,z)\propto\int^\infty_0 d\omega\,\omega^{\beta-1} T_{\ell_i}\left(s,t\right)\,,
\end{equation}
where $\omega$ is the center of mass energy and $T_{\ell_i}\left(s,t\right)$ is the flat-space amplitude stripped off the momentum conservation delta function and a kinematic phase.  For fixed $\{\ell_i\}$ the we have a function of two variables $(\beta, z)$, replacing the flat-space parameterization $(s,t)$. It was argued in \cite{Arkani-Hamed:2020gyp} that the function is analytic in $\beta$ except for integer values on the real axes. The poles for $\beta=-2\mathbb{Z}^+$ are controlled by the Wilson coefficients of the low energy EFT with the degree of the poles determined by the IR running. For $\beta=2\mathbb{Z}^+$, these encodes the polynomial suppression of the UV amplitude. For theories of quantum gravity that latter singularities are expected to be absent due to black hole productions. 

Note that while the four-point celestial amplitude is defined on the equator of the celestial sphere, its not a continuous function across the entire circle. Using the usual SL(2,$\mathbb{C}$) to fix three points to $(0,1,\infty)$, the equator is divided into three regions corresponding to $s,t,u$-channel kinematics respectively. Each channel is named after the positive Mandelstam variable, while the remaining two are negative in the physical region. Schematically, we have:
$$
\begin{tikzpicture}[scale=0.75]
\draw [line,ultra thick] (0,0) circle (2);
\draw (0,2) node[rectangle, color= red, opacity=1, fill, inner sep=1.5pt ,rotate=90] {~~};
\draw (1.73205,-1) node[rectangle, color= red, opacity=1, fill, inner sep=1.5pt, rotate=-30] {~~};
\draw (-1.73205,-1) node[rectangle, color= red, opacity=1, fill, inner sep=1.5pt,rotate=30] {~~};
\draw (0,2.5) node {0};
\draw (2.16506,-1.25) node {1};
\draw (-2.16506,-1.25) node {$\infty$};
\draw (0,-2.5) node {$\Psi^{12\leftrightarrow 34}$};
\draw (-2.75,1.25) node {$\Psi^{14\leftrightarrow23}$};
\draw (2.75,1.25) node {$\Psi^{13\leftrightarrow24}$};
\end{tikzpicture}$$  
where the superscript on $\Psi$ denotes the physical channel. Importantly, due to the $\beta$ dependence of the Mellin transform, there are non-trivial monodromies across the three branch points $(0,1,\infty)$, even for $\beta=2\mathbb{Z}$. Thus, the three functions are in fact distinct and are not analytically connected, as first observed in the context of three-dimensions~\cite{Lam:2017ofc}. We will first demonstrate that due to bulk factorization, each function will acquire an imaginary piece reflecting the presence of thresholds in the physical channel, i.e. the channel with positive center of mass energy. In particular, the imaginary part of the celestial amplitude can be computed by extending the original Mellin integration to a fan-like contour that enclosed the positive real axis:

$$\begin{tikzpicture}[scale=1]
\draw [->, color=gray] (0,-2) -- (0,2);
\draw [->, color=gray] (-3,0) -- (3,0);
\draw [->, color=gray,snake=zigzag, line after snake=1mm, segment amplitude=.5mm] (1.4,-.1) -- (3,-.1);
\node[draw, cross out, inner sep=1pt, thick] at (1.4,-.1) {};
\node[circle, opacity=1, fill, inner sep=.75pt] at (0.8,-.1) {};
\node[circle, opacity=1, fill, inner sep=.75pt] at (1,-.1) {};
\node[circle, opacity=1, fill, inner sep=.75pt] at (1.2,-.1) {};
\draw [->, color=gray,snake=zigzag, line after snake=1mm, segment amplitude=.5mm] (-1.4,.1) -- (-3,.1);
\node[draw, cross out, inner sep=1pt, thick] at (-1.4,.1) {};
\node[circle, opacity=1, fill, inner sep=.75pt] at (-0.8,.1) {};
\node[circle, opacity=1, fill, inner sep=.75pt] at (-1,.1) {};
\node[circle, opacity=1, fill, inner sep=.75pt] at (-1.2,.1) {};
\draw [->, color=gray,snake=zigzag, line after snake=1mm, segment amplitude=.5mm] (-.1,1.2) -- (-.1,2);
\node[draw, cross out, inner sep=1pt, thick] at (-.1,1.2) {};
\node[circle, opacity=1, fill, inner sep=.75pt] at (-.1,0.6) {};
\node[circle, opacity=1, fill, inner sep=.75pt] at (-.1,0.8) {};
\node[circle, opacity=1, fill, inner sep=.75pt] at (-.1,1) {};
\draw [->, color=gray,snake=zigzag, line after snake=1mm, segment amplitude=.5mm] (.1,-1.2) -- (.1,-2);
\node[draw, cross out, inner sep=1pt, thick] at (.1,-1.2) {};
\node[circle, opacity=1, fill, inner sep=.75pt] at (.1,-0.6) {};
\node[circle, opacity=1, fill, inner sep=.75pt] at (.1,-0.8) {};
\node[circle, opacity=1, fill, inner sep=.75pt] at (.1,-1) {};
\draw [line,-<-=.55] (0,0) -- (2.83,1);
\draw [line,->-=.55] (0,0) -- (2.83,-1);
\draw [line,->-=.5] (2.83,-1) arc(-19.46:19.46:3) ;
\node at (1.5,-0.9) {$I_2'$};
\node at (1.5,0.9) {$I_2$};
\node at (3.3,0) {$I_3'$};
\end{tikzpicture}$$

\noindent The contour integral captures the poles and discontinuities of the amplitude which can be expanded in the basis of orthogonal polynomials, Legendre polynomials for external scalars  and Jacobi polynomials for gauge boson or graviton amplitudes. The projection of these polynomials onto the celestial sphere are then nothing but the Poincar\'e partial waves $\Phi_{m,J}(\beta,\ell_i,z)$ of mass $m$ and spin $J$ introduced in \cite{Law:2020xcf}. Thus, factorization singularities of the flat-space amplitude, in the physical channel, is projected into the imaginary part of the celestial amplitude and given by a sum of Poincar\'e partial waves, schematically,
\ie
\textbf{Im}\,\Psi(\beta,\ell_i,z)=\sum_a p_a\Phi_{m_a,J_a}(\beta,\ell_i,z)\,.
\fe
If the external states are organized such that for the physical threshold corresponds to $a,b\rightarrow b,a$ process, then we further have $p_a>0$, a reflection of unitarity.

Recently, it was shown that the EFT coefficients are constrained through dispersion relations in a fashion that reflects an underlying positive geometry,  the EFThedron~\cite{Arkani-Hamed:2020blm}. Since as previously mentioned, for $\tilde{\mathcal{A}}_4$ the poles on the negative $\beta$ axes encodes the EFT coefficients, these must be expressible as some form of dispersion relation. However, while the usual flat-space dispersion relation involves imaginary pieces arising from thresholds in distinct channels, for the celestial amplitude the imaginary part at any point on the equator is given by thresholds in one channel along. To this end we analytic continue the celestial amplitudes outside their physical defining regions to the ``unphysical" regions. This allows us to establish the \emph{celestial dispersion relation}, given as:
\ie
{\pi\over 2}\,\underset{\beta \to -2n}{\bf Res}\left[\Psi^{12\leftrightarrow 34}(\beta,z)\right]=\textbf{Im}\,\left[\Psi^{12\leftrightarrow34}(\beta,z)+(-1)^n \Psi^{13\leftrightarrow24}(\beta,z)\right]\Big|_{\beta\to-2n}&&(z\ge 1)\,,
\fe 
where we've given the form in the region $z\ge 1$. Note that the RHS contains the imaginary part of both $s$ and $u$-channel functions, where each can be defined in terms of the fan-like contour integral. We've explicitly verified these results for massive scalar exchange, as well as open and closed string amplitudes.

This paper is organized as follows. Section~\ref{sec:2} reviews the kinematics of the four-point massless celestial amplitudes and the UV/IR behaviors onward. Section~\ref{sec:masless 4point scalars} computes explicitly the imaginary parts of the four-point massless celestial amplitudes, show their positive expansion in terms of the Poincar\'e partial waves, and check the crossing symmetry. Section~\ref{sec:string} studies the example of the imaginary part of the open and closed string celestial amplitudes. Section~\ref{sec:AnalyticContinuation} discusses the analytic continuation of the celestial amplitudes. Section~\ref{sec:CelestialDispersionRelation} derives the celestial dispersion relation and verifies it for the example of the open and closed string amplitudes.  Section~\ref{sec:conclusion} ends with a summary, further comments and future directions. 
Appendix~\ref{sec:four-point_helicity_amplitude} derives the general form of helicity amplitudes that satisfy the constraints from the Lorentz symmetry and momentum conservation. Appendix~\ref{sec:PMonSPS} reviews the action of the Poincar\'e generators on a massless single particle state. Appendix~\ref{sec:conformalpartialwaverepresentation} expands the celestial amplitude and the Poincar\'e partial waves in terms of the conformal partial waves. Appendix~\ref{app:3ptCA} computes the three-point structure constant that shows up in the conformal partial wave expansion.

\section{Review of the four-point celestial amplitude}\label{sec:2}
In this section, we review general properties of the four-point celestial amplitude. We will focus on massless amplitudes $\mathcal{A}$, where the  momenta are given as $p_i=\epsilon_i\omega_iq_i^\mu$, with $\epsilon=\pm$ denoting whether the particle is outgoing or incoming, and the null vector $q^\mu$ is parametrized by
\ie\label{eqn:NullMomentum}
q_i^\mu= (1+|z_i|^2,2{\rm Re}(z_i),2{\rm Im}(z_i),1-|z_i|^2)\,.
\fe
Later on, $z_i$ will be the complex stereographic coordinate on the celestial sphere. Thus the amplitude $\mathcal{A}$ instead of being a function of four momenta, is now a function of $(\omega_i,z_i)$. The celestial amplitude is then simply the Mellin transform of helicity amplitudes:
\ie\label{eqn:MellinTransfrom}
\tilde{\mathcal{A}}_{\Delta_i,\ell_i}(z_i,\bar{z}_i)=\Big(\prod_{i=1}^n\int_0^\infty\mathrm{d}\omega_i\omega_i^{\Delta_i-1}\Big)\mathcal{A}_{\ell_i}(\omega_i, z_i)\,,
\fe
where $\ell_i$ is the helicity of each leg.
\subsection{Space-time to Celestial sphere ``kinematics"}
Let us consider in detail the transformation of flat-space scattering amplitudes to the celestial sphere.  Since we will be interested in helicity amplitudes, it is natural to embed $(\omega_i,z_i)$ in the spinor variables $\epsilon_i\omega_iq_i^\mu (\sigma_\mu)=\lambda_i\tilde\lambda_i$, where:
\eq\label{Canonical}
\lambda_i=\epsilon_i\sqrt{2\omega_i}\left(\begin{array}{c}1 \\ z_i \end{array}\right), \quad \tilde\lambda_i=\sqrt{2\omega_i}\left(\begin{array}{c}1 \\ \bar{z}_i \end{array}\right)\,.
\eqe
The map is of course not unique, as any U(1) rotation $\lambda\rightarrow e^{i\theta}\lambda$ and $\tilde\lambda\rightarrow e^{{-}i\theta}\tilde\lambda$ preserves the same null vector. The fact that the first component of $\lambda$ ($\tilde{\lambda}$) is real, correspond to a specific choice of frame. In this ``canonical frame", the Lorentz invariant spinor brackets take the form:
\eq
\langle i j \rangle=\varepsilon^{ab}\lambda_{j,a}\lambda_{i,b}=2\epsilon_i\epsilon_j\sqrt{\omega_i\omega_j}(z_i{-}z_j),\quad  [i j]=\varepsilon^{\dot a\dot b}\tilde \lambda_{j,\dot a}\tilde \lambda_{i,\dot b}=2\sqrt{\omega_i\omega_j}(\bar{z}_i{-}\bar{z}_j)\,.
\eqe
Similarly the Mandelstam variable $s_{ij}=\langle ij\rangle[ji]$ is related to the distance $|z_{ij}|$ on the celestial sphere by
\begin{equation}\label{eqn:mandelstam}
s_{ij}=-(p_i+p_j)^2=-2p_i\cdot p_j=4\epsilon_i\epsilon_j\omega_i\omega_j|z_{ij}|^2\,.
\end{equation}

One can straight forwardly see that the ${\rm SL}(2,\bC)$ Lorentz transformation acting on the spinors translate to the M\"{o}bius transformation acting on the complex plane $z$:
\eq
\left(\begin{array}{cc} a & b \\c & d \end{array}\right)\lambda =e^{i\theta}\lambda',\quad z'=\frac{c+dz}{a+bz},\quad \omega'=\omega|a+bz|^2,\quad  e^{i\theta}=\frac{a+bz}{|a+bz|}\,.
\eqe
Importantly, while the spinor products are Lorentz invariants, when considered in terms of the celestial coordinates  they acquire a ``little group" phase under ${\rm SL}(2,{\mathbb C})$: 
\ie
&\langle i j \rangle'=2\epsilon_i\epsilon_j\sqrt{\omega'_i\omega'_j}(z'_i{-}z'_j){=}2e^{{-}i\theta_i}e^{{-}i\theta_j}\epsilon_i\epsilon_j\sqrt{\omega_i\omega_j}(z_i{-}z_j)\,,
\\
&[ i j ]'=2\sqrt{\omega'_i\omega'_j}(\bar z'_i{-}\bar z'_j){=}2e^{i\theta_i}e^{i\theta_j}\sqrt{\omega_i\omega_j}(\bar z_i{-}\bar z_j)\,.
\fe
The origin of the extra phase is simple: as seen in eq.(\ref{Canonical}) an arbitrary ${\rm SL}(2,{\mathbb C})$ transformation will rotate the spinors out of the canonical frame. Thus one requires a ``compensating transformation" to restore it back. Since the amplitude transforms covariantly under little group transformations, manifested as helicity weights for each leg, one immediately deduce that the amplitude transform under ${\rm SL}(2,{\mathbb C})$ as:
\eq\label{eqn:SL2ConA}
\mathcal{A}_{\ell_i}(\omega'_i, z'_i, \bar{z}'_i)=\Big(\prod_{j}e^{2\ell_j \theta_j}\Big)\mathcal{A}_{\ell_i}(\omega_i, z_i,\bar{z}_i)\,.
\eqe

The Mellin-transform in eq.(\ref{eqn:MellinTransfrom}) can then be viewed as changing from plane wave basis, to the conformal primary basis (or boost)~\cite{Pasterski:2016qvg,Pasterski:2017kqt}. As a result under SL(2,C) transformation $\tilde{\mathcal{A}}_{\Delta_i,J_i}$ transforms as:
\eq
\tilde{\mathcal{A}}_{\Delta_i,\ell_i}(z_i',\bar z_i')=\Big(\prod_j (a{+}bz_j)^{\Delta_j{+}\ell_j}(\bar{a}{+}\bar{b}\bar{z}_j)^{\Delta_j{-}\ell_j}\Big)\tilde{\mathcal{A}}_{\Delta_i,\ell_j}(z_i,\bar z_i)\,.
\eqe
Thus the $n$-point celestial amplitude $\tilde{\mathcal{A}}$ transforms like a $n$-point conformal correlator in two-dimensional conformal field theory
\begin{equation}
\langle O_{h_1,\bar{h}_1}(z_1,\bar{z}_1)\dots O_{h_n,\bar{h}_n}(z_n,\bar{z}_n)\rangle\,,
\end{equation}
where the left-moving and right-moving conformal dimensions $h_i$ and $\bar h_i$ are
\begin{equation}
h_i+\bar{h}_i=\Delta_i,\quad h_i-\bar{h}_i=\ell_i\,.
\end{equation}
The variables $\Delta_i$ in the Mellin transform \eqref{eqn:MellinTransfrom} and the helicities $\ell_i$ become the total scaling dimensions and spins.

In this paper, our main focus is on the 4-point amplitude. $\text{SL}(2,\mathbb{C})$ conformal symmetry constrains the amplitude to the form
\begin{equation}\label{eqn:generalCA}
\tilde{\mathcal{A}}_{\Delta_i,\ell_i}(z_i,\bar{z}_i)=\frac{\Big(\frac{z_{14}}{z_{13}}\Big)^{h_3-h_4}\Big(\frac{z_{24}}{z_{14}}\Big)^{h_1-h_2}\Big(\frac{\bar{z}_{14}}{\bar{z}_{13}}\Big)^{\bar{h}_3-\bar{h}_4}\Big(\frac{\bar{z}_{24}}{\bar{z}_{14}}\Big)^{\bar{h}_1-\bar{h}_2}}{z_{12}^{h_1+h_2}z_{34}^{h_3+h_4}\bar{z}_{12}^{\bar{h}_1+\bar{h}_2}\bar{z}_{34}^{\bar{h}_3+\bar{h}_4}}f_{\Delta_i,\ell_i}(z,\bar{z})\,,
\end{equation}
where the cross ratios $z$ and $\bar{z}$ and coordinate differences $z_{ij}$, $\bar{z}_{ij}$ are
\begin{equation}
z=\frac{z_{12}z_{34}}{z_{13}z_{24}}\quad,\bar{z}=\frac{\bar{z}_{12}\bar{z}_{34}}{\bar{z}_{13}\bar{z}_{24}},\quad   z_{ij}=z_i-z_j,\quad \bar{z}_{ij}=\bar{z}_i-\bar{z}_j\,.
\end{equation}
Translation invariance, or momentum conservation, further constrain the dependence on the cross-ratio~\cite{Stieberger:2018onx,Law:2019glh},
\begin{equation}\label{eqn:translation}
f_{\Delta_i,\ell_i}(z,\bar{z})=(z-1)^{\frac{\Delta_1-\Delta_2-\Delta_3+\Delta_4}{2}}\delta(iz-i\bar{z})\Psi(\mathbf{\Delta},\ell_i,z)
\end{equation}
where $\mathbf{\Delta}=\sum_i\Delta_i$. Due to the delta function $\delta(iz-i\bar z)$, the celestial amplitude to be supported on the equator of the celestial sphere. 

One can also derive \eqref{eqn:generalCA} and \eqref{eqn:translation} directly from the Mellin integral representation \eqref{eqn:MellinTransfrom}. First, as show in Appendix~\ref{sec:four-point_helicity_amplitude} by utilizing the momentum conservation and the ${\rm SL}(2,{\mathbb C})$ symmetry, the four-point helicity amplitude take the form as
\ie\label{eqn:4pt_helicity_amplitude}
\mathcal{A}_{\ell_i}(\omega_i, z_i)&=\delta^{(4)}(p_1+p_2+p_3+p_4)\frac{\Big(\frac{z_{14}\bar z_{13}}{\bar z_{14}z_{13}}\Big)^{\ell_3-\ell_4\over 2}\Big(\frac{z_{24}\bar z_{14}}{\bar z_{24}z_{14}}\Big)^{\ell_1-\ell_2\over 2}}{\left({z_{12}\over \bar z_{12}}\right)^{{\ell_1+\ell_2\over 2}}\left({z_{34}\over \bar z_{34}}\right)^{\ell_3+\ell_4\over 2}}T_{\ell_i}(s,t)\,,
\fe
where $s\equiv s_{12}$, $t=s_{14}$, and $u=s_{13}$ are the Mandelstam variables. 
The momentum conservation written in terms of the energies $\omega_i$ and the celestial sphere coordinates $z_i$ and $\bar z_i$ as
\begin{equation}\label{eqn:p4delta}
\delta\left(\sum_{i=1}^4\epsilon_i\omega_iq_i\right)=\frac{8}{\Lambda^2|\omega_4|}\delta(iz{-}i\bar{z})\delta\Big(\omega_1{+}\epsilon_1\epsilon_4\frac{\Lambda^2\omega_4}{z}\Big)\delta\Big(\omega_2{+}\epsilon_2\epsilon_4\frac{\Lambda^2\omega_4}{z(z{-}1)}\Big)\delta\Big(\omega_3{+}\epsilon_3\epsilon_4\frac{\Lambda^2\omega_4}{1-z}\Big)\,,
\end{equation}
where we have used the $\text{SL}(2,\mathbb{C})$ transformation to fix the coordinates $z_i$ to
\begin{equation}\label{eqn:conformalframe}
z_1=0,\quad z_2=z,\quad z_3=1,\quad z_4=\Lambda\gg1\,,
\end{equation} 
In this conformal frame, the celestial amplitude becomes
\ie\label{eqn:CAsimplified}
\tilde{\mathcal{A}}_{\Delta_i,\ell_i}(z_i,\bar{z}_i)
&=8 \Lambda ^{-2\Delta_4}|z-1|^{{1\over 2}(\Delta_1{-}\Delta_2{-}\Delta _3{+}\Delta_4)}|z|^{{-}\Delta _1{-}\Delta _2}\delta(iz{-}i\bar z)
\\
&\quad\times\theta({-}\epsilon_1\epsilon_4 z)\theta(\epsilon_2\epsilon_4 z(1{-}z))\theta(\epsilon_3\epsilon_4 (z{-}1)) 
\\
&\quad\times z^2|z-1|^{2{-}{{\bf\Delta}\over 2}}\int_0^\infty\mathrm{d}\tilde\omega_4\,\tilde\omega_4^{\mathbf{\Delta}{-}5}T_{\ell_i}\Big(\frac{4\tilde\omega_4^2}{z{-}1},{-}\frac{4\tilde\omega_4^2}{z}\Big)\,,
\fe
where $\tilde\omega_4=\Lambda^{-2}\omega_4$. Note that in the conformal frame \eqref{eqn:conformalframe} the $z$-dependent factor in front of $T(s,t)$ in \eqref{eqn:4pt_helicity_amplitude} reduces to $1$ by the delta functions. The Heaviside theta functions are there to ensuring the delta functions having support in the integration domain of $\omega_i$. Indeed \eqref{eqn:CAsimplified} reduces to \eqref{eqn:generalCA} and \eqref{eqn:translation} in the conformal frame \eqref{eqn:conformalframe} where one reads off $\Psi({\bf \Delta},\ell_i,z)$ as
\ie\label{eqn:PsiasInt}
\Psi({\bf \Delta},\ell_i,z)&=\theta({-}\epsilon_1\epsilon_4 z)\theta(\epsilon_2\epsilon_4 z(1{-}z))\theta(\epsilon_3\epsilon_4 (z{-}1)) 
\\
&\quad\times z^2|z-1|^{2{-}{{\bf\Delta}\over 2}}\int_0^\infty\mathrm{d}\tilde\omega_4\,\tilde\omega_4^{\mathbf{\Delta}{-}5}T_{\ell_i}\Big(\frac{4\tilde\omega_4^2}{z-1},-\frac{4\tilde\omega_4^2}{z}\Big)\,.
\fe

Now due to the step functions, depending on the choice of incoming legs the cross-ratio $z$ is constrained to different regions. Consider three distinct kinematic configuration distinguished by the incoming state being in $s, u$ or $t$-channel,
\ie\label{zregion}
&12\leftrightarrow34:\quad\epsilon_1=\epsilon_2=-\epsilon_3=-\epsilon_4\,,
\\
&13\leftrightarrow24:\quad\epsilon_1=\epsilon_3=-\epsilon_2=-\epsilon_4\,,
\\
&14\leftrightarrow23:\quad\epsilon_1=\epsilon_4=-\epsilon_2=-\epsilon_3\,.
\fe
The Heaviside theta functions in \eqref{eqn:PsiasInt} will constrain the celestial amplitudes with the three different kinematics to have supports on three separate intervals on the equator of the celestial sphere,
\ie\label{KRonequater}
&12\leftrightarrow34:\, z\ge1\,,\quad13\leftrightarrow24:\,0\le z\le 1\,,\quad14\leftrightarrow23:\, z\le 0\,.
\fe
It would be convenient to use the center of mass energy $\omega$ as the integration variable. We apply the changes of variables in the three different kinematics as
\ie\label{eqn:COV}
&12\leftrightarrow34:~ \omega^2=\frac{4\tilde\omega_4^2}{z-1}\,,\quad13\leftrightarrow24:~ \omega^2=\frac{4\tilde\omega_4^2}{z(1-z)}\,,\quad14\leftrightarrow23:~\omega^2=\frac{4\tilde\omega_4^2}{(-z)}\,.
\fe
The physical regions, changes of variables, and the corresponding parametrizations of the Mandelstam variables are summarized in Table~\ref{tab:kinematics}. Note that we can also identify $z=\frac{2}{1{-}\cos\theta}$, where $\theta$ is the scattering angle and the limit  $z\rightarrow \infty$ corresponds to the forward limit. 


\begin{table}
\begin{center}
 \begin{tabular}{|c|c|c|c|} 
  \hline
kinematics & $12\leftrightarrow 34$ & $13\leftrightarrow 24$ & $14\leftrightarrow 23$  \\
\hline
\multirow{2}{*}{physical region} & $z\ge 1$ & $1\ge z\ge 0$ & $0\ge z$ 
\\
& $s\ge0\ge u,t$ & $u\ge0\ge s,t$ & $t\ge0\ge s,u$
\\
\hline
$\omega$ & $2\tilde\omega_4\over \sqrt{z-1}$ & $2\tilde\omega_4\over \sqrt{z(1-z)}$ & $2\tilde\omega_4\over \sqrt{-z}$\rule{0pt}{2.6ex}\rule[-1.6ex]{0pt}{0pt}
\\
\hline
$(s,u,t)$ & $(\omega^2,-{1\over z}\omega^2,-{(z-1)\over z}\omega^2)$ & $(-z\omega^2,\omega^2,-(1-z)\omega^2)$ & $(-{(-z)\over 1-z}\omega^2,-{1\over 1-z}\omega^2,\omega^2)$  \\
 \hline
\end{tabular}
\end{center}
\caption{The physical regions, center of mass energy $\omega$ and Mandelstam variables in the three different kinematics.}
\label{tab:kinematics}
\end{table}

The celestial amplitude \eqref{eqn:PsiasInt} in these three kinematics are then defined as:
\ie\label{eqn:fMellinIntegrals}
\Psi^{12\leftrightarrow34}(\mathbf{\Delta},\ell_i,z)&={1\over 2^{{\bf \Delta}-7}}z^{2}\int^\infty_0 d\omega\,\omega^{{\bf \Delta} {-}5} T_{\ell_i}\left(\omega^2, {-}\frac{(z{-}1)}{z}\omega^2\right)&&(z\ge1)\,,
\\
\Psi^{13\leftrightarrow24}(\mathbf{\Delta},\ell_i, z)&={1\over 2^{{\bf \Delta}-7}}z^{\frac{\mathbf{\Delta}}{2}}\int^\infty_0 d\omega\,\omega^{{\bf \Delta}{-}5} T_{\ell_i}\left({-}z\omega^2, (z{-}1)\omega^2\right)&&(1\ge z\ge 0)\,,
\\
\Psi^{14\leftrightarrow23}(\mathbf{\Delta},\ell_i, z)&={1\over 2^{{\bf \Delta}-7}} (
{-}z)^{{{\bf \Delta}\over 2}}(1{-}z)^{2{-}\frac{\mathbf{\Delta}}{2}}\int^\infty_0 d\omega\,\omega^{{\bf \Delta}{-}5} T_{\ell_i}\left({z\over 1{-}z}\omega^2, \omega^2\right)&&(0\ge z)\,.
\fe
We stress that the celestial amplitude \textit{is not} given by a single function $\Psi(\mathbf{\Delta},z)$ defined on the equator. Rather, there are three separate functions $\Psi^{12\leftrightarrow34}$, $\Psi^{13\leftrightarrow24}$ and $\Psi^{14\leftrightarrow 23}$ that tile the equator. 

In short, the function $\Psi^{ij\leftrightarrow kl}(\mathbf{\Delta},\ell_i,z)$ will be related to the amplitude via 
\ie\label{Main}
\Psi^{ij\leftrightarrow kl}(\beta, \ell_i, z)&=B^{ij\leftrightarrow kl}(z)\int^\infty_0 d\omega\,\omega^{\beta{-}1} T_{\ell_i}^{ij\leftrightarrow kl}\left(\omega,z\right)\,,
\fe
where using the notation of~\cite{Arkani-Hamed:2020gyp} we introduce $\beta={\bf \Delta}{-}4$. $B^{ij\leftrightarrow kl}(z)$ denotes the prefactors in front of the integrals in \eqref{eqn:fMellinIntegrals}, and $T_{\ell_i}^{ij\leftrightarrow kl}(\omega,z)$ equals to $T_{\ell_i}(s,t)$ with the parameterizations given in Table~\ref{tab:kinematics}.

%
%
%

\subsection{Implications of UV/IR behavior of $T\left(\omega,z\right)$}
\label{sec:UV&IR}
As stressed in ~\cite{Arkani-Hamed:2020gyp}, the analytic property of the celestial amplitude in complex $\beta$ plane reflects the UV and IR properties of $T\left(\omega,z\right)$. Since we will be performing a Mellin transform with respect to the center of mass energy $\omega$, special attention will be paid to the region $\omega\rightarrow0$ and $\infty$ for which the integral might diverge. In limit $\omega\rightarrow 0$, we probe the IR limit of $T\left(\omega,z\right)$. Suppressing the massless logs for now, the amplitude takes the form
\eq\label{eqn:TIR}
T(\omega,z)|_{\omega\rightarrow 0}=T_{\rm massless}(\omega,z)+\sum_{p=0}^\infty g_{p}(z)\omega^{2p}\,.
\eqe 
In $s$-kinematics, where $z\geq 1$, $g_{p}(z)$ are polynomial functions of at most degree $2p$ in  $\frac{1{-}z}{z}$ reflecting the presence of contact interactions, i.e. higher dimension operators in the EFT description. The function $T_{\rm massless}(\omega,z)$ summarizes the contribution from the massless particle exchange, which contain poles in $\frac{1{-}z}{z}$ and is of degree $\omega^0$ or $\omega^2$ for photon and graviton exchange respectively. Thus we see that the low energy amplitude is essentially a polynomial expansion in $\omega^2$. These generate single poles in $\beta$ since
\eq
\Psi(\beta,z)\sim \int^\Lambda_0 d\omega\,\omega^{\beta{-}1}(\omega^{2p}){+}\cdots = \frac{\Lambda^{\beta{+}2p}}{\beta{+}2p}+\cdots\,,
\eqe
where we only consider the part of the integral where $\omega\in[0,\Lambda]$. Thus we see that $\Psi(\beta,z)$ will have simple poles at $\beta=0,-2,-4,\cdots$. When the massless loops are involved, the simple poles are then promoted to higher degree as discussed in~\cite{Arkani-Hamed:2020gyp}.

We now turn to the opposite limit, where $\omega\rightarrow \infty$ corresponding to fixed angle hard scattering, 
\ie\label{eqn:hard limit}
s\to {+}\infty\quad{\rm with}\quad {s\over t}={z\over 1-z} \quad {\rm fixed.}
\fe
We will exam two asymptotic behaviors of $T(s,t)$ in this limit,
\ie\label{eqn:asyTst}
T(\omega,z)\sim \begin{cases}
 g_{p}(z)\omega^{2p}{+}g_{p{-}1}(z)\omega^{2p{-}2}{+}\cdots& \text{power law},
\\
e^{-\omega^2 f(z)}&\text{exponential decay}.
\end{cases}
\fe
As discussed in~\cite{Arkani-Hamed:2020gyp}, the second scenario is expected for amplitudes in gravitational theories due to black hole production. Indeed this is the case for string theory, which we will discuss in detail later on. In such case $\Psi(\beta,z)$ is convergent for Re$[\beta]>0$. For the power law, following similar analysis as the IR region we again arrive at poles for $\beta={-}2p, {-}2p{+}2,{-}2p{+}4,\cdots$. Now we would like to have a celestial amplitude that is meromorphic in $\beta$, thus having a region of convergence. From the previous IR analysis, we see that $\Psi(\beta,z)$ will have poles at ${\rm Re}[\beta]\leq0$, this suggest that we must have $p<0$ such that there is a convergent region $0<{\rm Re}[\beta]<-p$.\footnote{Indeed for $p=0$ such as $\lambda \phi^4$ theory, the function $\Psi(\beta,z)$ cease to be meromorphic:
\eq
\int_{0}^\infty \frac{d\omega}{\omega} \omega^{i\epsilon}=\int_{-\infty}^{\infty}dx \;e^{i\epsilon x}=\delta(i\epsilon)\,.
\eqe
}

\subsection{Example: The massive scalar exchange}
\label{sec:massive_scalar_exchange}
As a simple example for the above analysis, consider a massless scalar $\phi$ coupled to a massive scalar $X$ via a cubic coupling $g\phi^2 X$. The tree level 4-point scattering amplitude is
\begin{equation}\label{eqn:TforMSE}
T\left(s,t\right)
=-g^2\left(\frac{1}{s-m^2+i\varepsilon}
+\frac{1}{u-m^2+i\varepsilon}
+\frac{1}{t-m^2+i\varepsilon}\right)\,.
\end{equation}
The Mellin integral \eqref{Main} converges when $\beta$ is bounded by
\ie
0\leq\beta\leq2\,.
\fe
Indeed for the $12\leftrightarrow34$ kinematics, by rewriting $\Psi^{12\leftrightarrow34}(\beta,z)$ as integrating from $\omega=-\infty$ to $\omega=\infty$,
\begin{equation}
\Psi^{12\leftrightarrow34}(\beta,z)=\frac{2^{3-\beta}z^2}{(1-e^{i\pi\beta})}\int^\infty_{-\infty} d\omega\,\omega^{\beta-1} T\left(\omega^2, -\frac{(z-1)}{z}\omega^2\right)
\end{equation}
we can close the contour upward and pick up the residues giving \cite{Nandan:2019jas}
\ie\label{eqn:FullScalarCA1234}
\Psi^{12\leftrightarrow34}_{\text{scalar}}(\beta,z)&={\pi g^2 \over \sin{\pi\beta\over 2}}\left(m\over 2\right)^{\beta-2} z^{2}\left[e^{{1\over 2}\pi i\beta}+z^{{\beta\over 2}}+\left(z\over z-1\right)^{{\beta\over 2}}\right]\quad (z\ge 1)\,.
\fe
Note that indeed starting from $\beta=2$, one has simple poles at $
\beta=2,4,6,\cdots$, reflecting the divergence in UV. Similarly, below $\beta=0$, one has simple poles at $\beta=0,-4,-6,\cdots$, reflecting the EFT coefficients.  The absence of $\beta=-2$ is due to the corresponding EFT operator vanishes on-shell, $s{+}t{+}u=0$.

Importantly, there is a non-trivial imaginary part,\footnote{Here and throughout this paper, we assume that $\beta$ is real when taking the imaginary part of $\Psi(\beta,z)$.}
\begin{equation}
\textbf{Im}\,\Psi^{12\leftrightarrow34}_{\text{scalar}}(\beta,z)=\pi g^2\left(m\over 2\right)^{\beta-2}  z^2=2^{3-\beta}\pi z^2 g^2\textbf{Res}_{\omega=m}\Big[\omega^{\beta-1}T(\omega,z)\Big]\,,
\end{equation}
which is non-zero for any value of $\beta$. Furthermore for $s$-channel kinematics, only the physical threshold in the $s$-channel propagator contributes to imaginary part, while the $t,u$ diagram only contribute to the real part. Thus we see that the imaginary part of the celestial amplitude encodes the information of bulk factorization. This will be the focus of the next section.

\section{The imaginary part of the celestial amplitude}\label{sec:masless 4point scalars}
Here we would like to pose the following question: given a celestial amplitude $\Psi(\beta,z)$, what are the properties that reflect its origin as a local flat-space scattering amplitude. Already in the massive scalar case we've seen that for the $s$-channel kinematics, the fact that a massive scalar was being exchanged is reflected in the imaginary part of the celestial amplitude. In this section we will systematically study this property.

\subsection{Bulk-locality to the imaginary part of  $\Psi(\beta,z)$}
\label{sec:ImPart}
The imaginary part of the amplitude is deeply rooted in causality, where the time ordered two point function introduces the $i\varepsilon$ prescription for the Feynman propagator. Indeed this is the origin of the imaginary piece of the scalar exchange $\Psi^{12\leftrightarrow34}(\beta,z)$, appearing in the $s$-channel where the internal particle can be interpreted as on-shell and moving forward in time. Thus to capture the imaginary piece, it will be useful to consider the Mellin transform as a contour integral on the complex $\omega$-plane.

Let us focus on the $12\leftrightarrow 34$ kinematics. From previous discussions, we have seen that for $\Psi^{12\leftrightarrow 34}(\beta,z)$ to be a meromorphic function, $T^{12\leftrightarrow 34}(\omega, z)$ must vanish as $\omega\rightarrow \pm \infty$, i.e. $T^{12\leftrightarrow 34}(\omega, z)$ behavior asymptotically as \eqref{eqn:asyTst} for $p<0$. In general, such an asymptotic behavior does not hold when $\omega$ approaches complex infinity, i.e. $\omega\to \infty\times e^{i\theta}$ for $\theta\neq 0\,,\pi$.

We will assume that the asymptotic behaviors in  \eqref{eqn:asyTst}, which is defined for real $\omega$, can be extended for a small range of $\arg \omega$. More precisely, we impose the following assumption on $T^{12\leftrightarrow 34}(\omega)$:

\paragraph{Boundedness:} $T^{12\leftrightarrow 34}(\omega,z)$ is bounded for some finite extension onto the the complex $\omega$, i.e.  
$T^{12\leftrightarrow 34}(\omega,z)\to 0$ as $|\omega|\to\infty$ with the argument
\ie\label{eqn:theta}
\arg \omega \in \big(-\theta^{12\leftrightarrow 34}_c,\theta^{12\leftrightarrow 34}_c\big)\cup \big(\pi-\theta^{12\leftrightarrow 34}_c,\pi+\theta^{12\leftrightarrow 34}_c\big)\,,
\fe
with a finite angle $\theta^{12\leftrightarrow 34}_c$ that in general depends on $z$. 

This assumption, while not rigorously proven here, can be argued from the fact that the amplitude is analytic in the complex plane except for the real and imaginary axes. Thus the convergence property can only change smoothly as we move away from the axes. Thus we expect that there always exists a finite neighborhood where $T$ remain convergent. As we will see, in the case of open and closed string amplitudes, $\theta_c$'s are in general finite and non-vanishing.

\begin{figure}[t]
\centering
\begin{tikzpicture}[scale=1]
\node at (-1.5, 1.5) {$\omega$-plane};
\draw [->, color=gray] (0,-2) -- (0,2);
\draw [->, color=gray] (-3,0) -- (3,0);
\draw [->, color=gray,snake=zigzag, line after snake=1mm, segment amplitude=.5mm] (1.4,-.1) -- (3,-.1);
\node[draw, cross out, inner sep=1pt, thick] at (1.4,-.1) {};
\node[circle, opacity=1, fill, inner sep=.75pt] at (0.8,-.1) {};
\node[circle, opacity=1, fill, inner sep=.75pt] at (1,-.1) {};
\node[circle, opacity=1, fill, inner sep=.75pt] at (1.2,-.1) {};
\draw [->, color=gray,snake=zigzag, line after snake=1mm, segment amplitude=.5mm] (-1.4,.1) -- (-3,.1);
\node[draw, cross out, inner sep=1pt, thick] at (-1.4,.1) {};
\node[circle, opacity=1, fill, inner sep=.75pt] at (-0.8,.1) {};
\node[circle, opacity=1, fill, inner sep=.75pt] at (-1,.1) {};
\node[circle, opacity=1, fill, inner sep=.75pt] at (-1.2,.1) {};
\draw [->, color=gray,snake=zigzag, line after snake=1mm] (-.1,1.2) -- (-.1,2);
\node[draw, cross out, inner sep=1pt, thick] at (-.1,1.2) {};
\node[circle, opacity=1, fill, inner sep=.75pt] at (-.1,0.6) {};
\node[circle, opacity=1, fill, inner sep=.75pt] at (-.1,0.8) {};
\node[circle, opacity=1, fill, inner sep=.75pt] at (-.1,1) {};
\draw [->, color=gray,snake=zigzag, line after snake=1mm, segment amplitude=.5mm] (.1,-1.2) -- (.1,-2);
\node[draw, cross out, inner sep=1pt, thick] at (.1,-1.2) {};
\node[circle, opacity=1, fill, inner sep=.75pt] at (.1,-0.6) {};
\node[circle, opacity=1, fill, inner sep=.75pt] at (.1,-0.8) {};
\node[circle, opacity=1, fill, inner sep=.75pt] at (.1,-1) {};
\draw [line,->-=.55] (0,0) -- (3,0);
\draw [line,-<-=.55] (0,0) -- (2.83,1);
\draw [line,->-=.5] (3,0) arc(0:19.46:3) ;
\node at (1.5,-0.4) {$I_1$};
\node at (1.5,0.9) {$I_2$};
\node at (3.2,0.5) {$I_3$};
\end{tikzpicture}
$\quad\quad\quad$
\begin{tikzpicture}[scale=1]
\node at (-1.5, 1.5) {$\omega$-plane};
\draw [->, color=gray] (0,-2) -- (0,2);
\draw [->, color=gray] (-3,0) -- (3,0);
\draw [->, color=gray,snake=zigzag, line after snake=1mm, segment amplitude=.5mm] (1.4,-.1) -- (3,-.1);
\node[draw, cross out, inner sep=1pt, thick] at (1.4,-.1) {};
\node[circle, opacity=1, fill, inner sep=.75pt] at (0.8,-.1) {};
\node[circle, opacity=1, fill, inner sep=.75pt] at (1,-.1) {};
\node[circle, opacity=1, fill, inner sep=.75pt] at (1.2,-.1) {};
\draw [->, color=gray,snake=zigzag, line after snake=1mm, segment amplitude=.5mm] (-1.4,.1) -- (-3,.1);
\node[draw, cross out, inner sep=1pt, thick] at (-1.4,.1) {};
\node[circle, opacity=1, fill, inner sep=.75pt] at (-0.8,.1) {};
\node[circle, opacity=1, fill, inner sep=.75pt] at (-1,.1) {};
\node[circle, opacity=1, fill, inner sep=.75pt] at (-1.2,.1) {};
\draw [->, color=gray,snake=zigzag, line after snake=1mm, segment amplitude=.5mm] (-.1,1.2) -- (-.1,2);
\node[draw, cross out, inner sep=1pt, thick] at (-.1,1.2) {};
\node[circle, opacity=1, fill, inner sep=.75pt] at (-.1,0.6) {};
\node[circle, opacity=1, fill, inner sep=.75pt] at (-.1,0.8) {};
\node[circle, opacity=1, fill, inner sep=.75pt] at (-.1,1) {};
\draw [->, color=gray,snake=zigzag, line after snake=1mm, segment amplitude=.5mm] (.1,-1.2) -- (.1,-2);
\node[draw, cross out, inner sep=1pt, thick] at (.1,-1.2) {};
\node[circle, opacity=1, fill, inner sep=.75pt] at (.1,-0.6) {};
\node[circle, opacity=1, fill, inner sep=.75pt] at (.1,-0.8) {};
\node[circle, opacity=1, fill, inner sep=.75pt] at (.1,-1) {};
\draw [line,-<-=.55] (0,0) -- (2.83,1);
\draw [line,->-=.55] (0,0) -- (2.83,-1);
\draw [line,->-=.5] (2.83,-1) arc(-19.46:19.46:3) ;
\node at (1.5,-0.9) {$I_2'$};
\node at (1.5,0.9) {$I_2$};
\node at (3.3,0) {$I_3'$};
\end{tikzpicture}
\caption{The analytic structure of the amplitude $T(\omega,z)$ on the complex $\omega$ plane. The branch cuts associated to the anomalous thresholds are not depicted in the figure, since they are away from the real $\omega$ axes, and would not contribute the the contour integrals when the angle between the $I_2$ and $-I_2'$ segments of the contour is small enough.}
\label{Fig:CScontour}
\end{figure}
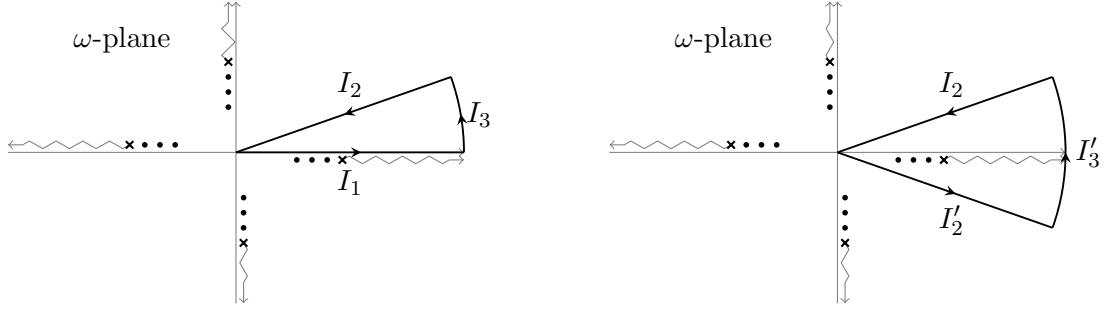

Let us exam the analytic structure of the amplitude $T^{12\leftrightarrow 34}(\omega,z)$ on the complex $\omega$-plane. By the bulk-locality, the exchange of single-particle states (of masses $m_i$) leads to $s$-channel poles located slight below or above the real axes (at $\omega=\pm m_i\pm i\epsilon$) due to Feynman $i\epsilon$.
At the loop level, the exchange of multi-particle states lead to branch cuts located slight below or above the real axes. The crossing images of the $s$-channel poles are the $t$-
 and $u$-channel poles (at $\omega=\pm i\sqrt{z\over z-1}m_i$ and $\omega=\pm i\sqrt{z}m_i$). The poles and branch cuts are shown schematically in figure~\ref{Fig:CScontour}. There could be other branch cuts away from the real and the imaginary axes, which do not correspond to the exchange of multi-particle states.\footnote{The branch points of such branch cuts are called the anomalous thresholds.} They are not depicted in the figures, as they would not play a role in our later computation.  Finally, the branch cut of the $\omega^{\beta-1}$ is chosen to be along the negative real axis, and is also not depicted in the figures.

Consider the fan-like contour displayed in the left of figure~\ref{Fig:CScontour}, with an angle less than $\theta_c$. As the infinity part $I_3$ vanishes, the absence of poles in the contour imply 
\begin{equation}\label{eqn:PsiinI2}
\Psi(\beta, z)=B(z)\int_{I_1}\mathrm{d}\omega\,\omega^{\beta{-}1}T(\omega,z)={-}B(z)\int_{I_2}\mathrm{d}\omega\,\omega^{\beta{-}1}T(\omega,z)\,.
\end{equation}
Now extend the contour symmetrically to the lower half plane, as shown in the right of figure~\ref{Fig:CScontour}. The new contour will then pick up the residue from the $i\varepsilon$ prescription of the propagators. Note that since $I_2'$ is just the reflection of $I_2$ along the real axes, the two simply have the opposite sign for the real part:
\ie\label{eqn:conjugate}
&\int_{I_2'}\mathrm{d}\omega\,\omega^{\beta{-}1}T(\omega,z)=-\Big[\int_{I_2}\mathrm{d}\omega\,\omega^{\beta{-}1}T(\omega,z)\Big]^*\,.
\fe
Furthermore, since the $i\varepsilon$ poles are away from the contour $I_2'$, and we are free to take the $\varepsilon\to 0$ limit, for which the amplitude $T(\omega,z)$ is a real function of $\omega$, i.e.
\ie\label{eqn:cxconj}
T(\omega,z)^*=T(\omega^*,z)\,.
\fe
Again since the integration along $I'_3$ vanishes, the imaginary part of $\Psi(\beta, z)$ is now simply given by the poles and branch cuts from the $i\varepsilon$ prescription
\ie\label{eqn:imaginary}
&\textbf{Im}\,\Psi(\beta, z)={-}{1\over 2i}B(z)\int_{I_2+I_2'}\mathrm{d}\omega\,\omega^{\beta{-}1}T(\omega,z)
\\
&=-B(z)\left\{\pi \sum_{i}\underset{\omega\to m_i}{\bf Res}\Big[\omega^{\beta{-}1}T(\omega,z)\Big]+\int_{M}^\infty\mathrm{d}\omega\,\omega^{\beta{-}1}\,\textbf{Disc}\,\left[T(\omega,z)\right]\right\}\,,
\fe
where $m_i$ are the position of the factorization poles, and $M$ is the branch point of the branch cuts, which is present due to loop effects.\footnote{If the convergent angle $\theta_0$ is great than $\frac{\pi}{2}$ (e.g. massive scalar exchange amplitude), one might worry that there will be poles coming from other channel contributing to the imaginary part in \eqref{eqn:imaginary}. In fact, because other channel poles are lying on the pure imaginary axis, and they are coming in pair. One can easily verify total residue sum is $0$ in the RHS contour of fig.~\ref{Fig:CScontour}. Thus the result is not changed even if the $\theta_0>\pi/2$ is bigger than $\pi/2$
}

In summary, we find that the imaginary part of the celestial amplitude is simply governed by the residue or (massive) discontinuity of the scattering amplitude.  For $s$-channel kinematics, we find that $\textbf{Im}[\Psi]$ is given by the projection of the  $s$-channel residue or discontinuity onto the celestial sphere as in eq.(\ref{eqn:imaginary}).

\subsection{Positivity in $\textbf{Im$[\Psi(\beta, z)]$}$}
\label{sec:masslessandspinningexchange}
If the external states are arranged as $a, b\rightarrow b,a$, where ($a,b$) represents potentially distinct species,  the residue and the discontinuity are positively expandable on Legendre polynomials for scalars and Jacobi polynomials for spinning states~\cite{Arkani-Hamed:2020blm}.  Thus the imaginary part of the celestial amplitude must be positively expanded on the Mellin-transform of these orthogonal polynomials.

Let us first focus on the case of scalar external states. At the tree-level, the imaginary part of the celestial amplitude only picks up the poles corresponding to the factorization channels. The associated residues in the three kinematic regimes are
\ie
&\text{$12\to34$}:\quad \textbf{Res}_{\omega=m}\Big[\omega^{\beta{-}1}\frac{m^{2J}P_J(\frac{u-t}{m^2})}{\omega^2-m^2}\Big]={m^{\beta{-}2{+}2J}\over2}\cdot P_{J}\left(z-2\over z\right)\,,
\\
&\text{$13\to24$}:\quad \textbf{Res}_{\omega=m}\Big[\omega^{\beta{-}1}\frac{m^{2J}P_J(\frac{s-t}{m^2})}{\omega^2-m^2}\Big]={m^{\beta{-}2{+}2J}\over2}\cdot P_{J}(1-2z),\\
&\text{$14\to23$}:\quad \textbf{Res}_{\omega=m}\Big[\omega^{\beta{-}1}\frac{m^{2J}P_J(\frac{u-s}{m^2})}{\omega^2-m^2}\Big]={m^{\beta{-}2{+}2J}\over2}\cdot P_{J}\Big(\frac{z+1}{z-1}\Big)\,.
\fe
Note that we have written the argument of the Legendre polynomials in terms or Mandelstam invariants in such a way that manifest the exchange symmetry in each channel. Thus from \eqref{eqn:imaginary} the imaginary part of the scalar celestial amplitude must be positively expanded on the following basis
\ie\label{eqn:ImCAs}
&\textbf{Im}\,\Psi^{12\leftrightarrow34}(\beta,z)=\pi z^2\sum_{a} p_aP_{J_a}\left(z-2\over z\right)&&(z\ge 1)\,,
\\
&\textbf{Im}\,\Psi^{13\leftrightarrow24}(\beta,z)=\pi z^{\frac{\beta}{2}+2}\sum_{a} p_a P_{J_a}(1-2z)&&(1\ge z\ge0)\,,
\\
&\textbf{Im}\,\Psi^{14\leftrightarrow23}(\beta,z)=\pi(-z)^{\frac{\beta}{2}+2}(1-z)^{-\frac{\beta}{2}}\sum_{a} p_aP_{J_a}\left(z+1\over z-1\right)&&(0\ge z)\,,
\fe
where $p_a=g_a^2 \left(m_a\over 2\right)^{\beta-3}m_a^{2J_a+1}$, and the index $a$ labels a state with quantum numbers $(m_a, J_a)$ coupling to the massless external states with strength $g_a$. While we have set up the equations for tree-level exchanges, extension to massive loops is straight forwards since the discontinuity along the branch cut also have a positive partial wave expansion. Thus for massive loops, we only need to change the discreet sum into a continuous integration. For example 
\eq\label{TtoL}
\sum_{a} p_aP_{J_a}\left(z-2\over z\right) \quad \rightarrow \quad \int_{M^2} dm^2 p_J(m^2) P_{J_a}\left(z-2\over z\right) 
\eqe
where $M^2$ denotes the beginning of the branch cut. This difference does not change our analysis, and thus for brevity from now on we will not make a distinction between the discrete sum and continuous integration.

Extension to spinning external states is straight forward. On the factorization pole, the residue polynomial is now:
\ie
T_{\ell_i}(s,t)|_{s\rightarrow m^2}&=-\frac{m^{2J}d_{\ell_{34},\ell_{12}}^{J}\left(\arccos(\frac{u-t}{m^2})\right) }{s-m^2+i\varepsilon},
\fe
where $\ell_{ij\pm kl}=(\ell_i-\ell_j)\pm(\ell_k-\ell_l)$, and $d^J_{\ell_{34},\ell_{12}}(\phi)$ is the Wigner (small) $d$-matrix which can be conveniently given in Jacobi polynomials,
\begin{equation}
    d_{\ell_{34},\ell_{12}}^{J}(\phi)=B^{J}_{12;34}\left(\sin{\phi\over2}\right)^{\ell_{12-34}}\left(\cos{\phi\over2}\right)^{\ell_{12+34}}\mathcal{J}^{\ell_{12-34},\ell_{12+34}}_{J-\ell_{12}}(\cos\phi),
\end{equation}
where $\mathcal{J}^{\alpha,\beta}_n$ is the Jacobi polynomial and the constant $B^{J}_{i_1i_2;i_3i_4}\equiv\sqrt{\frac{(J{+}\ell_{i_1i_2})!(J{-}\ell_{i_1i_2})!}{(J_i{+}\ell_{i_3i_4})!(J{-}\ell_{i_3i_4})!}}$. This gives the following representation of the imaginary part $\textbf{Im}\,[\Psi(\beta, z)]$ in the three different regions \eqref{KRonequater} on the equator:
\ie\label{eqn:ImCA}
&\textbf{Im}\,\Psi^{12\leftrightarrow34}(\beta,z)=\sum_{i} p_{i12;34}\;\;z^{2-\ell_{12}}(z-1)^{\frac{\ell_{12-34}}{2}}\mathcal{J}_{J_i-\ell_{12}}^{\ell_{12+34},\ell_{12-34}}\Big(\frac{z-2}{z}\Big)&&(z\ge 1)\,,
\\
&\textbf{Im}\,\Psi^{13\leftrightarrow24}(\beta,z)=\sum_{i} p_{i13;24}\;\;z^{\frac{\beta+\ell_{13+24}}{2}+2}(1-z)^{\frac{\ell_{13-24}}{2}}\mathcal{J}_{J_i-\ell_{13}}^{\ell_{13+24},\ell_{13-24}}(1-2z)&&(1\ge z\ge 0)\,,
\\
&\textbf{Im}\,\Psi^{14\leftrightarrow23}(\beta,z)=\sum_{i} p_{i14;23}\;\;\frac{(-z)^{\frac{\beta+\ell_{14+23}}{2}+2}}{(1-z)^{\frac{\beta}{2}+\ell_{23}}}\mathcal{J}_{J_i-\ell_{23}}^{\ell_{14+23},\ell_{14-23}}\left(\frac{z+1}{z-1}\right)&&(0\ge z)\,,
\fe
where $p_{i12;34}= \pi  g_i\left(m\over 2\right)^{\beta-3} m_i^{2J_i+1}B^{J_i}_{12;34}>0$. Indeed setting $\ell_i=0$ in $\eqref{eqn:ImCA}$ one recovers the scalar basis in $\eqref{eqn:ImCAs}$. The coefficient $g_i$ becomes positive if the external states are arranged as $\ell_3={-}\ell_2$ and $\ell_4={-}\ell_1$, i.e. such that the configuration corresponds to forward scattering. 

Thus for each kinematics, $\textbf{Im}\,\Psi$ has a positive expansion on the basis in eq.(\ref{eqn:ImCA}). Note that the fact that the Wigner $d$-matrix serves as an expansion basis for the flat-space amplitude is a reflection of factorization and Poincar\'e symmetry. Since the basis functions in eq.(\ref{eqn:ImCA}) is simply the projection of the Wigner $d$-matrix, they must be intimately tied to the ``image" of Poincar\'e symmetry on the celestial sphere. Indeed, in the following, we will find that these are precisely the Poincar\'e partial waves introduced in~\cite{Law:2020xcf}.

\subsection{Poincar\'e partial wave expansion}
\label{sec:PPW}
Let us first review the Poincar\'e partial waves introduced in~\cite{Law:2020xcf}, focusing on the celestial amplitude in the $12\leftrightarrow 34$ kinematics. It can be written as a inner product between the in and out states 
\ie\label{eqn:AmplitudeAsInner}
\tilde {\cal A}^{12\leftrightarrow 34}_{\Delta_i,\ell_i}(z_i,\bar z_i)& = \langle\Delta_3,z_3,\bar z_3,\ell_3,\Delta_4,z_4,\bar z_4,\ell_4|\Delta_1,z_1,\bar z_1,\ell_1,\Delta_2,z_2,\bar z_2,\ell_2\rangle\,.
\fe
The Hilbert space can be decomposed into irreducible unitary representations of the Poincar\'e algebra. According to Wigner's classification, they are the massive representations labeled by the mass $m$ and spin $J$, and the massless representations labeled by the helicity $\ell$. Consider the projectors
\ie
{\mathbb P}_{m,J}&={1\over 2J+1}\sum_{J_3=-J}^J|m,J,J_3\rangle \langle m,J,J_3|\,,
\\
{\mathbb P}_{\ell}&={1\over 2}\sum_{\epsilon=\pm}|\epsilon\ell\rangle \langle \epsilon\ell|\,,
\fe
which project onto a single massive or massless representation. The projectors ${\mathbb P}_{m,J}$ and ${\mathbb P}_{\ell}$ commute with the Poincar\'e generators $P^\mu$ and $M^{\mu\nu}$, i.e.
\ie\label{eqn:projectorPM}
&[P^\mu,{\mathbb P}_{m,J}]=0=[M^{\mu\nu},{\mathbb P}_{m,J}]\,,
\fe
and similarly for the massless projector ${\mathbb P}_{\ell}$. The massive and massless Poincar\'e partial waves are defined by inserting the projectors into the inner product \eqref{eqn:AmplitudeAsInner},
\ie
\tilde {\cal A}^{12\leftrightarrow 34}_{\Delta_i,\ell_i;m,J}(z_i,\bar z_i) &= \langle\Delta_3,z_3,\bar z_3,\ell_3,\Delta_4,z_4,\bar z_4,\ell_4|{\mathbb P}_{m,J}|\Delta_1,z_1,\bar z_1,\ell_1,\Delta_2,z_2,\bar z_1,\ell_2\rangle\,,
\\
\tilde {\cal A}^{12\leftrightarrow 34}_{\Delta_i,\ell_i;\ell}(z_i,\bar z_i) &= \langle\Delta_3,z_3,\bar z_3,\ell_3,\Delta_4,z_4,\bar z_4,\ell_4|{\mathbb P}_{\ell}|\Delta_1,z_1,\bar z_1,\ell_1,\Delta_2,z_2,\bar z_1,\ell_2\rangle\,.
\fe
The translation and Lorentz generators $P^\mu$ and $M^{\mu\nu}$ act on the massless single particle states as differential operators ${\cal P}^\mu$ and ${\cal M}^{\mu\nu}$ given explicitly in~\cite{Stieberger:2018onx}, which we list in Appendix~\ref{sec:PMonSPS}.  By \eqref{eqn:projectorPM}, the Poincar\'e partial waves satisfy the constraints
\ie\label{eqn:PPWCon}
&({\cal P}^\mu_1+{\cal P}^\mu_2-{\cal P}^\mu_3-{\cal P}^\mu_4)\tilde {\cal A}^{12\leftrightarrow 34}_{\Delta_i,\ell_i;m,J}(z_i,\bar z_i)=0\,,
\\
&({\cal M}^{\mu\nu}_1+{\cal M}^{\mu\nu}_2+{\cal M}^{\mu\nu}_3+{\cal M}^{\mu\nu}_4)\tilde {\cal A}^{12\leftrightarrow 34}_{\Delta_i,\ell_i;m,J}(z_i,\bar z_i)=0\,,
\fe
and similar for the massless Poincar\'e partial waves. Since the above constraints are the same constraints used in~\cite{Law:2019glh} to derive the form in \eqref{eqn:generalCA} and \eqref{eqn:translation} of the celestial amplitude, the Poincar\'e partial waves should also take the form as \eqref{eqn:generalCA} and \eqref{eqn:translation}.

Beside \eqref{eqn:PPWCon}, the Poincar\'e partial waves satisfy additional differential equations given by the Casimir operators of the Poincar\'e algebra. The Poincar\'e algebra has a quadratic and a quartic Casimir operators
\ie
&P_\mu P^\mu\,,\quad W_\mu W^\mu\,,
\fe
where $W^\mu={1\over 2}\epsilon^{\mu\nu\rho\sigma}M_{\nu\rho}P_\sigma$ is the Pauli-Lubanski pseudo-vector. The two Casimir operators have eigenvalues $-m^2$ and $m^2J(J+1)$ when acting on a state with mass $m$ and spin $J$, i.e.
\ie
P_\mu P^\mu \left| m, J,J_3\right>&=-m^2\left| m, J,J_3\right>\,,
\\
W_\mu W^\mu \left| m, J,J_3\right>&=m^2J(J+1)\left| m, J,J_3\right>\,.
\fe
By inserting the Casimir operator $P^\mu P_\mu$ into the inner product form of the Poincar\'e partial wave $\tilde {\cal A}^{12\leftrightarrow 34}_{\Delta_i,\ell_i;m,J}(z_i,\bar z_i)$, and using the formulae \eqref{eqn:PoincareGeneratorsInDifferentials} and \eqref{eqn:cPdiff}, one find a differential equation
\ie\label{eqn:P^2onA}
&({\cal P}_1+{\cal P}_2)^\mu({\cal P}_1+{\cal P}_2)_\mu\tilde {\cal A}^{12\leftrightarrow 34}_{\Delta_i,\ell_i;m,J}(z_i,\bar z_i)
\\
&=\langle\Delta_3,z_3,\bar z_3,\ell_3,\Delta_4,z_4,\bar z_4,\ell_4| {\mathbb P}_{m,J}P^\mu P_\mu|\Delta_1,z_1,\bar z_1,\ell_1,\Delta_2,z_2,\bar z_1,\ell_2\rangle
\\
&=-m^2\tilde {\cal A}^{12\leftrightarrow 34}_{\Delta_i,\ell_i;m,J}(z_i,\bar z_i)\,.
\fe
Similarly, the Casimir operator $W^\mu W_\mu$ leads to another differential equation
\ie\label{eqn:W^2onA}
({\cal W}_1+{\cal W}_2)^\mu({\cal W}_1+{\cal W}_2)_\mu\tilde {\cal A}^{12\leftrightarrow 34}_{\Delta_i,\ell_i;m,J}(z_i,\bar z_i)=m^2 J(J+1)\tilde {\cal A}^{12\leftrightarrow 34}_{\Delta_i,\ell_i;m,J}(z_i,\bar z_i)\,,
\fe
where ${\cal W}^\mu={1\over 2}\epsilon^{\mu\nu\rho\sigma}{\cal M}_{\nu\rho}{\cal P}_\sigma$. 
Factoring out some conformal factors similar to \eqref{eqn:generalCA} and \eqref{eqn:translation} as
\ie
\tilde{\mathcal{A}}^{12\leftrightarrow 34}_{\Delta_i,\ell_i;m,J}(z_i,\bar{z}_i)&=\frac{\Big(\frac{z_{14}}{z_{13}}\Big)^{h_3-h_4}\Big(\frac{z_{24}}{z_{14}}\Big)^{h_1-h_2}\Big(\frac{\bar{z}_{14}}{\bar{z}_{13}}\Big)^{\bar{h}_3-\bar{h}_4}\Big(\frac{\bar{z}_{24}}{\bar{z}_{14}}\Big)^{\bar{h}_1-\bar{h}_2}}{z_{12}^{h_1+h_2}z_{34}^{h_3+h_4}\bar{z}_{12}^{\bar{h}_1+\bar{h}_2}\bar{z}_{34}^{\bar{h}_3+\bar{h}_4}}
\\
&\quad\times(z-1)^{\frac{\Delta_1-\Delta_2-\Delta_3+\Delta_4}{2}}\delta(iz-i\bar{z})\Phi^{12\leftrightarrow 34}_{m,J}(\mathbf{\Delta},\ell_i,z)\,,
\fe
\eqref{eqn:P^2onA} reduces to the differential equation on the total dimension $\bf\Delta$ and the cross ratio $z$
\ie\label{eqn:P^2Casimir}
&-4 e^{2\partial_{\bf \Delta}} \Phi^{12\leftrightarrow 34}_{m,J}({\bf\Delta},\ell_i,z)=-m^2\Phi^{12\leftrightarrow 34}_{m,J}({\bf\Delta},\ell_i,z)\,.
\fe
Similarly, \eqref{eqn:W^2onA} gives the differential equation
\ie
&\left[{({1\over 4}\ell_{12+34}^2-4)z^2+(10-\ell_{12}\ell_{34})z-6\over z-1}+(3z-4)z\partial-(z-1)z^2\partial^2\right]\Phi^{12\leftrightarrow 34}_{m,J}({\bf\Delta},\ell_i,z)
\\
&=J(J+1)\Phi^{12\leftrightarrow 34}_{m,J}({\bf\Delta},\ell_i,z)\,.
\fe
Let us briefly comment on the Casimir equations for the massless Poincar\'e partial waves. By the same derivation, we find the Casimir operator $P^\mu P_\mu$ gives the differential equation \eqref{eqn:P^2Casimir} with $m^2=0$. However, the differential operator on the right hand side of \eqref{eqn:P^2Casimir} is invertible. Hence, the massless Poincar\'e partial waves should be zero identically.

One could repeat the above analysis for the celestial amplitude in the $13\leftrightarrow 24$ and $14\leftrightarrow 23$ kinematics, and derive the corresponding differential equations for the Poincar\'e partial waves $\Phi^{13\leftrightarrow 24}_{m,J}({\bf\Delta},\ell_i,z)$ and $\Phi^{14\leftrightarrow 23}_{m,J}({\bf\Delta},\ell_i,z)$. The solutions to the Casimir equations in the three kinematics \eqref{KRonequater} are (again 
$\beta=\mathbf{\Delta}-4$)
\ie\label{eqn:poincarepartialwave}
\Phi_{m,J}^{\ell_i}(\beta,z)=\left(\frac{m}{2}\right)^{\beta}\sqrt{\frac{(2J{+}1)}{m}}\begin{cases}
B^J_{12,34}z^{2{-}\ell_{12}}(z{-}1)^{\frac{\ell_{12{-}34}}
{2}}\mathcal{J}_{J{-}\ell_{12}}^{\ell_{12{+}34},\ell_{12{-}34}}\Big(\frac{z{-}2}{z}\Big)&(z\ge1)\,,
\\
B^J_{13,24}z^{\frac{\beta{+}\ell_{13{+}24}}{2}{+}2}(1{-}z)^{\frac{\ell_{13{-}24}}{2}}\mathcal{J}_{J{-}\ell_{13}}^{\ell_{13{+}24},\ell_{13{-}24}}(1{-}2z)&(1\ge z\ge0)\,,
\\
B^J_{14,23}\frac{(-z)^{\frac{\beta+\ell_{14+23}}{2}+2}}{(1-z)^{\frac{\beta}{2}+\ell_{23}}}\mathcal{J}_{J-\ell_{23}}^{\ell_{14+23},\ell_{14-23}}\left(\frac{z+1}{z-1}\right)&(0\ge z)\,.
\end{cases}
\fe
One see exactly that the Poincar\'e partial waves \eqref{eqn:poincarepartialwave} are identical to the expansion basis in \eqref{eqn:ImCA} in each kinematics threshold up to normalization factors.

Thus combining everything, we conclude that the imaginary part of the celestial amplitude has a positive expansion on the massive Poincar\'e partial wave, namely,
\begin{equation}\label{eqn:PPWE}
    \textbf{Im}\,\Psi(\beta,\ell_i,z)=\sum_{a} p_{a}  \Phi_{m_a, J_a}(\beta,\ell_i,z) , \quad p_a>0\,.
\end{equation}
Note that since the Jacobi polynomials are orthogonal polynomials, the Poincar\'e partial waves also satisfy orthogonality relations \cite{Law:2020xcf}, and the coefficients $ p_{a}$ are unique. We emphasize that the above is valid for tree-level exchanges and massive internal loops. Once simply do the trivial translation in eq.(\ref{TtoL}). The generalization to supersymmetric theories is straightforward. Once again the expansion polynomial is given by the projection of the susyspinning polynomials on to the celestial sphere. These polynomials are again Jacobi polynomials, but with the helicities indicating that of the highest weighted state in a susy multiplet. For details see~\cite{Liu:2020fgu}.


\paragraph{Crossing symmetry:}
Let us consider the implications of crossing symmetry on the celestial amplitude. For a crossing symmetric amplitude, we would expect the three kinematic region of the celestial amplitude to be related to each other. Indeed, we have:  
\ie\label{Crossing}
\begin{cases}
\text{$s\leftrightarrow u$}\quad\Psi^{13\leftrightarrow24}(\beta, z)=z^{{\beta\over2}+4} \,\Psi^{12\leftrightarrow34}(\beta, {1\over z})& (1\ge z\ge 0)\,,
 \\
\text{$s\leftrightarrow t$}\quad\Psi^{12\leftrightarrow34}(\beta, z)=\left(\frac{z}{1-z}\right)^{{\beta\over 2}+2}\Psi^{14\leftrightarrow23}(\beta, {1-z})&( z\ge1)\,,
\\
\text{$u\leftrightarrow t$}\quad\Psi^{13\leftrightarrow24}(\beta, z)=(z-1)^2\Psi^{14\leftrightarrow23}(\beta, \frac{z}{z-1})& (1\ge z \ge 0)\,.
\end{cases}
\fe
for the celestial amplitude of any permutation invariant theory. Furthermore, since the residue of each channel must respect the exchange symmetry of the legs on one side of the factorization, this implies that the imaginary part of $\Psi(\beta, z)$ must further satisfy the following ``self-dual" like identities:
\ie\label{SelfDual}
\begin{cases}
 \textbf{Im}\,\Psi^{13\leftrightarrow24}(\beta, z)=\left(\frac{z}{1-z}\right)^2\textbf{Im}\,\Psi^{13\leftrightarrow24}(\beta,1-z)& (1\ge z \ge 0)\,,
\\
\textbf{Im}\,\Psi^{14\leftrightarrow23}(\beta, z)=(-1)^{\beta\over2}z^{{3\beta\over2}+4}\textbf{Im}\,\Psi^{14\leftrightarrow23}(\beta,{1\over z})& (0\ge z)\,,
\\
\textbf{Im}\,\Psi^{12\leftrightarrow34}(\beta, z)=(z-1)^2\textbf{Im}\,\Psi^{12\leftrightarrow34}\left(\beta,\frac{z}{z-1}\right)& (z\ge1)\,.
\end{cases}
\fe
It is straightforward to verify that the Poincar\'e partial waves in \eqref{eqn:poincarepartialwave} respect the crossing equations (\ref{Crossing}) and (\ref{SelfDual}).


\section{Open and closed string celestial amplitudes}
\label{sec:string}
We have already seen how the imaginary part of the celestial amplitude encodes the information of poles and branch cuts on complex $\omega$ plane. Furthermore, as discussed in Section \ref{sec:AnalyticContinuation}, the integrand can be analytic continued to other unphysical $z$ domains. However, the UV convergence of the integral relies on certain regions of $\omega$ plane.  In this section, we will first use type-I and II superstring amplitudes:\footnote{Here we have neglected the $F^4$ and $R^4$ pre-factor for simplicity. Their effect can be incorporated straightforwardly. }
\eq
\textrm{type{-}I}:\quad \frac{\Gamma[{-}s]\Gamma[{-}t]}{\Gamma[1{+}u]},\quad \quad \textrm{type{-}II}:\quad \frac{\Gamma[{-}s]\Gamma[{-}t]\Gamma[{-}u]}{\Gamma[1{+}s]\Gamma[1{+}u]\Gamma[1{+}t]}\,,
\eqe
 as examples to determine the convergent regions on $\omega$ plane for all $z$ dependance. As indicated in Figure~\ref{fig:openstringconvergent}, they are in general $z$-dependent for open strings while enjoys $z$-independent for closed strings. Then we will discuss how bulk-locality encoded in the imaginary part of tree-level celestial string amplitude in physical $z$ domains. Finally, when we discuss the celestial dispersion relation in Section \ref{sec:CelestialDispersionRelation} we will verify those relations with celestial string amplitudes in Section \ref{sec:CelestialDispersionString}.

\subsection{Convergent regions in string theory}
As is widely known, in the fixed-angle scatterings, string amplitudes are expected to decay exponentially at large real $\omega$. However, the suppression is in general not true in the whole complex $\omega$ plane. Therefore in this sub-section we will study this problem for open and closed string amplitudes. Writing $\omega=re^{i\theta}$, the $r\to\infty$ limit of string amplitudes behave as:
\begin{equation}
T(r^2e^{2i\theta},z)\sim\exp\Big[g(\theta,z)r^2+\mathcal{O}(\log r)\Big]\,.
\end{equation}
Hence, for given $z$, the amplitude is suppressed in the limit $r\to\infty$ when
\ie\label{eqn:Reg<0}
\textbf{Re}\left[g(\theta,z)\right]<0\,,
\fe
which, as one will see, gives the convergent region
\ie\label{eqn:convergent_region_string}
\theta\in(-\theta_{c},\theta_{c})\cup(\pi-\theta_c,\pi+\theta_c)\,,
\fe
where the angle $\theta_{c}$ is the root of $\textbf{Re}[g(\theta,z)]=0$ when $\theta\in(0,{\pi\over 2})$. Let us consider the open and closed string separately,
\begin{itemize}
\item {\bf Open string}
\newline The gluon amplitude in the type-I superstring theory is
\begin{equation}\label{eqn:openT}
T_{\rm open}(s,u)=\frac{\Gamma(-s)\Gamma(-t)}{\Gamma(1-s-t)}\,.
\end{equation}
In the $12\to34$ kinematic region $(z>1)$, and restricting to $\theta\in(0,{\pi\over 2})$, we have
\ie\label{eqn:openrlarge}
g(\theta,z)=
e^{2 i \theta}\frac{\left[(z{-}1) (\log (z-1){+}i \pi ){-}z \log (z)\right]}{z}\,.
\fe
The inequality \eqref{eqn:Reg<0} gives the following equation for the angle $\theta_c$,
\begin{equation}
\cot(2\theta_{c})=\frac{\pi(1-z)}{z\log z-(z-1)\log(z-1)}\,.
\end{equation}
The angle $\theta_c$ as a function of $z$ is plotted in figure~\ref{fig:openstringconvergent}. One can see that starting from $z=1$, the angle $\theta_c$ is gradually increasing from $\frac{\pi}{4}$ to $\frac{\pi}{2}$ as $z\to\infty$.
\begin{figure}
\centering
\includegraphics[width=0.5\textwidth]{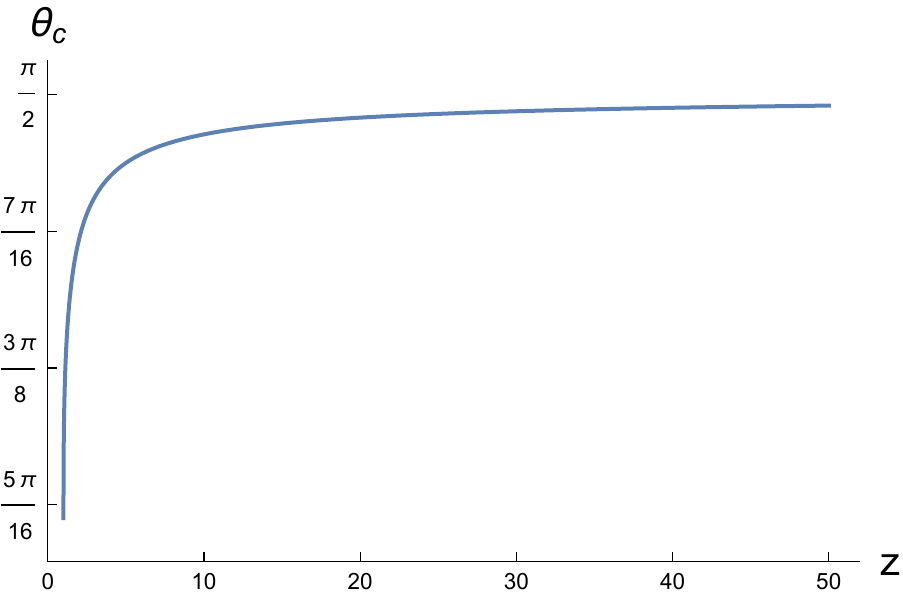}
\caption{The angle $\theta_c$ for the open string amplitude in the $12\leftrightarrow 34$ kinematics as a function of $z$.}
\label{fig:openstringconvergent}
\end{figure}
\newline
Also one can perform the same analysis for general angle $\theta\in(-\pi,\pi)$, and find that the convergent regions in the other quadrants are also specified by the same angle $\theta_c$ and given by \eqref{eqn:convergent_region_string}. One can further repeat the same analysis for the other kinematics $13\leftrightarrow 24$ and $14\leftrightarrow 23$. The results are summarized in Table~\ref{tab:Stringregions}.


\begin{table}[H]
 \begin{tabular}{|c|c|c|} 
  \hline
 $\theta_{c}^{12\leftrightarrow 34}(z)\quad (z\ge1)$ & $\theta_{c}^{13\leftrightarrow 24}(z)\quad (1\ge z\ge0)$ & $\theta_{c}^{14\leftrightarrow 23}(z)\quad (0\ge z)$  \\
 \hline
$\frac{1}{2}\cot^{-1}\Big[\frac{\pi(1-z)}{z\log z-(z-1)\log(z-1)}\Big]$ & ${\pi\over 4}$ & $\frac{1}{2}\cot^{-1}\Big[\frac{\pi z}{(1-z)\log(1-z)+z\log(-z)}\Big]$ \\ 
 \hline
\end{tabular}
\caption{Characteristic angles in open string.}
\label{tab:Stringregions}
\end{table}

\item {\bf Closed string}
\newline The graviton amplitude in the type-II superstring is 
\begin{equation}\label{eqn:closedT}
T_{\rm closed}(s,u)=\frac{\Gamma(-s)\Gamma(-t)\Gamma(-u)}{\Gamma(1+s)\Gamma(1+t)\Gamma(1+u)}
\end{equation}
It turns out that the large $r$ behavior of the closed string amplitude is much simpler than the open string case. In the $12\leftrightarrow34$ kinematic region $(z\ge1)$, we find
\begin{equation}
g(\theta,z)=2e^{2 i \theta }\frac{(z-1)\log(z-1)-z\log z}{z},\quad(0<\theta<\pi/2)\,.
\end{equation}
The condition for the real part to be negative is
\begin{equation}
\cos(2\theta)>0\Rightarrow0<\theta<\frac{\pi}{4}\,.
\end{equation}
We can repeat the same analysis for general angle $\theta\in(-\pi,\pi)$ and for the other kinematics $13\leftrightarrow 24$ and $14\leftrightarrow 23$. The convergent regions for all cases are given by \eqref{eqn:convergent_region_string} with the angle $\theta_c$ equals to $\pi\over 4$.

\end{itemize}

\subsection{$\textbf{Im}\,\Psi(\beta,z)$ in string theory}
Taking $\beta\in\mathbb{R}$, from \eqref{eqn:imaginary}, $\textbf{Im}\,\Psi(\beta,z)$ is given as a sum of the residues of the poles (and also an integral of the discontinuity) near the positive real axis. For both open and closed string amplitudes, the poles near the positive real axis are located at $\omega=\sqrt{n}$ for $n\in{\mathbb Z}_{>0}$, and we have
\begin{equation}
\textbf{Im}\,\Psi(\beta,z)=-\pi B(z)\sum_{n=1}^\infty
(\sqrt{n})^{\beta{-}1}\underset{\omega_n=\sqrt{n}}{\bf Res}\Big[T(\omega_n,z)\Big]\,\quad(\beta>\beta_*)
\end{equation}
In the following, we will verify this formula by directly comparing the truncated sum with the numerical result of the Mellin integration:
\begin{itemize}
\item {\bf Open String}
\newline The massive poles of the open string amplitude \eqref{eqn:openT} are located at $s,t=1,2,\dots,n$. And the corresponding residues for $s$-channel are,
\begin{equation}
\textbf{Res}_{s=n}[T_{\text{open}}(s,t)]=\frac{\prod_{k=1}^{n-1}(t+k)}{n!}\,.
\end{equation}
The residues of the $t$-channel poles are given by exchanging $s$ with $t$. Now we can compute $\textbf{Im}\Psi(\beta,z)$. In $12\to34$ kinematic region, using eq.\eqref{eqn:imaginary}, the imaginary part will only receive contribution from $s$-channel massive poles located on the positive-$\omega$ axis at $\omega_n=\sqrt{n}$ with the residue
\begin{equation}
\underset{\omega=\sqrt{n}}{\bf Res}\Big[T^{12\to34}_{\text{open}}\Big(\omega^2,-\frac{1}{z}\omega^2\Big)\Big]=\frac{\left[n \left(\frac{1}{z}-1\right)+1\right]_{n-1}}{2 \sqrt{n} n!}\,.
\end{equation}
Thus, the imaginary part of the open string celestial amplitude is
\begin{align}\label{eqn:im1234open}
\textbf{Im}\,\Psi^{12\leftrightarrow 34}_{\text{open}}(\beta,z)&=-\pi B^{12\to34}(z)\sum_{n=1}^\infty (\sqrt{n})^{\beta-1}\underset{\omega=\sqrt{n}}{\bf Res}\Big[T^{12\to34}_{\text{open}}\Big(\omega^2,-\frac{1}{z}\omega^2\Big)\Big]\notag\\
&=-\pi 2^{3-\beta}z^2\sum_{n=1}^\infty(\sqrt{n})^{\beta-2}\frac{\left[n \left(\frac{1}{z}-1\right)+1\right]_{n-1}}{2 n!}\quad(z>1)\,,
\end{align}
One can perform the same computation for $14\to23$, the result is,
\begin{align}
\textbf{Im}\,\Psi^{14\leftrightarrow 23}_{\text{open}}(\beta,z)&=-\pi B^{14\to23}(z)\sum_{n=1}^\infty (\sqrt{n})^{\beta-1}\underset{\omega=\sqrt{n}}{\bf Res}\Big[T^{14\to23}_{\text{open}}\Big(\frac{z}{1-z}\omega^2,-\frac{1}{z-1}\omega^2\Big)\Big]\notag\\
&=-\pi2^{3-\beta}(-z)^{\frac{\beta}{2}+2}(1-z)^{-\frac{\beta}{2}}\sum_{n=1}^\infty(\sqrt{n})^{\beta-2}\frac{\left(\frac{n z}{1-z}+1\right)_{n-1}}{2 n!}\quad(z<0)\,.
\end{align}
Note that $\textbf{Im}\,\Psi^{13\leftrightarrow 24}_{\text{open}}(\beta,z)=0$ because the open string amplitude $T_{\rm open}(s,t)$ has no $u$-channel pole. The comparison of the truncated sum of eq.\eqref{eqn:im1234open} with the result from the numerical Mellin integral \eqref{Main} is given in Figure~\ref{fig:Imopen}. One can see that $\textbf{Im}\,\Psi^{12\leftrightarrow 34}_{\text{open}}(\beta,z)$ diverges in the limits $z\to\infty$ and $z\to1$. The divergent at $z\to\infty$ is due to the overall prefactor $B^{12\to34}(z)=\pi z^2$, while the divergent at $z\to1$ is due to the non-convergence of the residue sum reflecting the massless $1/t$ singularity in that limit. Note that the divergence occurs for positive $\beta$. For negative $\beta$ the imaginary part is finite as we will see in Section~\ref{eqn:stringOPElimit}.



\begin{figure}[htbp]
\centering    
\begin{minipage}{7cm}
	\centering          
	\includegraphics[scale=0.45]{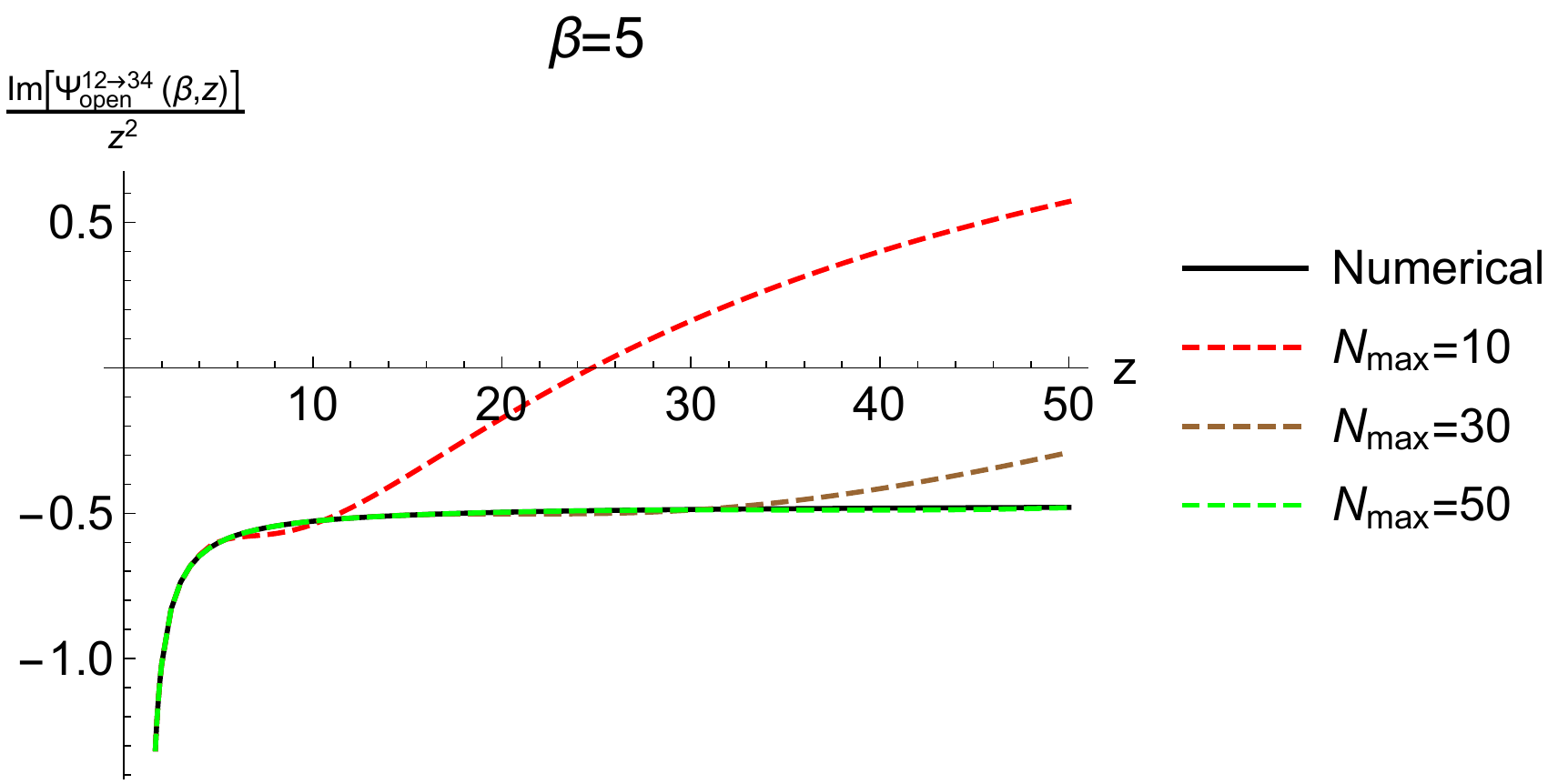}   	\end{minipage}
	\qquad
\begin{minipage}{7cm}
	\centering      
	\includegraphics[scale=0.45]{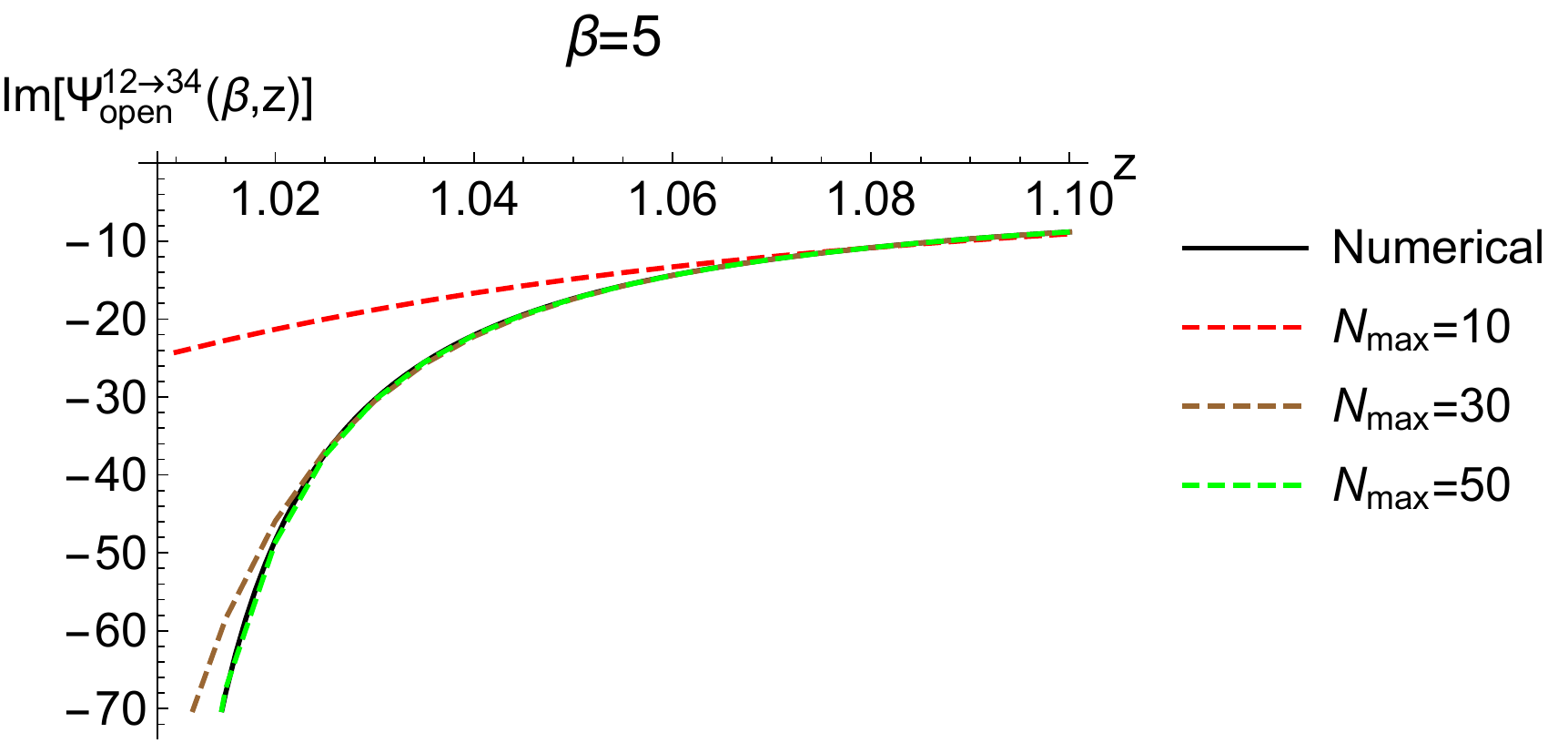}   	\end{minipage}
\caption{Here we present the comparison of numerical evaluation of Mellin integral and residue sum formula eq.\eqref{eqn:im1234open} for the imaginary part of celestial open string amplitude. LHS and RHS shows the behavior of $\textbf{Im}\,\Psi^{12\leftrightarrow 24}_{\text{open}}(\beta,z)$ in $z\to\infty$ and $z\to1$} 
\label{fig:Imopen} 
\end{figure}

\item {\bf Closed String}
\newline
In the closed string case, the residues at $s=1,2,\dots n$ are
\begin{equation}
\textbf{Res}_{s=n}[T_{\text{closed}}(s,t)]=\frac{\prod_{k=1}^{n-1}(t+k)^2}{(n!)^2}
\end{equation}
Again, let us focus on the $12\leftrightarrow 34$ kinematics and the region $z\ge 1$. Using eq.\eqref{eqn:imaginary}, the residue of the pole at $\omega=\sqrt{n}$ is
\begin{equation}
\underset{\omega=\sqrt{n}}{\bf Res}\,\Big[T_{\rm closed}(\omega^2,\frac{1-z}{z}\omega^2)\Big]=\frac{\left[\left(1+\frac{1-z}{z}n\right)_{n-1}\right]^2}{2\sqrt{n}n (n!)^2}\,.
\end{equation}
Thus, the imaginary part of the massive contribution is given by
\begin{equation}\label{eqn:imclosed}
\textbf{Im}\,\Psi_{\text{closed}}^{12\leftrightarrow34}(\beta,z)=-\pi 2^{3-\beta}z^2\sum_{n=1}^\infty\frac{(\sqrt{n})^{\beta-2}}{2}\left(\frac{\left[n \left(\frac{1}{z}-1\right)+1\right]_{n-1}}{n!}\right)^2
\end{equation}
The comparison with result from the numerical integral is given in figure~\ref{fig:Imclosed}. Similar with the open string, the divergences in the $z\to1$ and $z\to\infty$ limits are implied by the massless $t,u$ poles.
\newline
Using the crossing symmetry, we find the closed string celestial amplitudes for the other two kinematics
\ie\label{eqn:closedResSumOther2}
\textbf{Im}\,\Psi_{\text{closed}}^{13\leftrightarrow24}(\beta,z)&=-\pi2^{3-\beta}z^{\frac{\beta}{2}+2}\sum_{n=1}^\infty\frac{(\sqrt{n})^{\beta-2}}{2}\left[\frac{\left(\frac{n}{z-1}+1\right)_{n-1}}{n!}\right]^2\,,
\\
\textbf{Im}\,\Psi_{\text{closed}}^{14\leftrightarrow23}(\beta,z)&=-\pi2^{3-\beta}(-z)^{\frac{\beta}{2}+2}(1-z)^{-\frac{\beta}{2}}\sum_{n=1}^\infty\frac{(\sqrt{n})^{\beta-2}}{2}\left[\frac{\left(\frac{n z}{1-z}+1\right)_{n-1}}{n!}\right]^2\,.
\fe
\begin{figure}[htbp]
\centering    
\begin{minipage}{7cm}
	\centering          
	\includegraphics[scale=0.45]{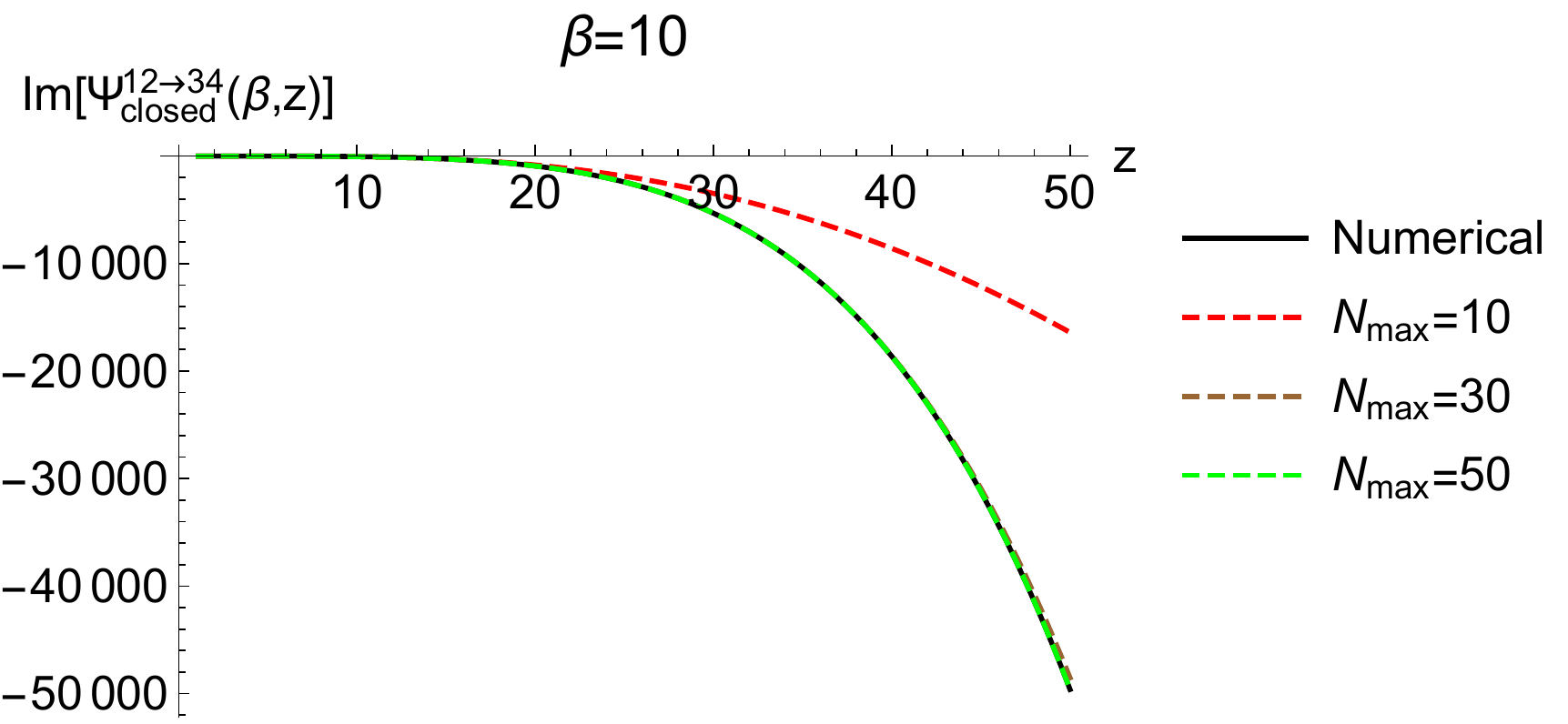}   	\end{minipage}
	\qquad
\begin{minipage}{7cm}
	\centering      
	\includegraphics[scale=0.45]{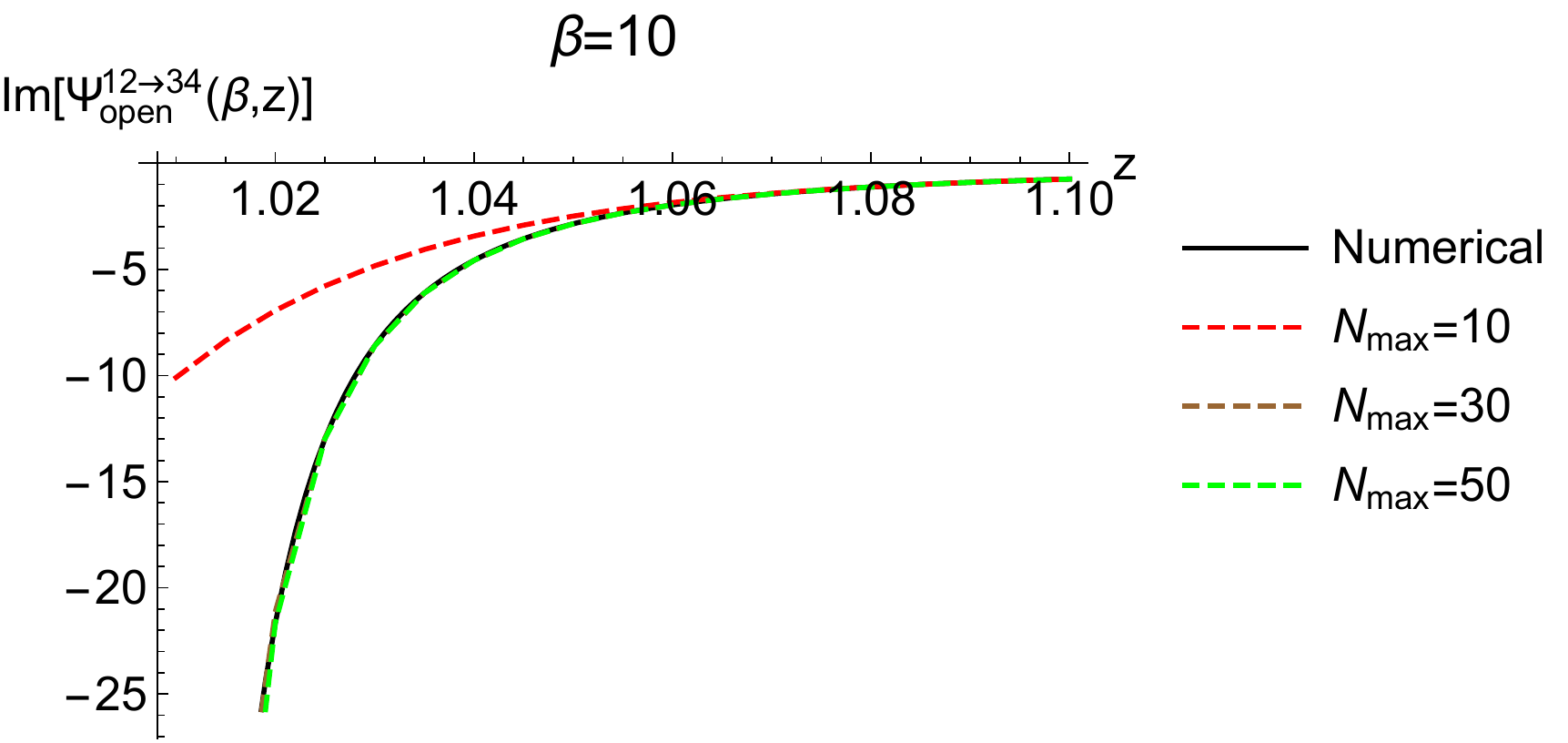}   	\end{minipage}
\caption{Here we present the comparison of numerical evaluation of Mellin integral and residue sum formula eq.\eqref{eqn:imclosed} for the imaginary part of celestial closed string amplitude. LHS and RHS shows the behavior of $\textbf{Im}\,\Psi^{12\leftrightarrow 24}_{\text{closed}}(\beta,z)$ in $z\to\infty$ and $z\to1$} 
\label{fig:Imclosed} 
\end{figure} 
\end{itemize}

\section{Analytic continuation in $z$}
\label{sec:AnalyticContinuation}
As reviewed in sec.\ref{sec:2}, the celestial four-point amplitude consists of three different functions defined in three distinct regions on the real circle of the complex sphere. One can ask if the three functions can be analytically connected to each other. Consider the simple massive scalar exchange example, and compare eq.(\ref{eqn:FullScalarCA1234}) with 
\ie
\Psi^{13\leftrightarrow24}(\beta,z)&={\pi g^2\over \sin {\pi \beta\over 2}}\left(\frac{m}{2}\right)^{\beta-2} z^{2}\left[1{+}e^{{1\over 2}\pi i \beta}z^{\beta\over 2}{+}\left(z\over 1-z\right)^{{\beta\over 2}}\right]&& (1\ge z \ge 0)\,,
\\
\Psi^{14\leftrightarrow23}(\beta,z)&={\pi g^2\over \sin {\pi \beta\over 2}}\left(\frac{m}{2}\right)^{\beta-2} z^{2}\left[1{+}\left(-z\right)^{\beta\over2}+e^{{1\over 2}\pi i \beta}\left(\frac{-z}{1-z}\right)^{\beta\over2}\right]&& (0\ge z)\,.
\fe
It is straight forward to see that the functions cannot be analytically related.

There are, however, reasons we might want to consider the analytic continuation of a celestial amplitude from the physical region as specified in Table~\ref{tab:kinematics} to the unphysical ones. Recall that the presence of bulk factorization is imprinted in the imaginary part of of celestial sphere, where in each kinematic region only the channel with the physical threshold is projected. Thus for fixed $z$, to access the information of all factorization channels through the imaginary part, we will need to analytic continue $\Psi$ of other kinematic configurations to their unphysical regions. Such analytic continuation will be important  when we consider dispersion relations for the celestial amplitude in Section~\ref{sec:CelestialDispersionRelation}.

\subsection{The scalar example}\label{sec:ScalarExchange}
Before turning to the general analysis, we will use the massive scalar exchange example to illustrate the last point. To continue $\Psi^{12\leftrightarrow34}_{\rm scalar}(\beta,z)$ outside the physical region $z>1$, we need to be careful about potential monodromies as we continue across the boundaries at $z=1$ and $z=\infty$. Continuing across $z=1$ by $z-1\to e^{-\pi i}(1-z)$ and across $z=\infty$ by $z\to  e^{\pi i}(-z)$, we obtain
\ie\label{eqn:scalar1234AC}
\Psi^{12\leftrightarrow34}_{\rm scalar}(\beta,z)&={\pi g^2 \over \sin{\pi\beta\over 2}}\left(m\over 2\right)^{\beta-2} z^{2}\begin{cases}e^{{1\over 2}\pi i\beta}+z^{{\beta\over 2}}+e^{{1\over 2}\pi i\beta}\left(z\over 1-z\right)^{{\beta\over 2}}& (1> z> 0)\,,
\\
e^{{1\over 2}\pi i\beta}+e^{{1\over 2}\pi i\beta} (-z)^{{\beta\over 2}}+\left(z\over z-1\right)^{{\beta\over 2}}& (0> z)\,.
\end{cases}
\fe
The imaginary part of $\Psi_{\rm scalar}^{12\leftrightarrow34}(\beta,z)$ in the new (unphysical) regions are
\ie\label{AnaIm}
\textbf{Im}\,\Psi^{12\leftrightarrow34}_{\text{scalar}}(\beta,z)&=\pi g^2\left(m\over 2\right)^{\beta-2}z^2 \begin{cases}
1+\left(z\over 1-z\right)^{\beta\over 2}&(1> z> 0)\,,
\\
1+\left(-z\right)^{\beta\over 2}\hspace{2.7375cm}{}&(0> z)\,,
\end{cases}
\\
&=2^{3-\beta}\pi z^2 g^2\begin{cases}
\underset{\omega\to m, \,\sqrt{z\over 1-z}m}{\bf Res}\Big[\omega^{\beta-1}T^{12\leftrightarrow34}_{\text{scalar}}(\omega,z)\Big]&(1> z> 0)\,,
\\
\underset{\omega\to m,\,\sqrt{-z}m}{\bf Res}\Big[\omega^{\beta-1}T^{12\leftrightarrow34}_{\text{scalar}}(\omega,z)\Big]&(0> z)\,.
\end{cases}
\fe
We see that the analytically continued result receives contribution from the $s$- and $t$-channel poles in $1> z> 0$ (the ``$u$"-region), and $s$- and $u$-channel poles in $0> z$ (the ``$t$"-region). Note that the analytic continuation prescription here is uniquely fixed by demanding that the analytic continued amplitudes match irrespective of which physical region it originated from. More precisely,
\ie
&\Psi^{13\leftrightarrow24}_{\text{scalar}}(\beta,z) \propto\Psi^{14\leftrightarrow23}_{\text{scalar}}(\beta,z) && (z> 1)\,,
\\
&\Psi^{14\leftrightarrow23}_{\text{scalar}}(\beta,z) \propto\Psi^{12\leftrightarrow34}_{\text{scalar}}(\beta,z) && (1> z> 0)\,,
\\
&\Psi^{12\leftrightarrow34}_{\text{scalar}}(\beta,z) \propto\Psi^{13\leftrightarrow24}_{\text{scalar}}(\beta,z) && (0> z)\,,
\fe
where the proportionality coefficients being $z$-independent phases. Explicit comparison fixes the phase to be $(1,e^{-{1\over 2}i\pi \beta}, e^{{1\over 2}i\pi \beta})$ respectively.

\subsection{General analysis}
\label{sec:analytic_continuation_general}
Having gone through the scalar exercise, we are now ready to consider the continuation for general amplitudes.  At the level of the integrand, the analytic continuation on $z$ corresponds to the analytic continuation on the Mandelstam variables from the physical to the ``unphysical" kinematic regions, as shown in Table~\ref{tab:regions}:
\begin{table}[H]
\begin{center}
 \begin{tabular}{|c|c|c|c|} 
  \hline
 & $12\leftrightarrow 34$ & $13\leftrightarrow 24$ & $14\leftrightarrow 23$  \\
 \hline
 $z> 1$ & $s> 0> u,t$ & $u,t>0> s$ & $u,t>0> s$ \\ 
 \hline
$1> z> 0$ &  $s,t> 0> u$ & $u>0> s,t$ & $s,t>0> u$ \\ 
 \hline
$0> z$ &  $s,u> 0> t$ & $s,u>0> t$ & $t>0> s,u$ \\ 
 \hline
\end{tabular}
\end{center}
\caption{The physical and ``unphysical" kinematic regions.}
\label{tab:regions}
\end{table}
The physical regions are those that have only one positive Mandelstam, sitting on the diagonal entries. Take $\Psi^{12\leftrightarrow 34}$ as an example. The analytic continuation to unphysical regions in $z$, say $0> z$, leads to an unphysical kinematic setup ($s,u> 0> t$) and is distinct from $\Psi^{14\leftrightarrow 23}$ originated from the physical setup $t>0> s,u$. Thus it is natural that the three distinct functions on the equator cannot be analytically continued to each other. On the other hand, continuing $\Psi^{13\leftrightarrow24}$ and $\Psi^{14\leftrightarrow23}$ to their common unphysical region, i.e. $z> 1$, do yield the same Mandelstam region  ($u,t>0> s$) and it is natural to identify their analytic continuation. The precise phase can be read off from known amplitudes. Thus we impose the following conditions on the analytic continued celestial amplitudes
\ie\label{eqn:AnalyCon}
&\Psi^{13\leftrightarrow24}(\beta,z) = \Psi^{14\leftrightarrow23}(\beta,z) && (z> 1)\,,
\\
&\Psi^{14\leftrightarrow23}(\beta,z) =e^{-{1\over 2}i\pi \beta} \Psi^{12\leftrightarrow34}(\beta,z) && (1> z> 0)\,,
\\
&\Psi^{12\leftrightarrow34}(\beta,z) = e^{{1\over 2}i\pi \beta}  \Psi^{13\leftrightarrow24}(\beta,z) && (0> z)\,,
\fe
where the phases $e^{\pm{1\over 2}i\pi\beta}$ are due to analytic continuing the prefactor $B(\beta,z)$ in \eqref{Main}. These conditions could be viewed as the crossing symmetry for celestial amplitudes.\footnote{We thank the anonymous SciPost referee for emphasizing this point.} Indeed they are satisfied by the massive scalar exchange. The above is only valid for massless scalar external states. For massless spinning states, crossing will be more complicated since in general it will be related to different helicity sectors. We leave this to future work.

\begin{figure}[t]
\centering
\begin{tikzpicture}[scale=1]

\draw [fill={rgb, 255:red, 204; green, 229; blue, 255}  ,fill opacity=0.4, draw opacity=0 ]  (3,2) -- (0,0) -- (3,-2) ;
\draw [fill={rgb, 255:red, 204; green, 229; blue, 255 }  ,fill opacity=0.4, draw opacity=0 ] (-3,2) -- (0,0) -- (-3,-2) ;

\draw [->, color=gray] (0,-3) -- (0,3);
\draw [->, color=gray] (-3,0) -- (3,0);

\node[circle, opacity=1, fill, inner sep=.75pt] at (1.2,-.1) {};
\node[circle, opacity=1, fill, inner sep=.75pt] at (-1.2,.1) {};

\node[circle, color= blue, opacity=1, fill, inner sep=.75pt] at (-.1,1) {};
\node[circle, color= red, opacity=1, fill, inner sep=.75pt] at (-.1,2) {};

\node[circle, color= blue, opacity=1, fill, inner sep=.75pt] at (.1,-1) {};
\node[circle, color= red, opacity=1, fill, inner sep=.75pt] at (.1,-2) {};

\draw [line,->-=.55] (0,0) to ( 3, 0 );

\node at (2,0.3) {$I_1$};

\end{tikzpicture}
$\quad\quad\quad$
\begin{tikzpicture}[scale=1]

\draw [fill={rgb, 255:red, 255; green, 204; blue, 204  }  ,fill opacity=0.4, draw opacity=0 ]  (2,3) -- (0,0) -- (-2,3)  ;
\draw [fill={rgb, 255:red, 255; green, 204; blue, 204 }  ,fill opacity=0.4, draw opacity=0 ]   (2,-3) -- (0,0) -- (-2,-3)   ;

\draw [->, color=gray] (0,-3) -- (0,3);
\draw [->, color=gray] (-3,0) -- (3,0);

\node[circle, opacity=1, fill, inner sep=.75pt] at (1.2,-.1) {};
\node[circle, opacity=1, fill, inner sep=.75pt] at (-1.2,.1) {};

\node[circle, color= blue, opacity=1, fill, inner sep=.75pt] at (-.1,1) {};
\node[circle, color= red, opacity=1, fill, inner sep=.75pt] at (2,-.1) {};

\node[circle, color= blue, opacity=1, fill, inner sep=.75pt] at (.1,-1) {};
\node[circle, color= red, opacity=1, fill, inner sep=.75pt] at (-2,.1) {};

\draw [line,-<-=.55] (0,0) to ( 1.75, 3 );
\draw [line,->-=.55] (0,0) to ( 1.75, -3 );

\node at (.75,2) {$I_2$};
\node at (.75,-2) {$I_2'$};

\end{tikzpicture}

\caption{The blue (pink) regions is the convergent angular region $\Theta(z)$ for $z=1+\epsilon>1$ ($ z= 1-\epsilon<1$). The black, blue or red dot is a representative for the $s$-, $t$- or $u$-channel poles at $\omega = \pm m$, $\omega=\pm i\sqrt{z\over z-1}m_i$ or $\omega=\pm i\sqrt{z}m_i$, respectively.}
\label{fig:AC}
\end{figure}
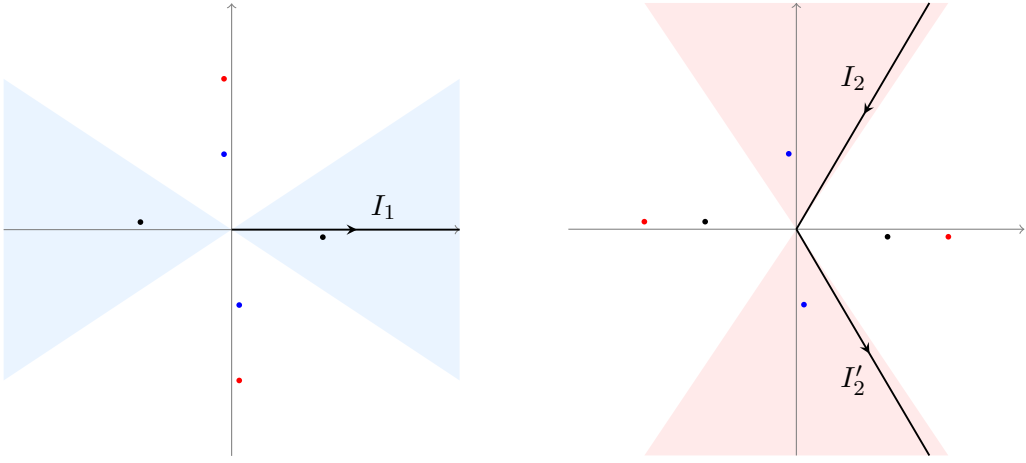

Note that to analytic continue, we need to ensure that the celestial amplitude exists along the path of continuation. More precisely, the celestial amplitude $\Psi(\beta,z)$ given by the integral of the flat-space amplitude $T(\omega,z)$ must be convergent along the path. As an example, consider continuing $\Psi^{12\leftrightarrow 34}$ from its physical region $z> 1$ to $1> z> 0$. More explicitly, our path of continuation is given by 
\ie\label{eqn:analytic_continutaiton_path}
&P_1:~\text{from}~\text{any}~z>1~\text{to}~z=1+\epsilon\,,
\\
&P_2:~z=1+\epsilon e^{-i\phi}~\text{for}~\phi~\text{from}~0~\text{to}~\pi\,,
\\
&P_3:~\text{from}~z=1-\epsilon~\text{to any}~z\in(0,1)\,,
\fe
where $\epsilon>0$ is a small number.
Previously  in Section~\ref{sec:ImPart}, we introduced the boundedness condition around \eqref{eqn:theta}. For convenience, we restate the condition at here:
\ie
\lim_{|\omega|\to 0} T^{12\leftrightarrow 34}(\omega,z)=0\quad{\rm for}\quad z>1
\fe
when
\ie\label{eqn:Thetaz>1}
&\arg\omega\in \Theta^{12\leftrightarrow 34}(z)\equiv\big(-\theta^{12\leftrightarrow 34}_c(z),\theta^{12\leftrightarrow 34}_c(
z)\big)
\\
&\hspace{3.5cm}\cup \big(\pi-\theta^{12\leftrightarrow 34}_c(z),\pi+\theta^{12\leftrightarrow 34}_c(z)\big)\,,
\fe
where we have emphasized the $z$ dependence by writing $\theta^{12\leftrightarrow 34}_c(z)$. 
Now, let us consider the region $1>z>0$. Using the identity $T^{12\leftrightarrow 34}(\omega,z) = T^{13\leftrightarrow 24}({i\omega\over \sqrt{z}},z)$, we find that $T^{12\leftrightarrow 34}(\omega,z)$ vanishes as $|\omega|\to \infty$ with the argument
\ie\label{eqn:Theta1>z>0}
&\arg\omega\in \Theta^{12\leftrightarrow 34}(z)\equiv  \Big({\pi\over 2}-\theta^{13\leftrightarrow 24}_c(z),{\pi\over 2}+\theta^{13\leftrightarrow 24}_c(
z)\Big)
\\
&\hspace{3.5cm}\cup \Big(-{\pi\over 2}-\theta^{13\leftrightarrow 24}_c(z),-{\pi\over 2}+\theta^{13\leftrightarrow 24}_c(
z)\Big) \quad{\rm for}\quad 1>z>0\,.
\fe
In general, the region \eqref{eqn:Theta1>z>0} does not contain the angle $\arg\omega=0$. Hence, we need to deform the integration contour continuously along the way of the analytic continuation $P_2$ in \eqref{eqn:analytic_continutaiton_path}. For such continuous contour deformation to exist, the convergent region $\Theta(z)$ should exist and continuously move from \eqref{eqn:Thetaz>1} to \eqref{eqn:Theta1>z>0} when going along the path $P_2$. This is illustrated in figure~\ref{fig:AC}, where the region $\Theta(z)$ for $z>1$ ($1>z>0$) is drawn as the blue (pink) region. For the massive scalar exchange the amplitude is convergent for all $\arg\omega\in[0,2\pi)$, and one do not need to deform the integration contour. For string theory amplitudes, the blue region ($\Theta^{12\leftrightarrow 34}(z)$ for $z>1$) is computed in sec.\ref{sec:string} and the pink region ($\Theta^{12\leftrightarrow 34}(z)$ for $1>z>0$) can be obtained by the formula \eqref{eqn:Theta1>z>0} using the $\theta^{13\leftrightarrow 24}_c$ given in Table~\ref{tab:Stringregions}. 
As we will see in sec.\ref{sec:analytic_continuation_string} the convergent region $\Theta(z)$ deforms continuously from the blue to the pink regions when going along the path $P_2$. 

With this understanding, we can now carry out the analytic continuation. Begin with $\Psi^{12\leftrightarrow 34}$ in the physical region $z> 1$ defined by the Mellin integral \eqref{Main}, which is an integral along the contour $I_1$ in figure~\ref{fig:AC}. 
When going along the path \eqref{eqn:analytic_continutaiton_path}, the integration contour should deform accordingly, such that it is always inside the corresponding angular region $\Theta(z)$.  When arriving the destination $z\in (0,1)$, the integration contour becomes $-I_2$ in figure~\ref{fig:AC}. Finally, under the analytic continuation, the $t$-channel poles (represented by the red dots in figure~\ref{fig:AC}) are rotated from the imaginary axis to the real axis, while the $s$- and $u$-channel poles (represented by the black and blue dots in figure~\ref{fig:AC}) remain near the real and the imaginary axes, respectively.

Now, let us consider the imaginary part of the celestial amplitude $\textbf{Im}\,\Psi^{12\leftrightarrow 34}(\beta,z)$ after analytically continuing to $1> z > 0$. Similar to the discussion in Section~\ref{sec:ImPart}, it can be computed by the integral along the contour $I_2+I_2'$ as shown in figure~\ref{fig:AC}. However, in this case, we cannot close the contour by adding the $I_3'$ piece as in figure~\ref{Fig:CScontour}, because the $I_3'$ contour extends outside the pink angular region, and the contour integral diverges. When applied to the case of the massive scalar exchange amplitude studied in Section~\ref{sec:massive_scalar_exchange}, we have a better asymptotic property that
\ie\label{eqn:BestBoundedC}
\Theta(z)=[0,2\pi)\quad \forall z\in{\mathbb C}\,.
\fe
In such case, $\textbf{Im}\,\Psi^{12\leftrightarrow 34}(\beta,z)$ in the region $1> z> 0$ can also be computed by the fan-like contour $I_2+I_2'+I_3'$, which picks up the residues and discontinuities of the poles and branch cuts inside the contour. The result is
\ie\label{eqn:imaginaryAC}
&\textbf{Im}\,\Psi^{12\leftrightarrow 34}_{>\Lambda}(\beta, z)
\\
&=-B^{12\leftrightarrow 34}(z)
\pi \sum_{i}\underset{\omega\to m_i,\,\sqrt{z\over 1-z}m_i}{\bf Res}\Big[\omega^{\beta{-}1}T^{12\leftrightarrow 34}(\omega,z)\Big]
\\
&\quad+\left(\int_{M}^\infty+\int_{\sqrt{z\over 1-z} M}^\infty\right)\mathrm{d}\omega\,\omega^{\beta{-}1}\,\textbf{Disc}\,\left[T^{12\leftrightarrow 34}(\omega,z)\right]
\quad (1> z>0)\,.
\fe
We see that the imaginary part of \eqref{eqn:scalar1234AC} indeed agrees with the formula \eqref{eqn:imaginaryAC}.


\section{Celestial dispersion relation}
\label{sec:CelestialDispersionRelation}
The analyticity properties of a scattering amplitude $T(s,t)$ can be summarized by dispersion relations. When the Mandelstam variable $u=u_0$ is fixed with the absolute value $|u_0|$ smaller than the multi-particle production threshold $M$, the dispersion relation for the scattering amplitude reads\footnote{We assume the asymptotic behavior of the amplitude as $T(s,-s-u_0)\to 0$ when $|s|\to\infty$. For amplitudes with worse asymptotic behavior like $T(s,-s-u_0)\to |s|^N$ when $|s|\to\infty$, similar dispersion relations can be written down with $N$ unknown coefficients known as subtraction constants.}
\ie\label{eqn:dispersion}
T(s,t)&=-\sum_i {g^2_i\over s-m_i^2}-\sum_i {g^2_i\over t-m_i^2}
\\
&\quad+ {1\over \pi}\int^{\infty}_{M^2}ds' {{\bf Disc}_s\, \left[T(s',t) \right]\over s' - s}+{1\over \pi}\int_{M^2}^{\infty}dt'{{\bf Disc}_t\, \left[T(s,t') \right]\over t' - t}\,,
\fe
where the discontinuity ${\bf Disc}_s[T(s,t)]$ is defined by
\ie
{\bf Disc}_s[T(s,t)]={1\over 2i}\left(T(s+i\epsilon,t)-T(s-i\epsilon,t)\right)\,,
\fe
and similarly for ${\bf Disc}_t[T(s,t)]$. By unitarity, they are positively expandable on appropriate orthogonal polynomials. Along the line of S-matrix bootstrap, the dispersion relations have been used recently to constrain the EFT coefficients -- the expansion coefficients of the amplitude $T(s,t)$ at $s=u=t=0$ \cite{Arkani-Hamed:2020blm,Bellazzini:2020cot,Tolley:2020gtv,Caron-Huot:2020cmc,Caron-Huot:2021rmr}.

On the celestial sphere, on one hand, as we have seen in Section~\ref{sec:masless 4point scalars}, the poles and discontinuities that appear on the right hand side of the dispersion relation \eqref{eqn:dispersion} contribute to the imaginary part of the celestial amplitude. On the other hand, as elucidated in~\cite{Arkani-Hamed:2020gyp} and also discussed in Section~\ref{sec:UV&IR}, the EFT coefficients show up as the residues of the poles in the celestial amplitude on the complex $\beta$-plane at negative even integers. Therefore, it is natural to expect a relation between the residues of the poles at $\beta=-2n$ for $n\in{\mathbb Z}_{\le0}$ and the imaginary part of the celestial amplitude, given by the projection of the dispersion relation \eqref{eqn:dispersion} on the celestial sphere. However, the dispersion relation \eqref{eqn:dispersion} cannot be directly translated to the celestial sphere by the Mellin transform \eqref{Main}, since for generic $z$, the $\omega$-integral extends to the region where all the Mandelstam variables are large and the dispersion relation \eqref{eqn:dispersion} does not hold. Keeping $u=u_0$ fixed and finite corresponds to the OPE limit (colinear limit) of the celestial amplitude, for example, the $z\to\infty$ limit in the $12\leftrightarrow 34$ kinematics.

From the celestial amplitude of the massive scalar exchange \eqref{eqn:FullScalarCA1234}, we observe the following relation between the residues of the poles at negative even integer $\beta$ and the imaginary part of the celestial amplitude
\ie\label{eqn:celestialdispersion}
{\pi\over 2}\,\underset{\beta \to -2n}{\bf Res}\left[\Psi_{\text{scalar}}^{12\leftrightarrow 34}(\beta,z)\right]=\textbf{Im}\,\left[\Psi_{\text{scalar}}^{12\leftrightarrow34}(\beta,z)+(-1)^n\Psi_{\text{scalar}}^{13\leftrightarrow24}(\beta,z)\right]\Big|_{\beta\to-2n}&&(z\ge 1)\,,
\fe
and similar relations in the other regions. Importantly, the celestial amplitude $\Psi_{\text{scalar}}^{13\leftrightarrow 24}(-2n,z)$ on the right hand side is given by the analytic continuation from the physical defining region $1> z > 0$ to the region $z> 1$, as discussed in Section~\ref{sec:ScalarExchange}. Note that we could have chosen the analytic continuation of $\Psi_{\text{scalar}}^{14\leftrightarrow23}$ instead, since  from eq.(\ref{eqn:AnalyCon}), $\Psi_{\text{scalar}}^{13\leftrightarrow24}(\beta,z)=\Psi_{\text{scalar}}^{14\leftrightarrow23}(\beta,z)$ when $z\ge 1$.

More general celestial amplitudes satisfy a similar celestial dispersion relation,
\ie\label{eqn:CDRconj}
{\pi\over 2}\,\underset{\beta \to -2n}{\bf Res}\left[\Psi^{12\leftrightarrow 34}(\beta,z)\right]=\textbf{Im}\,\left[\Psi^{12\leftrightarrow34}(\beta,z)+(-1)^n \Psi^{13\leftrightarrow24}(\beta,z)\right]\Big|_{\beta\to-2n}&&(z\ge 1)\,.
\fe
In the above, we take $n\geq1$ if gravity is non-dynamical and $n>1$ when graviton exchange are involved. The celestial amplitude $\Psi^{13\leftrightarrow 24}(-2n,z)$ on the right hand side of \eqref{eqn:CDRconj} is given by the analytic continuation from the physical region $1\ge z \ge 0$ to the region $z\ge 1$, as discussed in Section~\ref{sec:AnalyticContinuation}.

\begin{figure}[t]
\centering
\begin{tikzpicture}[scale=1]
\node at (-1.5, 1.5) {$\omega$-plane};
\draw [->, color=gray] (0,-2) -- (0,2);
\draw [->, color=gray] (-3,0) -- (3,0);
\draw [->, color=gray,snake=zigzag, line after snake=1mm, segment amplitude=.5mm] (1.4,-.1) -- (3,-.1);
\node[draw, cross out, inner sep=1pt, thick] at (1.4,-.1) {};
\node[circle, opacity=1, fill, inner sep=.75pt] at (0.8,-.1) {};
\node[circle, opacity=1, fill, inner sep=.75pt] at (1,-.1) {};
\node[circle, opacity=1, fill, inner sep=.75pt] at (1.2,-.1) {};

\draw [->, color=gray,snake=zigzag, line after snake=1mm, segment amplitude=.5mm] (-1.4,.2) -- (-3,.2);
\node[draw, cross out, inner sep=1pt, thick] at (-1.4,.2) {};
\node[circle, opacity=1, fill, inner sep=.75pt] at (-0.8,.2) {};
\node[circle, opacity=1, fill, inner sep=.75pt] at (-1,.2) {};
\node[circle, opacity=1, fill, inner sep=.75pt] at (-1.2,.2) {};

\draw [->, color=gray,snake=zigzag, line after snake=1mm, segment amplitude=.5mm] (-.1,1.2) -- (-.1,2);
\node[draw, cross out, inner sep=1pt, thick] at (-.1,1.2) {};
\node[circle, opacity=1, fill, inner sep=.75pt] at (-.1,0.6) {};
\node[circle, opacity=1, fill, inner sep=.75pt] at (-.1,0.8) {};
\node[circle, opacity=1, fill, inner sep=.75pt] at (-.1,1) {};
\draw [->, color=gray,snake=zigzag, line after snake=1mm, segment amplitude=.5mm] (.1,-1.2) -- (.1,-2);
\node[draw, cross out, inner sep=1pt, thick] at (.1,-1.2) {};
\node[circle, opacity=1, fill, inner sep=.75pt] at (.1,-0.6) {};
\node[circle, opacity=1, fill, inner sep=.75pt] at (.1,-0.8) {};
\node[circle, opacity=1, fill, inner sep=.75pt] at (.1,-1) {};

\draw [->, color=gray,snake=zigzag, line after snake=1mm, segment amplitude=.5mm] (0,0) -- (-3,0);
\draw [line,->-=.55] (0,0.1) to ( -3, 0.1 );
\draw [line,-<-=.55] (0,-0.1) to ( -3, -0.1 );
\draw [line] (0,-0.1) arc(-90:90:0.1) ;

\node at (0,-2.88) {};
\end{tikzpicture}
$\quad\quad\quad$
\begin{tikzpicture}[scale=1]
\draw [->, color=gray] (0,-2) -- (0,2);
\draw [->, color=gray] (-3,0) -- (3,0);
\draw [->, color=gray,snake=zigzag, line after snake=1mm, segment amplitude=.5mm] (1.4,-.1) -- (3,-.1);
\node[draw, cross out, inner sep=1pt, thick] at (1.4,-.1) {};
\node[circle, opacity=1, fill, inner sep=.75pt] at (0.8,-.1) {};
\node[circle, opacity=1, fill, inner sep=.75pt] at (1,-.1) {};
\node[circle, opacity=1, fill, inner sep=.75pt] at (1.2,-.1) {};
\draw [->, color=gray,snake=zigzag, line after snake=1mm, segment amplitude=.5mm] (-1.4,.1) -- (-3,.1);
\node[draw, cross out, inner sep=1pt, thick] at (-1.4,.1) {};
\node[circle, opacity=1, fill, inner sep=.75pt] at (-0.8,.1) {};
\node[circle, opacity=1, fill, inner sep=.75pt] at (-1,.1) {};
\node[circle, opacity=1, fill, inner sep=.75pt] at (-1.2,.1) {};
\draw [->, color=gray,snake=zigzag, line after snake=1mm, segment amplitude=.5mm] (-.1,1.2) -- (-.1,2);
\node[draw, cross out, inner sep=1pt, thick] at (-.1,1.2) {};
\node[circle, opacity=1, fill, inner sep=.75pt] at (-.1,0.6) {};
\node[circle, opacity=1, fill, inner sep=.75pt] at (-.1,0.8) {};
\node[circle, opacity=1, fill, inner sep=.75pt] at (-.1,1) {};
\draw [->, color=gray,snake=zigzag, line after snake=1mm, segment amplitude=.5mm] (.1,-1.2) -- (.1,-2);
\node[draw, cross out, inner sep=1pt, thick] at (.1,-1.2) {};
\node[circle, opacity=1, fill, inner sep=.75pt] at (.1,-0.6) {};
\node[circle, opacity=1, fill, inner sep=.75pt] at (.1,-0.8) {};
\node[circle, opacity=1, fill, inner sep=.75pt] at (.1,-1) {};

\draw [line,->-=.55,red] (.5,0) to[distance=.35cm,out=90,in=-180] (2.99,0.26);
\draw [line,-<-=.55,red] (.5,0) to[distance=.35cm,out=-90,in=-180] (2.99,-0.26);
\draw [line,-<-=.55,blue] (-.5,0) to[distance=.35cm,out=90,in=0] ( -2.99, 0.26 );
\draw [line,->-=.55,blue] (-.5,0) to[distance=.35cm,out=-90,in=0] ( -2.99, -0.26 );
\draw [line,->-=.5,green] ( 2.99, 0.2615 ) arc(5:45:3) ;
\draw [line,->-=.5,green] ( -2.12, 2.12 ) arc(135:175:3) ;
\draw [line,->-=.5,green] ( -2.99, -0.2615 ) arc(185:225:3) ;
\draw [line,->-=.5,green] ( 2.12, -2.12 ) arc(-45:-5:3) ;

\draw [line,->-=.5,purple] (2.12, 2.12) to (0,.3);
\draw [line,-<-=.5,purple] (-2.12, 2.12) to (0,.3);
\draw [line,->-=.5,brown] (-2.12, -2.12) to (0,-.3);
\draw [line,->-=.5,brown] (2.12, -2.12) to (0,-.3);

\draw [line] (0,0) circle (4pt) ;
\draw [-<-=.58] (.05,.14) ;
\node at (.325,0.225) {${\cal C}_0$};
\node at (2.3,2.3) {${\cal C}_0'$};
\draw [line,->=.58] (.87,.5) to (1.3,.75);
\draw [line,->=.58] (-.87,.5) to (-1.3,.75);
\draw [line,->=.58] (.87,-.5) to (1.3,-.75);
\draw [line,->=.58] (-.87,-.5) to (-1.3,-.75);
\node[circle, opacity=1, fill, inner sep=.75pt] at (0,0) {};
\end{tikzpicture}
\caption{{\bf Left:} The contour ${\cal C}_{(-\infty,0]}$. {\bf Right:} Pulling the contour ${\cal C}_0$ to ${\cal C}_0'$, which further decomposes into the red contour ${\cal D}_0$, the blue contour ${\cal D}_1$, the purple contour ${\cal D}_2$, the brown contour ${\cal D}_3$, and finally the green contours.}
\label{Fig:C0contour}
\end{figure}
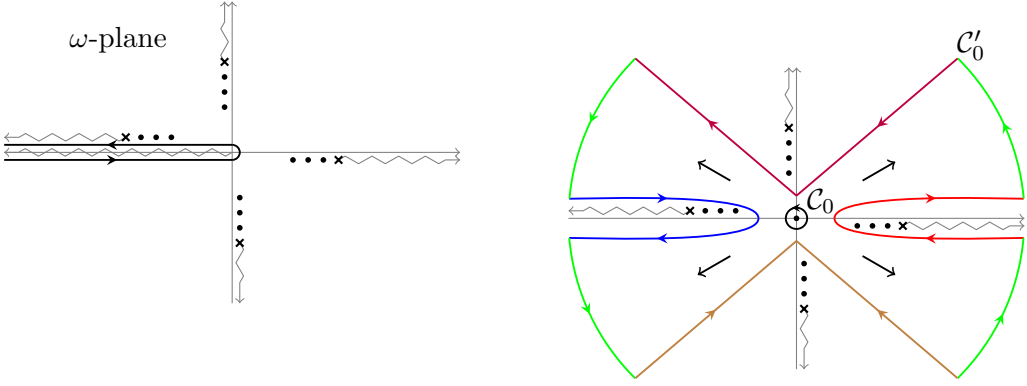

Let us now derive \eqref{eqn:CDRconj}. First we note that the original Mellin integral \eqref{Main}  can be written as a contour integral along the negative real axes:
\ie
\Psi^{12\leftrightarrow 34}(\beta,z)
={1\over 2i}\csc(\pi\beta) B^{12\leftrightarrow 34}(z)\oint_{{\cal C}_{(-\infty,0]}} d\omega\, \omega^{\beta-1}T^{12\leftrightarrow 34}(\omega,z)\,,
\fe
where the contour ${\cal C}_{(-\infty,0]}$ is a thin counterclockwise contour along the negative real axes that picks up the discontinuity across the branch cut associated with the $\omega^{\beta-1}$ factor, as shown in the left of figure~\ref{Fig:C0contour}. Note that we've used $i\epsilon$ to push the cuts associated with $T$ on to the complex $\omega$ plane. This identity can be understood by noting that 
\ie
{\bf Disc}[\omega^{\beta-1}]={1\over 2i}\left[(|\omega|e^{i\pi})^{\beta-1}-(|\omega|e^{-i\pi})^{\beta-1}\right]=-\sin(\pi\beta)|\omega|^{\beta-1}\quad{\rm when}\quad \omega\in(-\infty,0]\,.
\fe
via. change of variable, we recover the original Mellin integral. This formula makes the pole structure of the celestial amplitude manifest as the $\csc(\pi\beta)$ pre-factor, and the contour integral along ${\cal C}_{(-\infty,0]}$ converges on the entire $\beta$-plane. With this formula, the celestial amplitude $\Psi(\beta,z)$ can be analytically continued to $\beta\le0$. The pre-factor $\csc(\pi\beta)$ has simple poles at $\beta\in{\mathbb Z}_{\le 0}$, for which $\omega^{\beta-1}$ yields a (multi)pole at $\omega=0$. Thus when computing the residue for $\beta\in{\mathbb Z}_{\le 0}$, the contour ${\cal C}_{(-\infty,0]}$ is contracted to a small circle ${\cal C}_0$ centered at the origin. We then arrive at\footnote{The residues of the poles at $\beta=-2n-1$ all equal to zero, because $T(\omega,z)$ is an even function in $\omega$.}
\ie
\underset{\beta \to -2n}{\bf Res}\left[\Psi^{12\leftrightarrow 34}(\beta,z)\right]={1\over2\pi i}B(z)\oint_{{\cal C}_0} d\omega\,\omega^{-2n-1}T^{12\leftrightarrow 34}(\omega,z)\,.
\fe

To proceed, let us impose on $T^{12\leftrightarrow 34}(\omega,z)$ the boundedness condition introduced previously around \eqref{eqn:theta} and also the following condition:

\paragraph{Analyticity:} The amplitude $T^{12\leftrightarrow 34}(\omega,z)$ is analytic inside the fan-shaped contour ${\cal C}_0'$ as shown in figure~\ref{Fig:C0contour}, with the extension angle determined by $\theta^{12\leftrightarrow 34}_c$ in \eqref{eqn:theta}.

The poles and branch cuts located near the real or imaginary axis on the $\omega$-plane correspond to the physical thresholds, i.e. the exchange of single or multi-particle states. At loop levels, there are potentially also poles or branch cuts from anomalous thresholds located away from the real or imaginary axis. The analyticity assumption is amount to forbid their appearance inside ${\cal C}_0'$. 
Now, let us pull the contour ${\cal C}_0$ to the ${\cal C}_0'$. Due to the boundedness assumption, the arcs of the contour ${\cal C}_0'$ at infinite (the green contours) do not contribute to the integral. 

The remaining part of the contour ${\cal C}_0'$ decomposes into four pieces: the red contour ${\cal D}_0$, the blue contour ${\cal D}_1$, the purple contour ${\cal D}_2$, and the brown contour ${\cal D}_3$. The contour integral along ${\cal D}_1$ can be simply related to the contour integral along ${\cal D}_0$ and using \eqref{eqn:imaginary}, we get
\ie
\frac{\pi z^2}{2\pi i}\int_{{\cal D}_1}\mathrm{d}\omega\,\omega^{-2n-1}T^{12\to34}(\omega,z)&=e^{-2n\pi i}\frac{\pi z^2}{2\pi i}\int_{{\cal D}_0}\mathrm{d}\omega\,\omega^{-2n-1}T^{12\to34}(\omega,z)
\\
&=\frac{1}{\pi}\mathbf{Im}[\Psi^{12\to34}(\beta,z)]\,.\label{eqn:Im1234int}
\fe
Next, performing the change of variable $\omega=i\sqrt{z}\omega'$, we find:
\begin{equation}
\frac{\pi z^2}{2\pi i}\int_{{\cal D}_2+{\cal D}_3}\mathrm{d}\omega\,\omega^{-2n-1}T^{12\to34}(\omega,z)=\frac{(-1)^n}{2i}z^{2-n}\int_{{\cal D}_2'+{\cal D}_3'}\mathrm{d}\omega'(\omega')^{-2n-1}T^{13\to24}(\omega',z)\,,
\end{equation}
where ${\cal D}_2'$ and ${\cal D}_3'$ are the contours given by rotating the contours ${\cal D}_2$ and ${\cal D}_3$ clockwise around origin by $\pi/2$. The contour integral along ${\cal D}_3'$ can be simply related to the contour integral along ${\cal D}_2'$, which equals to the $I_2 + I_2'$ in Figure~\ref{fig:AC} contour that computes the imaginary part of the analytic continued celestial amplitude. More explicitly, we have
\ie
\frac{z^{2-n}\pi}{2i}\int_{{\cal D}_3'}\mathrm{d}\omega'(\omega')^{-2n-1}T^{13\to24}(\omega,z)&=e^{-2\pi n i}\frac{z^{2-n}\pi}{2i}\int_{{\cal D}_2'}\mathrm{d}\omega'(\omega')^{-2n-1}T^{13\to24}(\omega,z)
\\
&=\mathbf{Im}[\Psi^{13\to24}(\beta,z)]\,.
\label{eqn:Im1324z>1}
\fe
Putting everything together, we find the celestial dispersion relation  \eqref{eqn:CDRconj}.

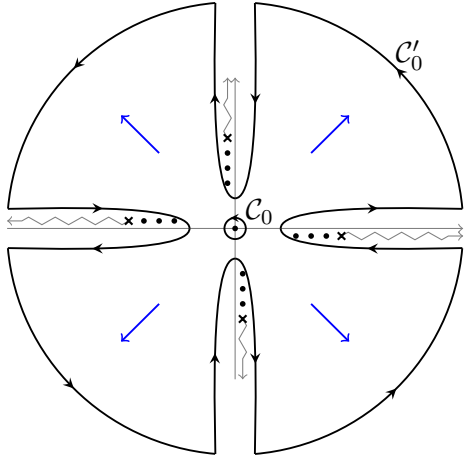
\begin{figure}[t]
\centering
\begin{tikzpicture}[scale=1]
\draw [->, color=gray] (0,-2) -- (0,2);
\draw [->, color=gray] (-3,0) -- (3,0);
\draw [->, color=gray,snake=zigzag, line after snake=1mm, segment amplitude=.5mm] (1.4,-.1) -- (3,-.1);
\node[draw, cross out, inner sep=1pt, thick] at (1.4,-.1) {};
\node[circle, opacity=1, fill, inner sep=.75pt] at (0.8,-.1) {};
\node[circle, opacity=1, fill, inner sep=.75pt] at (1,-.1) {};
\node[circle, opacity=1, fill, inner sep=.75pt] at (1.2,-.1) {};
\draw [->, color=gray,snake=zigzag, line after snake=1mm, segment amplitude=.5mm] (-1.4,.1) -- (-3,.1);
\node[draw, cross out, inner sep=1pt, thick] at (-1.4,.1) {};
\node[circle, opacity=1, fill, inner sep=.75pt] at (-0.8,.1) {};
\node[circle, opacity=1, fill, inner sep=.75pt] at (-1,.1) {};
\node[circle, opacity=1, fill, inner sep=.75pt] at (-1.2,.1) {};
\draw [->, color=gray,snake=zigzag, line after snake=1mm, segment amplitude=.5mm] (-.1,1.2) -- (-.1,2);
\node[draw, cross out, inner sep=1pt, thick] at (-.1,1.2) {};
\node[circle, opacity=1, fill, inner sep=.75pt] at (-.1,0.6) {};
\node[circle, opacity=1, fill, inner sep=.75pt] at (-.1,0.8) {};
\node[circle, opacity=1, fill, inner sep=.75pt] at (-.1,1) {};
\draw [->, color=gray,snake=zigzag, line after snake=1mm, segment amplitude=.5mm] (.1,-1.2) -- (.1,-2);
\node[draw, cross out, inner sep=1pt, thick] at (.1,-1.2) {};
\node[circle, opacity=1, fill, inner sep=.75pt] at (.1,-0.6) {};
\node[circle, opacity=1, fill, inner sep=.75pt] at (.1,-0.8) {};
\node[circle, opacity=1, fill, inner sep=.75pt] at (.1,-1) {};

\draw [line,->-=.55] (.6,0) to[distance=.35cm,out=90,in=-180] (2.99,0.26);
\draw [line,-<-=.55] (.6,0) to[distance=.35cm,out=-90,in=-180] (2.99,-0.26);
\draw [line,->-=.55] (0,-.4) to[distance=.35cm,out=0,in=90] ( 0.26, -2.99 );
\draw [line,-<-=.55] (0,-.4) to[distance=.35cm,out=180,in=90] ( -0.26, -2.99 );
\draw [line,-<-=.55] (0,.4) to[distance=.35cm,out=0,in=-90] ( 0.26, 2.99 );
\draw [line,->-=.55] (0,.4) to[distance=.35cm,out=180,in=-90] ( -0.26, 2.99 );
\draw [line,-<-=.55] (-.6,0) to[distance=.35cm,out=90,in=0] ( -2.99, 0.26 );
\draw [line,->-=.55] (-.6,0) to[distance=.35cm,out=-90,in=0] ( -2.99, -0.26 );
\draw [line,->-=.5] ( 2.99, 0.2615 ) arc(5:85:3) ;
\draw [line,->-=.5] ( 0.2615, -2.99 ) arc(-85:-5:3) ;
\draw [line,->-=.5] ( -0.2615, 2.99 ) arc(95:175:3) ;
\draw [line,->-=.5] ( -2.99, -0.2615 ) arc(185:265:3) ;

\draw [line] (0,0) circle (4pt) ;
\draw [-<-=.58] (.05,.14) ;
\node at (.325,0.225) {${\cal C}_0$};
\node at (2.3,2.3) {${\cal C}_0'$};
\draw [line,->=.58,blue] (1,1) to (1.5,1.5);
\draw [line,->=.58,blue] (-1,1) to (-1.5,1.5);
\draw [line,->=.58,blue] (1,-1) to (1.5,-1.5);
\draw [line,->=.58,blue] (-1,-1) to (-1.5,-1.5);
\node[circle, opacity=1, fill, inner sep=.75pt] at (0,0) {};
\end{tikzpicture}
\caption{Contour deformation from ${\cal C}_0$ to ${\cal C}_0'$.}
\label{Fig:C0contour1}
\end{figure}

Note that for the amplitude of the massive scalar exchange \eqref{eqn:TforMSE}, since the amplitude vanishes in all direction, we can instead deform the contour to ${\cal C}_0'$ in Figure~\ref{Fig:C0contour1}. Therefore, the integral picks up contributions from the poles and branch cuts near the real axis and gives the imaginary part of $\Psi^{12\leftrightarrow 34}(\beta,z)$ by \eqref{eqn:imaginary}. The contribution from the imaginary axis is
\ie
&-B(z)\left\{\pi \sum_{i}\underset{\omega\to i\sqrt{z}m_i,\,i\sqrt{z\over z-1}m_i}{\bf Res}\Big[\omega^{-2n{-}1}T^{12\leftrightarrow 34}(\omega,z)\Big]\right.
\\
&\quad\quad\quad\quad\left.+\left(\int_{i\sqrt{z}M}^{i\infty}+\int_{i\sqrt{z\over z-1}M}^{i\infty}\right)\mathrm{d}\omega\,\omega^{-2n{-}1}\,\textbf{Disc}\,\left[T^{12\leftrightarrow 34}(\omega,z)\right]\right\}\,,
\fe
which equals to $\textbf{Im}\,[(-1)^n\Psi^{13\leftrightarrow24}(-2n,z)]$ or $\textbf{Im}\,[(-1)^n\Psi^{14\leftrightarrow23}(-2n,z)]$, by the formulae similar to \eqref{eqn:imaginaryAC} with the appropriate changes of the integration variable, $\omega\to i\sqrt{z}\omega$ or $\omega\to i\sqrt{z\over z-1}\omega$.

\subsection{Celestial dispersion relation in string theory}
\label{sec:CelestialDispersionString}
Let us do the numerical check of our conjecture \eqref{eqn:CDRconj} for the celestial dispersion relation, by considering the open and closed string celestial amplitudes. Specifically, we will give the analytic form of the residues at $\beta=-10$ and $\beta=-12$ poles and numerically evaluate $\text{\bf Im}\,\Psi^{12\leftrightarrow34}$ and $\text{\bf Im}\,\Psi^{13\leftrightarrow24}$ through the fan-like integrals as indicated in \eqref{eqn:Im1234int} and \eqref{eqn:Im1324z>1}. Finally, we will plot the results above and show they indeed match.
\begin{itemize}
\item{Numeric results of $\beta=-10$}
    \newline The analytic forms of residues at the $\beta=-10$ poles of open and closed string celestial amplitude are,
    \ie
    &\frac{\pi}{2}\underset{\beta \to -10}{\bf Res}\left[\Psi_{\text{open}}^{12\leftrightarrow 34}(\beta,z)\right]\,
    \\
   &=\frac{512\pi}{45z^3}\left[-\pi^4z(-1+z)(4+7z(-1+z))\zeta(3)-60\pi^2z(-1+z)(1+z(-1+z))\zeta(5)\right.\,
   \\
   &\left.\quad\quad\quad\quad-360(1+z(-1+z))^2\zeta(7)\right]\,,
   \\
    &\frac{\pi}{2}\underset{\beta \to -10}{\bf Res}\left[\Psi_{\text{closed}}^{12\leftrightarrow 34}(\beta,z)\right]=-\frac{16384\pi(-1+z)\left[1+z(-1+z)\right]\zeta(3)\zeta(5)}{z^2}.
    \fe
where $\zeta(x)$ is the Riemann zeta function.
The comparison with the RHS of (\ref{eqn:CDRconj}) is shown in Figure~\ref{fig:matchbeta-10}.
\begin{figure}[htbp]
\centering    
\begin{minipage}{7cm}
	\centering          
	\includegraphics[scale=0.60]{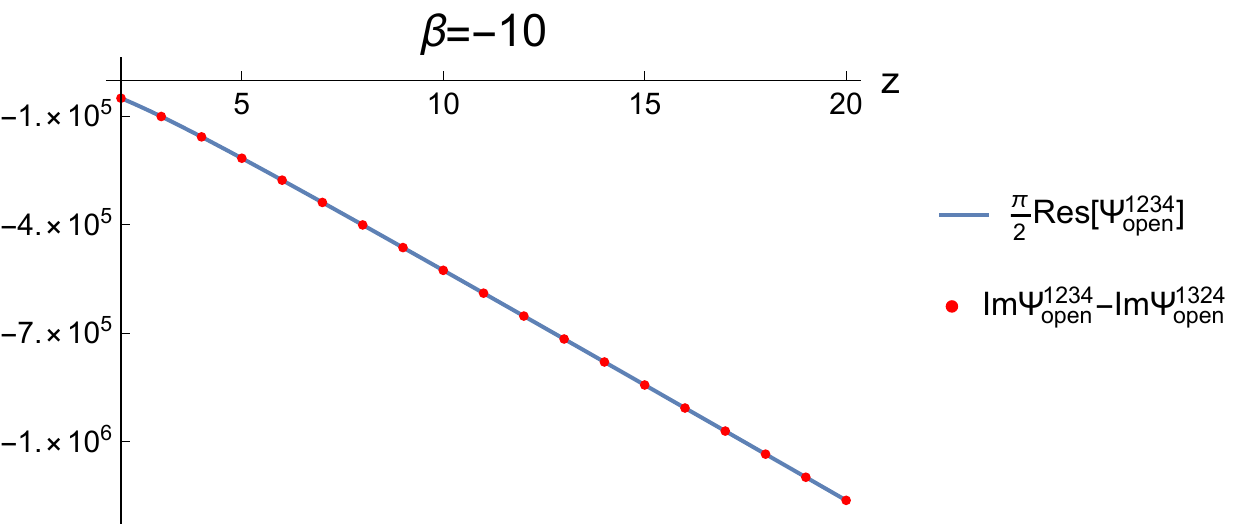}   	\end{minipage}
	\qquad
\begin{minipage}{7cm}
	\centering      
	\includegraphics[scale=0.60]{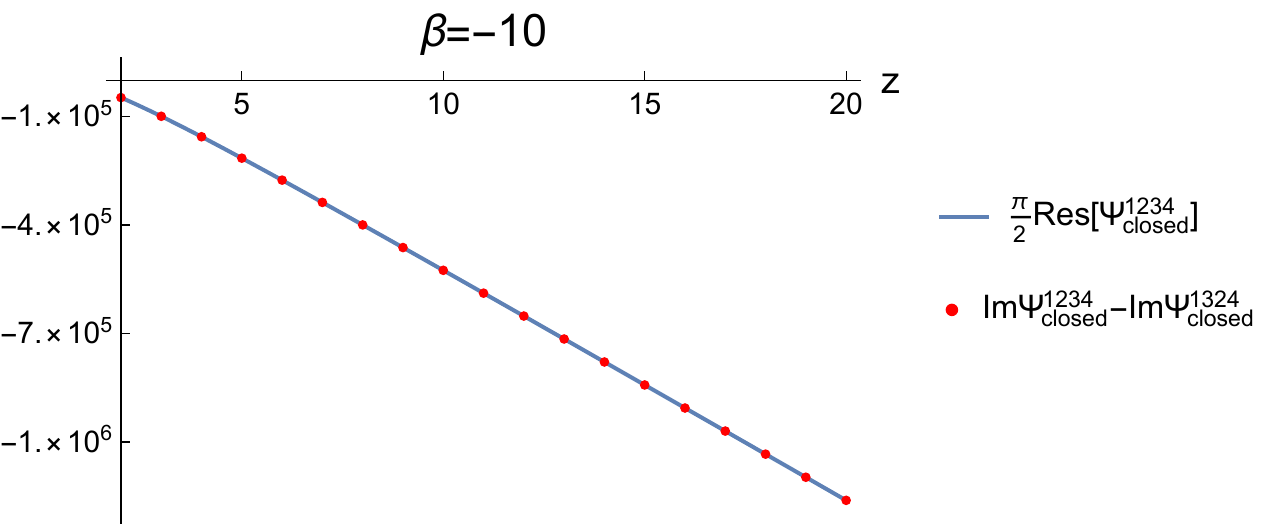}   	\end{minipage}
\caption{The comparison between the analytic formula for the residue at $\beta=-10$ and the numerical evaluation of the imaginary part for the open and closed string celestial amplitudes.
}
\label{fig:matchbeta-10} 
\end{figure} 
 \item{Numeric results of $\beta=-12$}
 \newline
The analytic forms of residues at the $\beta=-12$ poles of open and closed string celestial amplitude are,
\ie
 &\frac{\pi}{2}\underset{\beta \to -12}{\bf Res}\left[\Psi_{\text{open}}^{12\leftrightarrow 34}(\beta,z)\right]\,
    \\
  &=\frac{128\pi}{14175z^4}\left.[\pi^8\left(-192-z(-1+z)(912+z(-1+z)(1256+381z(-1+z)))\right)\,
  \right.\\
  &\quad\quad\quad\quad\quad\left.-151200\pi^2\zeta(3)^2(-1+z)^2z^2-1814400\zeta(3)\zeta(5)z(-1+z)(1+z(-1+z))\right]\,,
\\
&\frac{\pi}{2}\underset{\beta \to -12}{\bf Res}\left[\Psi_{\text{closed}}^{12\leftrightarrow 34}(\beta,z)\right]\,
\\
&=\frac{256\pi}{945z^4}\left[3\psi^{(8)}(1)+9z(-1+z)\left(\psi^{(8)}(1)-4480z(-1+z)(2\zeta(3)^3+(10+3z(-1+z)\zeta(9)))\right)\right]\,,
\fe
where $\psi^{(n)}(z)$ is the polygamma function. The comparison is given in Figure~\ref{fig:matchbeta-12}.

\begin{figure}[htbp]
\centering    
\begin{minipage}{7cm}
	\centering          
	\includegraphics[scale=0.60]{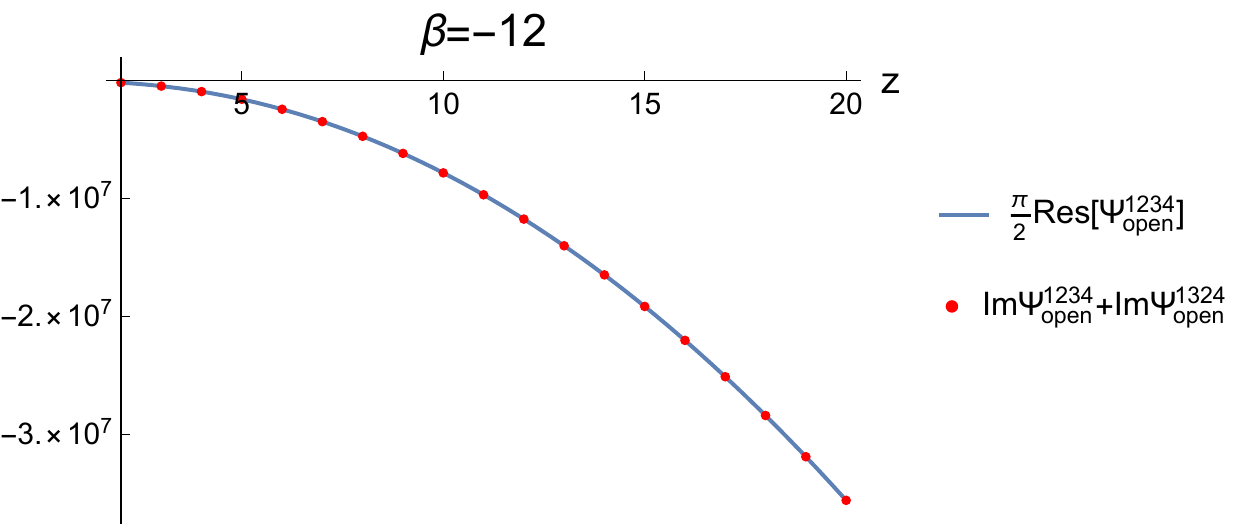}   	\end{minipage}
	\qquad
\begin{minipage}{7cm}
	\centering      
	\includegraphics[scale=0.60]{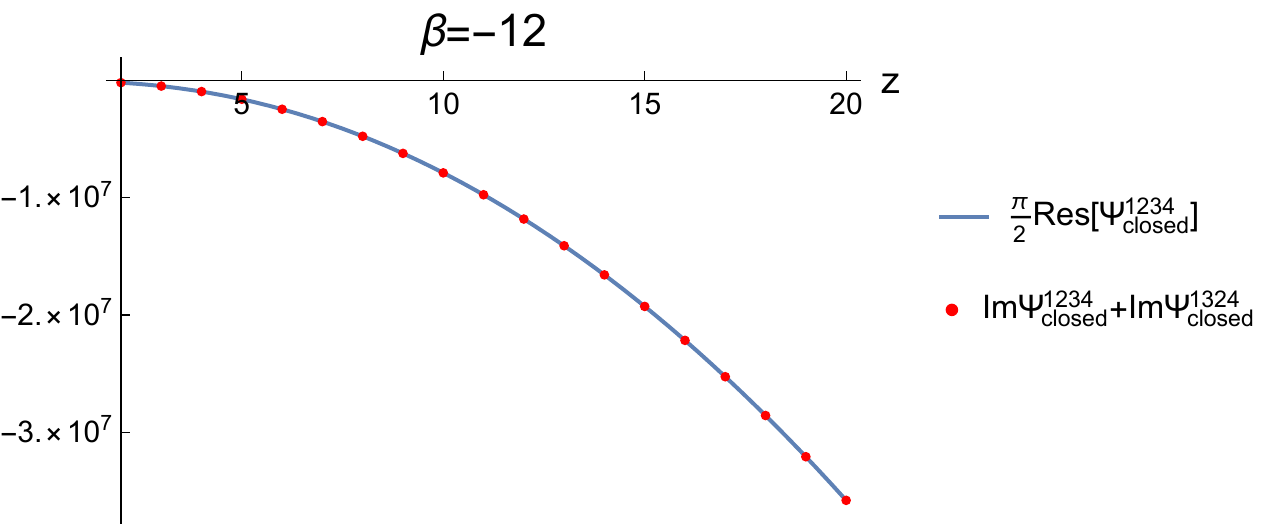}  \end{minipage}
\caption{
The comparison between the analytic formula for the residue at $\beta=-10$ and the numerical evaluation of the imaginary part for the open and closed string celestial amplitudes.
}
\label{fig:matchbeta-12} 
\end{figure} 
\end{itemize}

\subsection{OPE limit}
\label{eqn:stringOPElimit}
In the celestial dispersion relation, on the right hand side, the first term $\textbf{Im}\,\Psi^{12\leftrightarrow 34}(\beta,z)$ equals to a sum over the contributions from the $s$-channel exchange as discussed in Section~\ref{sec:masless 4point scalars}. However, in general, the second term $\textbf{Im}\,\Psi^{13\leftrightarrow 24}(\beta,z)$ cannot be expressed as a sum over contributions from exchange process. The obstruction is due to the fact that in general the amplitude $T^{13\leftrightarrow 24}(\omega,z)$ does not converge in the $\omega\to\infty$ limit when $z>1$, and thereby we cannot closed the $I_2+I_2'$ contour in Figure~\ref{fig:AC}.\footnote{The analysis around Figure~\ref{fig:AC} is for analytic continuing $T^{12\leftrightarrow 34}(\omega,z)$ from $z>1$ to $1>z>0$, but we can repeat the same analysis for analytic continuing $T^{13\leftrightarrow 24}(\omega,z)$ from $1>z>0$ to $z>1$, and the Figure~\ref{fig:AC} still applies upon reinterpreted the black, blue and red dots as representatives for the $u$-, $s$-, and $t$-channel poles.} This is exactly the case for the open and closed string amplitudes, as we have seen in the above analysis that $\textbf{Im}\,\Psi^{13\to24}_{\rm open/closed}(\beta,z)$ cannot be expressed as a sum over residues.

However, the high energy behaviors of the string scattering amplitudes become nicer in the OPE limit (co-linear limit). This would allow us to close the $I_2+I_2'$ contour for the leading terms in the OPE. In the following analysis, we choose to work with $\textbf{Im}\,\Psi^{14\to23}_{\rm open/closed}(\beta,z)$ by noting the relations \eqref{eqn:AnalyCon}.
Let us expand the open string amplitude \eqref{eqn:openT} in the OPE limit $z\to \infty$ as
\ie
&T^{14\leftrightarrow 23}_{\rm open}(\omega,z) =T_{\rm open}\left({z\over 1-z}\omega^2,\omega^2\right) =\sum_{m=0}^\infty  T^{14\leftrightarrow 23,(m)}_{\rm open}(\omega) z^{-m}\,.
\fe
From the explicit amplitude \eqref{eqn:openT}, we find that in the high energy limit $\omega\to\infty$ the expansion coefficients behave as
\ie
&T^{14\leftrightarrow 23,(m)}_{\rm open}(\omega)\sim \omega^{2m-2}\csc(\pi\omega^2)\log^{m}(\omega)\,.
\fe
The $I_2+I_2'$ contour in Figure~\ref{fig:AC} can be closed at infinity when $\beta\le -2m+2$. Hence, the leading terms in the expansion of $\textbf{Im}\Psi^{14\to23}_{\rm open}(\beta,z)$ can be computed by a sum over the residue of the $t$-channel poles. For example, we have 
\ie
\textbf{Im}\left[\Psi^{14\to23}_{\rm open}(\beta,z)\right]\Big|_{\beta\to 0} &= -\pi B^{14\to23}(z)\bigg[\sum^\infty_{n=1}\frac{(-1)^{n+1}}{2 n^2} + \sum^\infty_{n=1}\frac{(-1)^{n+1} (\psi ^{(0)}(n)+\gamma )}{2 n}z^{-1}+{\cal O}(z^{-2})\bigg]
\\
&=-\pi B^{14\to23}(z)\bigg[\frac{\pi ^2}{24}-\frac{1}{4} \log ^2(2) z^{-1}+{\cal O}(z^{-2})\bigg]\,.
\fe

Similarly, the closed string amplitude \eqref{eqn:closedT} expanded in the OPE limit $z\to \infty$ gives
\ie
&T^{14\leftrightarrow 23}_{\rm closed}(\omega,z) =T_{\rm closed}\left({z\over 1-z}\omega^2,\omega^2\right) =\sum_{m=-1}^\infty  T^{14\leftrightarrow 23,(m)}_{\rm closed}(\omega) z^{-m}\,.
\fe
In the high energy limit $\omega\to\infty$, we find that the expansion coefficients behave as
\ie
&T^{14\leftrightarrow 23,(m)}_{\rm closed}(\omega)\sim\omega^{2m-4}\log^{m+1}(\omega)\,.
\fe
The $I_2+I_2'$ contour can be closed when $\beta< -2m+4$. Again, the leading terms in the expansion of $\textbf{Im}\Psi^{14\to23}_{\rm closed}(\beta,z)$ can be computed by a sum over the residue of the $t$-channel poles. For example, we have 
\ie
\textbf{Im}\left[\Psi^{14\to23}_{\rm open}(\beta,z)\right]\Big|_{\beta\to 0} &= -\pi B^{14\to23}(z)\bigg[\sum^\infty_{n=1}\frac{1}{2 n^3} + \sum^\infty_{n=1}\frac{\psi ^{(0)}(n)+\gamma }{n^2}z^{-1}+{\cal O}(z^{-2})\bigg]
\\
&=-\pi B^{14\to23}(z)\bigg[\frac{\zeta (3)}{2}+\zeta (3) z^{-1}+{\cal O}(z^{-2})\bigg]\,.
\fe

One can similarly compute the imaginary part of the open and closed string celestial amplitudes in the OPE limit for other non-positive integer $\beta$ and verify the celestial dispersion relation for the leading orders in the large $z$ expansion.

\section{Conclusion and outlook} \label{sec:conclusion}

In this paper, we studied the relations between the bulk-locality, unitarity and analyticity of the celestial amplitudes. 

\begin{enumerate}
\item We showed that the imaginary part of the celestial amplitude can be expressed as the integral along the fan-like contour (the right plot in Figure~\ref{Fig:CScontour}) that encloses the positive real axis on the complex plane of the center of mass energy $\omega$.  The exchanges of single- and multi-particle states produce poles and branch cuts near the positive real axis and contribute to the fan-like contour integral. This demonstrates that the imaginary part of the celestial amplitude is given by bulk factorization singularities. 

\item By unitarity, the imaginary part of the celestial amplitude can be positively expanded in the basis of Legendre or Jacobi polynomials for scalar or spinning particle amplitudes. The projection of these orthogonal polynomials onto the celestial sphere can be matched to the Poincar\'e partial waves that satisfy the massive Casimir equation of the Poincar\'e algebra.

\item The four-point celestial amplitude from three distinct physical kinematic configurations tiles the equator of the celestial sphere. On the boundary of each region that functions on the two sides are not continuously connected. Instead we studied the analytic continuation of the celestial amplitude from the physical regions (the intervals listed in Table~\ref{tab:kinematics}) to the to the unphysical region. 

\item Assuming specific high energy behaviour for the flat-space amplitude for complex scattering angles, we prove a celestial version of the dispersion relation \eqref{eqn:CDRconj}. This allows us to relate the residues of the poles of the celestial amplitude at negative even integers on the complex $\beta$-plane, which is yields the EFT coefficients, to the imaginary part of the celestial amplitude and its analytic continuation.

\item These results are verified in the context of  the open and closed string amplitudes.

\end{enumerate}

In defining the fan contour, we've assumed that the amplitude is well behaved as $\omega\rightarrow \infty $ even as $\omega$ is continued on to the complex plane with a small angle. While we've shown that this is holds for scalar exchange and string theory amplitudes, it will be desirable to derive such behaviour from some general principle. 

The celestial dispersion relation is reminiscent of the Zamolodchikov's recursive representation for the conformal blocks \cite{Zamolodchikov:1985ie,Zamolodchikov:1995aa,Penedones:2015aga}. When tuning the conformal dimension of a primary operator to some special values, certain descendant operators become zero norm and decouple from the conformal multiplet.  This phenomenon reflects on the conformal block as poles on the complex conformal dimension plane whose residues are the conformal blocks formed by the zero norm descendants. In our celestial dispersion relation \eqref{eqn:CDRconj}, we see that the residue of the poles at negative even integer $\beta$ are the imaginary part of the celestial amplitude, which can be further expanded in terms of the Poincar\'e partial waves. It would be interesting to see if the origin of these poles can be traced by to the exchange of some zero norm states in the celestial CFT.

\vskip 1cm 

\acknowledgments 
We would like to thank Shu-Heng Shao and Yu-Chi Hou for enlightening discussions. Cm C would like to thank the hospitality of NTU high-energy theory group. Cm C is partly supported by National Key R\&D Program of China (NO. 2020YFA0713000). Yt H, and Zx H is supported by MoST Grant No. 109-2112-M-002 -020 -MY33. Yt H is also supported by Golden Jade fellowship.
\appendix 

\section{Four-point helicity amplitude}\label{sec:four-point_helicity_amplitude}
Let us consider a four-point amplitude of massless particles with helicities $\ell_i$ for $i=1,\cdots,\,4$. The amplitude can be expressed as a function of the angle brackets $\langle ij\rangle$ and the square brackets $[ij]$. The Mandelstam variables are products of angle and square brackets, $s_{ij}=\langle ij\rangle[ji]$. Hence, we can instead choose to express the amplitude as a function of the angle bracket $\langle ij\rangle$ and $s$, $t$. There are relations among these variables. First, we have the Schouten identity
\ie\label{eqn:Schouten_identity}
\langle 41\rangle\langle 23\rangle + \langle 42\rangle\langle 31\rangle + \langle 43\rangle\langle 12\rangle=0\,.
\fe
We also have relations from the momentum conservation
\ie\label{eqn:Momentum_Conservation}
&{\langle 23\rangle\over \langle 13\rangle}(-s-t) + {\langle 24\rangle\over \langle 14\rangle}t=0\,,\quad
&{\langle 12\rangle\over \langle 32\rangle}t + {\langle 14\rangle\over \langle 34\rangle}s=0\,,\quad
&{\langle 12\rangle\over \langle 42\rangle}(-s-t) + {\langle 13\rangle\over \langle 43\rangle}s=0\,.
\fe
We can rewrite the left hand side of the above equation as a $3\times 2$ matrix acting on the vector $(s,t)$, and the Schouten identity implies that all the second minors of matrix vanish. Hence, the relations \eqref{eqn:Momentum_Conservation} all linearly relate to each others.

Using \eqref{eqn:Schouten_identity} and \eqref{eqn:Momentum_Conservation}, we can write the amplitude as a function of the variables
\ie
\langle12\rangle\,,\quad \langle13\rangle\,,\quad \langle14\rangle\,,\quad \langle34\rangle\,,\quad s\,,\quad t\,,
\fe
or equivalently as a function of the variables
\ie
&{z_{12}\over \bar z_{12}}=-\epsilon_1\epsilon_2{\langle12\rangle^2\over s}\,,\quad {z_{13}\over \bar z_{13}}=\epsilon_1\epsilon_3{\langle13\rangle^2\over s+t}\,,\quad {z_{14}\over \bar z_{14}}=-\epsilon_1\epsilon_4{\langle14\rangle^2\over t}\,,
\\
& {z_{34}\over \bar z_{34}}=-\epsilon_3\epsilon_4{\langle34\rangle^2\over s}\,,\quad s\,,\quad t\,.
\fe
Now, using the constraint from the ${\rm SL}(2,{\mathbb C})$ symmetry \eqref{eqn:SL2ConA}, we find
\ie
\mathcal{A}_{\ell_i}(\omega_i, z_i)& = \left({z_{12}\over \bar z_{12}}\right)^{-\ell_2}\left({z_{13}\over \bar z_{13}}\right)^{-\ell_1+\ell_2-\ell_3+\ell_4\over 2}\left({z_{14}\over \bar z_{14}}\right)^{-\ell_1+\ell_2+\ell_3-\ell_4\over 2}\left({z_{34}\over \bar z_{34}}\right)^{\ell_1-\ell_2-\ell_3-\ell_4\over 2}
\\
&\quad\times \delta^{(4)}(p_1+p_2+p_3+p_4)T(s,t)
\\
&=\delta^{(4)}(p_1+p_2+p_3+p_4)\frac{\Big(\frac{z_{14}\bar z_{13}}{\bar z_{14}z_{13}}\Big)^{\ell_3-\ell_4\over 2}\Big(\frac{z_{24}\bar z_{14}}{\bar z_{24}z_{14}}\Big)^{\ell_1-\ell_2\over 2}}{\left({z_{12}\over \bar z_{12}}\right)^{{\ell_1+\ell_2\over 2}}\left({z_{34}\over \bar z_{34}}\right)^{\ell_3+\ell_4\over 2}}T(s,t)\,,
\fe
where in the second equality we have used the identities \eqref{eqn:Momentum_Conservation}.

\section{Poincar\'e generators on single particle states}
\label{sec:PMonSPS}

The Poincar\'e generators $P^\mu$ and $M^{\mu\nu}$ acting on a massless single particle state $|\Delta,z,\ell\rangle$ in the conformal primary basis as
\ie\label{eqn:PoincareGeneratorsInDifferentials}
P^\mu |\Delta,z,\ell\rangle = {\cal P}^\mu|\Delta,z,\ell\rangle\,,\quad M^{\mu\nu} |\Delta,z,\ell\rangle ={\cal M}^{\mu\nu}|\Delta,z,\ell\rangle\,,
\fe
where ${\cal P}^\mu$ and ${\cal M}^{\mu\nu}$ are differential operators, whose explicit form are given by~\cite{Stieberger:2018onx}
\ie
{\cal M}^{01}&={i\over 2}\left[(\bar z^2-1)\bar\partial +(z^2-1)\partial+2(\bar h\bar z+hz)\right]\,,
\\
{\cal M}^{02}&=-{1\over 2}\left[(\bar z^2+1)\bar\partial -(z^2+1)\partial+2(\bar h\bar z-hz)\right]\,,
\\
{\cal M}^{03}&=i(\bar z\bar \partial+z\partial+\bar h+h)\,,
\\
{\cal M}^{12}&=-\bar z\bar \partial+z\partial-\bar h+h\,,
\\
{\cal M}^{13}&={i\over 2}\left[(\bar z^2+1)\bar\partial +(z^2+1)\partial+2(\bar h\bar z+hz)\right]\,,
\\
{\cal M}^{23}&=-{1\over 2}\left[(\bar z^2-1)\bar\partial - (z^2-1)\partial+2(\bar h\bar z - hz)\right]\,,
\fe
and
\ie\label{eqn:cPdiff}
{\cal P}^\mu=2q^\mu e^{\partial_\Delta}\,.
\fe

\section{Analytic continuation of the string amplitudes}
\label{sec:analytic_continuation_string}
\begin{figure}[htbp]
\centering    
\begin{minipage}{7cm}
	\centering          
	\includegraphics[scale=0.5]{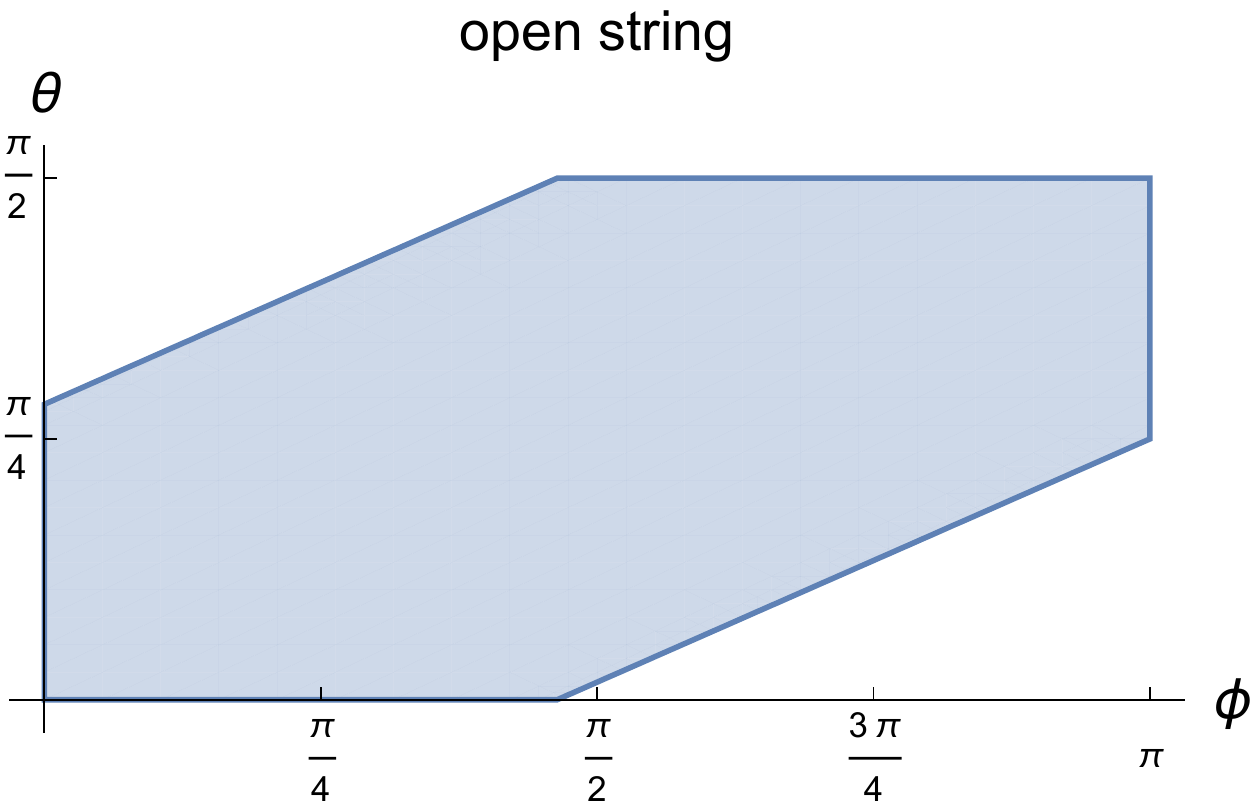}   	\end{minipage}
	\qquad
\begin{minipage}{7cm}
	\centering      
	\includegraphics[scale=0.5]{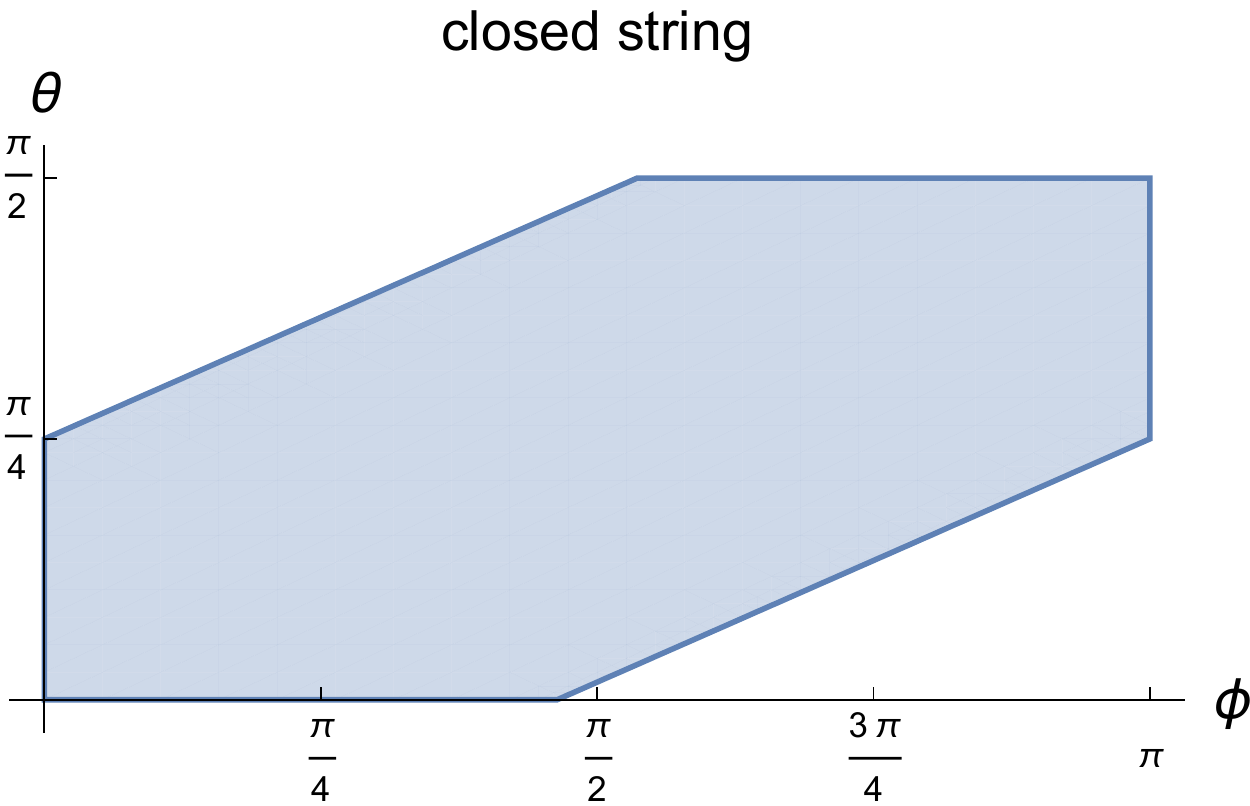}   	\end{minipage}
\caption{The convergence region $\Theta^{12\leftrightarrow 34}(z)$ of the open and closed string amplitudes for $\epsilon=10^{-6}$.} 
\label{fig:StringAmpConv} 
\end{figure}

In this appendix, we apply the analytic continuation procedure in Section~\ref{sec:analytic_continuation_general} to the open and closed string amplitudes. As discussed in Section~\ref{sec:analytic_continuation_general}, the integration contour of the celestial amplitude are being continuously deformed when going along the path \eqref{eqn:analytic_continutaiton_path}. The integration contour should asymptote to the angle inside the convergent region $\Theta^{12\leftrightarrow 34}(z)$. Hence, we need to ensure that the convergent region varies continuously along the path \eqref{eqn:analytic_continutaiton_path}. Let us parametrize the string amplitudes \eqref{eqn:openT} and \eqref{eqn:closedT} as
\ie
s=\omega^2\,,\quad t = -{z-1\over z}\omega^2\,,\quad\omega=re^{i\theta}\,,\quad z=1+\epsilon  e^{-i\phi}\,.
\fe
The amplitudes vanish in the limit $r\to\infty$ when the angles $\theta$ and $\phi$ are inside the region plotted in Figure~\ref{fig:StringAmpConv}. We see that we can indeed find continuous deformations of the integration contours which are always inside the convergent regions.

\section{Conformal partial wave representation}
\label{sec:conformalpartialwaverepresentation}

The four-point celestial amplitude \eqref{eqn:generalCA} can be expanded in the conformal partial wave basis as
\ie
f_{\Delta_i,\ell_i}(z,\bar{z})=\sum^\infty_{\ell=-\infty}\int_{0}^{\infty}{\mathrm{d}\nu\over 2\pi}\, {p_\ell(\Delta_i,\ell_i;\nu) \over n_\ell(\nu)}\Psi^{\rm conf.}_{\Delta_i,\ell_i;1+i\nu,\ell}(z,\bar{z})\,,
\fe
where $ \Psi^{\rm conf.}_{\Delta_i,\ell_i;1+i\nu,\ell}(z,\bar{z})$ is the conformal partial wave normalized such that 
\ie
{1\over 2}\int {d^2z\over |z|^4}\Psi^{\rm conf.}_{\Delta_i,\ell_i;1+i\nu,\ell}(z,\bar{z})\Psi^{\rm conf.}_{1-\Delta_i,-\ell_i;1-i\nu',-\ell'}(z,\bar{z})=n_\ell(\nu)\times 2\pi\delta_{\ell,\ell'}\delta(\nu-\nu')\,,
\fe
for external dimensions in the principal series, i.e. $\Delta_i\in 1+i{\mathbb R}$. The normalization $n_\ell(\nu)$ is\footnote{We follow the convention in~\cite{Murugan:2017eto,Simmons-Duffin:2017nub}.}
\ie
n_\ell(\nu)={ 2\pi^3\over \ell^2+\nu^2}\,.
\fe

For the scalar amplitude with the tree-level massive scalar exchange in the $12\leftrightarrow34$ kinematics, the imaginary part of the expansion coefficient $p_\ell(\Delta_i;\nu)$ factorizes as~\cite{Lam:2017ofc}
\ie\label{eqn:pfac}
\textbf{Im}\,p_\ell(\Delta_i;\nu)=\delta_{\ell,0}\pi^2 m^2  C_{\Delta_1,\Delta_2;1+i\nu}C_{\Delta_3,\Delta_4;1-i\nu}\,,
\fe
where $C_{\Delta_1,\Delta_2;\Delta_3}$ is the 3-point coefficient of the celestial amplitude of two massless and one massive scalars. Using the shadow representation of the conformal partial wave, the factorization \eqref{eqn:pfac} is equivalent to
\ie\label{eqn:CPW1}
&\textbf{Im}\,\tilde{\mathcal{A}}_{\Delta_i}^{12\leftrightarrow34}(z_i,\bar{z}_i)=\int_{0}^{\infty}{\mathrm{d}\nu\over 2\pi}\,{\pi^2 m^2\over n_\ell(\nu)}\int \mathrm{d}^2z
\\
&\quad\quad\quad\quad\quad\quad\times\,\tilde{\mathcal{A}}_{3,\Delta_1,\Delta_2;1+i\nu}(z_1,\bar{z}_1,z_2,\bar{z}_2,z,\bar{z})\tilde{\mathcal{A}}_{3,\Delta_3,\Delta_4;1-i\nu}(z_3,\bar{z}_3,z_4,\bar{z}_4,z,\bar{z})\,.
\fe
where $\tilde{\mathcal{A}}_{\Delta_1,\Delta_2;\Delta_3}(z_i,\bar{z}_i)$ is the three-point celestial amplitude,
\ie
\tilde{\mathcal{A}}_{\Delta_1,\Delta_2;\Delta_3}(z_i,\bar{z}_i)={C_{\Delta_1,\Delta_2;\Delta_3}\over |z_{12}|^{\Delta_1+\Delta_2-\Delta_3}|z_{13}|^{\Delta_1+\Delta_3-\Delta_2}|z_{23}|^{\Delta_2+\Delta_3-\Delta_1}}\,.
\fe
In this appendix, we generalize the above result to the four-point scalar celestial amplitude with the tree-level massive spin-$J$ particle exchange. We show that the imaginary part of such an amplitude factorizes as an integral of a product of two three-point celestial amplitudes of two massless scalars and one massive spin-$J$ particle,
\ie\label{eqn:CPW1}
&\textbf{Im}\,\tilde{\mathcal{A}}_{\Delta_i,J}^{12\leftrightarrow34}(z_i,\bar{z}_i)=\pi^2 m^{2J+2}J!\sum_{\ell=-J}^J\int_{0}^{\infty}{\mathrm{d}\nu\over 2\pi}\,{\mu_{J,\ell}(\nu)\over 2^{|\ell|} n_\ell(\nu)}\int \mathrm{d}^2z
\\
&\quad\quad\quad\quad\quad\quad\times\,\tilde{\mathcal{A}}^{(J)}_{3,\Delta_1,\Delta_2;1+i\nu,\ell}(z_1,\bar{z}_1,z_2,\bar{z}_2,z,\bar{z})\tilde{\mathcal{A}}^{(J)}_{3,\Delta_3,\Delta_4;1-i\nu,-\ell}(z_3,\bar{z}_3,z_4,\bar{z}_4,z,\bar{z})\,,
\fe
where $\mu_{J,\ell}(\nu)$ is
\ie
\mu_{J,\ell}(\nu)&=(-1)^{|\ell|}{2^{J-|\ell|}(|\ell|+1)_{J-|\ell|}(|\ell|+{1\over 2})_{J-|\ell|}\over (J-|\ell|)!(2|\ell|+1)_{J-|\ell|}}\,.
\fe

Let us first focus on the right hand side of \eqref{eqn:CPW1}. The 3-point celestial amplitude is given by the integral~\cite{Law:2020tsg}
\ie\label{eqn:Mellin3-pt}
\tilde{\mathcal{A}}^{(J)}_{3,\Delta_i,\Delta,\ell}(z_i,\bar{z}_i)&=\left(\prod^2_{i=1}\int_0^\infty d\omega_i\,\omega_i^{\Delta_i-1}\right) \int\frac{dy}{y^3}dz'd\bar{z}' \; 
\\
&\quad\times\sum_{b=-J}^{J}G^{(J)}_{\ell,b}(\hat k; z,\bar z) A_{3,b}(p_1,p_2,k)\delta^4(k+p_1+p_2)\,,
\fe
where $A_{3,b}(p_1,p_2,k)$ is the three-point amplitude explicitly given by
\begin{equation}\label{eqn:CA3pt1}
A_{3,b}(p_1,p_2,k)= p_{12}^{\mu_1}\dots p_{12}^{\mu_J}\epsilon_{b,\mu_1\mu_2\dots\mu_J}\,,
\end{equation}
with all momenta outgoing. $\epsilon_{b,\mu_1\mu_2\dots\mu_J}$ is the polarization tensor for spin-$s$ particle. The massless momenta $p_1$, $p_2$ are parametrized as before by $p_i=-\omega_i q_i$ and \eqref{eqn:NullMomentum}. The massive momentum $k$ is parametrized by
\ie
&k=m\hat k\,,&&\hat k^\mu ={1\over 2y} (1+y^2+|z'|^2,2{\rm Re}(z'),2{\rm Im}(z'),1-y^2-|z'|^2)\,.
\fe
$G^{(J)}_{\ell,b}(\hat k; z,\bar z)$ is the integration weight matrix that relates the spin-$J$ massive irreducible representations of the little group on $\mathbb{R}^{1,3}$ to the spin-$\ell$ representations of the conformal group on the celestial sphere. The integration weight matrix $G^{(J)}_{\ell,b}(\hat k; z,\bar z)$ is computed in~\cite{Law:2020tsg}
\ie\label{eqn:IWM}
\sum_{b=-J}^JG_{\ell,b}^{(J)}(\hat k; z,\bar z)\epsilon_b^{\mu_1\mu_2\cdots \mu_J}&=\frac{1}{J! \left(\frac{1}{2}\right)_J}K^{\mu_1} K^{\mu_2}\cdots K^{\mu_J} G_{\ell,\Delta}^{(J)}(\hat{k}, Y; z,\bar{z})\Big|_{Y=0}\,,
\\
G_{\ell,\Delta}^{(J)}(\hat{k}, Y; z,\bar{z})&=\begin{cases} \frac{(Y\cdot q)^{J{-}\ell}\left[(\hat{k}\cdot q)(Y\cdot \partial_{\bar z} q){-}(\hat{k}\cdot \partial_{\bar z} q)(Y\cdot q)\right]^{\ell}}{(-\hat{k}\cdot q)^{\Delta+J}}&{\rm for}\quad\ell\ge0\,,
\\
\frac{(Y\cdot q)^{J+\ell}\left[(\hat{k}\cdot \partial_z q)(Y\cdot q)-(\hat{k}\cdot q)(Y\cdot \partial_z q)\right]^{-\ell}}{(-\hat{k}\cdot q)^{\Delta+J}}&{\rm for}\quad\ell<0\,,
\end{cases}
\\
K_\mu(\hat k,Y)&={1\over 2}\left[\partial_{Y^\mu}+\hat k_{\mu} (\hat k\cdot \partial_Y)\right]+(Y\cdot\partial_Y)\partial_{Y^\mu}
\\
&\quad+\hat k_{\mu}(Y\cdot\partial_Y)(\hat k\cdot \partial_Y)-{1\over 2}Y_\mu\left(\partial_Y^2+(\hat k\cdot\partial_Y)^2\right)\,,
\fe
where the null vector $q$ is parametrized by $z$ and $\bar z$ as in \eqref{eqn:NullMomentum}. Plugging this formula into \eqref{eqn:CA3pt1}, we find
\ie\label{eqn:CA3pt2}
\tilde{\mathcal{A}}^{(J)}_{3,\Delta_i,\Delta,\ell}(z_i,\bar{z}_i)&=\frac{1}{J!(\frac{1}{2})_J}\left(\prod^2_{i=1}\int_0^\infty d\omega_i\,\omega_i^{\Delta_i-1}\right) \int\frac{dy}{y^3}dz'd\bar{z}'
\\
&\quad\times\delta^4(p+p_1+p_2)(p_{12}\cdot K)^JG_{\ell,\Delta}^{(J)}(\hat{k},Y;z,\bar{z})\,.
\fe
Now, the right hand side of \eqref{eqn:CPW1} can be simplified as
\ie\label{eqn:integral2}
&\frac{\pi^2 m^{2J+2}J!}{\left(J!(\frac{1}{2})_J\right)^2}\left(\prod^4_{i=1}\int_0^\infty d\omega_i\,\omega_i^{\Delta_i-1}\right)\sum_{\ell=-J}^J\int_0^{\infty}{\mathrm{d}\nu\over 2\pi}{\mu_{J,\ell}(\nu)\over 2^{|\ell|}n_\ell(\nu)}\left(\prod_{i=1}^2\int\frac{\mathrm{d}y_i}{y_i^3}\mathrm{d}^2z_i'\right)\int \mathrm{d}^2 z
\\
&\times\delta^{(4)}(p_1+p_2+k_1)\delta^{(4)}(p_3+p_4-k_2)
\\
&\times \left[\left(p_{12}\cdot K(\hat k_1,Y_1)\right)^J G_{\ell,1+i\nu}^{(J)}(\hat{k}_1,Y_1;z,\bar{z})\right]\left[\left(p_{34}\cdot K(\hat k_2,Y_2)\right)^J G_{-\ell,1-i\nu}^{(J)}(\hat{k}_2,Y_2;z,\bar{z})\right]\Big|_{Y_1=Y_2=0}
\\
&=\frac{\pi^2 m^{2J+2}J!}{4\pi \left(J!(\frac{1}{2})_J\right)^2}\left(\prod^4_{i=1}\int_0^\infty d\omega_i\,\omega_i^{\Delta_i-1}\right)\int\frac{\mathrm{d}y_1}{y_1^3}\mathrm{d}^2z_1'\delta^{(4)}(p_1+p_2+k_1)\delta^{(4)}(p_3+p_4-k_1)
\\
&\quad\times \left(p_{12}\cdot K(\hat k_1,{Y_1})\right)^J\left(p_{34}\cdot K(\hat k_1,{Y_2})\right)^J(Y_1\cdot Y_2)^J\Big|_{Y_1=Y_2=0}
\\
&=\pi m^{2J}\left(\prod^4_{i=1}\int_0^\infty d\omega_i\,\omega_i^{\Delta_i-1}\right)\delta^{(4)}(p_1+p_2+p_3+p_4)\delta(s-m^2)P_J\Big(\frac{u-t}{m^2}\Big)\,,
\fe
where in the second equality, we have used the orthogonality condition~\cite{Law:2020tsg}
\begin{equation}
\sum_{\ell=-J}^J\int_{-\infty}^\infty\mathrm{d}\nu{\mu_{J,\ell}(\nu)\over 2^{|\ell|} n_\ell(\nu)}\int\mathrm{d}^2z\,G_{\ell,1+i\nu}^{(J)}(\hat{k}_1,Y_1;z,\bar{z})G_{-\ell,1-i\nu}^{(J)}(\hat{k}_2,Y_2;z,\bar{z})=\delta(\hat{k}_1,\hat{k}_2)(Y_1\cdot Y_2)^s\,,
\end{equation}
where the delta function $\delta(\hat{k}_1,\hat{k}_2)$ is defined by
\ie
\int_0^\infty{dy_2\over y_2^3}\int d^2 z_2'\delta(\hat{k}_1,\hat{k}_2)F(\hat k_2)=F(\hat k_1)\,.
\fe
In the third equality of \eqref{eqn:integral2}, we have used the identities
\ie
&\int\frac{\mathrm{d}y_1}{y_1^3}\mathrm{d}^2z_1'\delta^{(4)}(p_1+p_2+k_1)=\frac{4}{m^2}\delta(s-m^2)\,,
\\
&\left(p_{12}\cdot K(\hat k_1,{Y_1})\right)^J\left(p_{34}\cdot K(\hat k_1,{Y_2})\right)^J(Y_1\cdot Y_2)^J\Big|_{Y_1=Y_2=0}=J!\left(\frac{(2J-1)!!}{2^{J}}\right)^2P_J\Big(\frac{u-t}{m^2}\Big)\,.
\fe

Next, let us look at the left hand side of \eqref{eqn:CPW1}. The four-point celestial amplitude can be computed using the Mellin integral \eqref{eqn:MellinTransfrom} and the formula \eqref{eqn:4pt_helicity_amplitude} with $\ell_i=0$,
\ie\label{eqn:LHSfacspin}
\textbf{Im}\,\tilde{\mathcal{A}}_{\Delta_i,J}^{12\leftrightarrow34}(z_i,\bar{z}_i)&=\left(\prod^4_{i=1}\int_0^\infty d\omega_i\,\omega_i^{\Delta_i-1}\right)\delta^{(4)}(p_1+\cdots+p_4)
\\
&\quad\times m^{2J}\textbf{Im}\,\left[-\frac{P_{J}\left(\frac{u-t}{m^2}\right)}{s-m^2+i\epsilon}
-\frac{P_{J}\left(\frac{s-t}{m^2}\right)}{u-m^2+i\epsilon}
-\frac{P_{J}\left(\frac{u-s}{m^2}\right)}{t-m^2+i\epsilon}\right]
\\
&=\pi m^{2J} \left(\prod^4_{i=1}\int_0^\infty d\omega_i\,\omega_i^{\Delta_i-1}\right)P_{J}\left(\frac{u-t}{m^2}\right)\delta(s-m^2)\delta^{(4)}(p_1+\cdots+p_4)\,.
\fe
\eqref{eqn:integral2} and \eqref{eqn:LHSfacspin} matches exactly; hence, the factorization formula \eqref{eqn:CPW1} follows.

By the conformal symmetry, the three-point celestial amplitude $\tilde{\mathcal{A}}^{(J)}_{3,\Delta_1,\Delta_2;\Delta,\ell}(z_i,\bar{z}_i)$ takes the form as
\ie
\tilde{\mathcal{A}}^{(J)}_{3,\Delta_1,\Delta_2;\Delta,\ell}(z_i,\bar{z}_i)={C^{(J)}_{\Delta_1,\Delta_2;\Delta,\ell}\over |z_{12}|^{\Delta_1+\Delta_2-(\Delta-\ell)}|z_{13}|^{\Delta_1+(\Delta-\ell)-\Delta_2}|z_{23}|^{\Delta_2+(\Delta-\ell)-\Delta_1}}\left(z_{12}\over z_{13}z_{23}\right)^\ell\,.
\fe
Plugging this into \eqref{eqn:CPW1} and using the form \eqref{eqn:generalCA} of the four-point celestial amplitude, we find
\begin{equation}\label{eqn:CPWspin}
\textbf{Im}\,f_{\Delta_i,J}^{12\leftrightarrow34}(z,\bar{z})=\pi^2 m^{2J+2}J!\sum_{\ell=-J}^J\int_{0}^{\infty}{\mathrm{d}\nu\over 2\pi}{\mu_{J,\ell}(\nu)\over 2^{|\ell|} n_\ell(\nu)}C_{\Delta_1,\Delta_2;1+i\nu,\ell}^JC_{\Delta_3,\Delta_4;1-i\nu,-\ell}^{-J}\Psi_{\Delta_i,\ell_i;1+i\nu,\ell}^{\rm conf.}(z,\bar{z})\,.
\end{equation}
Note that with the prefactor \eqref{eqn:translation}, the left hand side of \eqref{eqn:CPWspin} is exactly the scalar Poincare partial wave \eqref{eqn:poincarepartialwave}, i.e.
\begin{equation}
\textbf{Im}\,f_{\Delta_i,\ell}^{12\leftrightarrow34}(z,\bar{z})=(z-1)^{\frac{\Delta_1-\Delta_2-\Delta_3+\Delta_4}{2}}\delta(iz-i\bar{z})\Psi_{m,l}^{12\leftrightarrow34}(\mathbf{\Delta},z)\,.
\end{equation}
Thus, \eqref{eqn:CPWspin} gives a conformal partial wave representation of the scalarPoincar\'e partial wave! 

Finally, in \cite{Law:2020tsg}, the three-point coefficients $C_{\Delta_1,\Delta_2;\Delta_3}^J$ for $J=0,\,1,\,2$ are computed, and recursion relations for the general three-point coefficients are derived. In Appendix~\ref{app:3ptCA}, we compute the general three-point coefficients by directly evaluating the Mellin integral \eqref{eqn:CA3pt2}.


\section{Computation of $C_{\Delta_1,\Delta_2;\Delta,\ell}^{(J)}$}\label{app:3ptCA}
Let us compute the structure constant $C_{\Delta_1,\Delta_2;\Delta,\ell}^{(J)}$ by explicitly working out the Mellin integral \eqref{eqn:CA3pt2}. The momentum conservation delta function gives
\ie
y={m\over 2(\omega_1+\omega_2)}\,,\quad z' = {z_1\omega_1+z_2\omega_2\over \omega_1+\omega_2}\,,\quad \bar z' = {\bar z_1\omega_1+\bar z_2\omega_2\over \omega_1+\omega_2}\,, \quad \omega_1 \omega_2={m^2\over 4|z_{12}|^2}\,,
\fe
and the Jacobian
\ie
&|{\rm Jacobian}|= {8m |z_{12}|^4\omega_2^2\over (m^2+ 4|z_{12}|^2 \omega_2^2)^3}\,.
\fe
The inner products that appear in the $(p_{12}\cdot K)^{J}G_{\ell,\Delta}^{(J)}(\hat k, Y;z,\bar z)$ are summarized as
\ie
& 2p_{12}\cdot q=-4|z-z_2|^2\omega_2+{m^2 |z-z_1|^2\over |z_{12}|^2\omega_2}\,,\quad  2 k\cdot q=-4|z-z_2|^2\omega_2-{m^2 |z-z_1|^2\over |z_{12}|^2\omega_2}\,,
\\
&p_{12}^2=m^2\,,\quad k\cdot  p_{12}=0\,,
\\
&(k\cdot q)( p_{12}\cdot \partial_{\bar z} q)-(k\cdot \partial_{\bar z} q)(q\cdot  p_{12})=-{2m ( z- z_1)( z- z_2)\over  z_{12}}\,.
\fe
We find
\ie\label{eqn:p23K}
 p_{12}\cdot K=\left( N_Y+{1\over 2}\right) p_{12}\cdot \partial_Y -{1\over 2}( p_{12}\cdot Y)\left(\partial_Y^2+(\hat k\cdot\partial_Y)^2\right),
\fe
where $N_Y = Y\cdot \partial_Y$ simply counts the number of $Y$. We have the following identities
\ie
\partial_Y^2G_{\ell,\Delta}^{(J)}(\hat k, Y;z,\bar z)&=0\,,
\\
(\hat k\cdot\partial_Y)^2G_{\ell,\Delta}^{(J)}(\hat k, Y;z,\bar z) &= (J-|\ell|)(J-|\ell|-1)G_{\ell,\Delta}^{(J-2)}(\hat k, Y;z,\bar z)\,.
\fe

Using these formulae, we find
\ie
&(p_{12}\cdot K)^JG^{(J)}_{\ell,\Delta}(\hat k, Y;z,\bar z)\Big|_{Y=0}
\\
&={1\over 2^J} \sum^{J\over 2}_{n=0} (-1)^n (2J-1-2n)!! (2n-1)!! {J\choose 2n}
\\
&\quad\quad\times m^{2n}(p_{12}\cdot\partial_Y)^{J-2n} (\hat k_{12}\cdot\partial_Y)^{2n} G^{(J)}_{\ell,\Delta}(\hat k, Y;z,\bar z)\Big|_{Y=0}
\\
&={1\over 2^J} \sum^{J\over 2}_{n=0} (-1)^n (2J-1-2n)!! (2n-1)!!  {J\choose 2n}
\\
&\quad\quad\times m^{2n}{(J-|\ell|)!(J-2n)!\over (J-|\ell|-2n)!}G^{(J-2n)}_{\ell,\Delta}(\hat k,p_{12},z,\bar z)\,.
\fe
Let us assume $\ell\ge 0$, and consider the integral
\ie
&\left(\prod^2_{i=1}\int_0^\infty d\omega_i\,\omega_i^{\Delta_i-1}\right) \int\frac{dy}{y^3}dz'd\bar{z}'\delta^4(p+p_1+p_2)G_{\ell,\Delta}^{(J-2n)}(\hat{k},p_{12};z,\bar{z})
\\
&=2^{2-\Delta_1-\Delta_2-\Delta} m^{\Delta_1+\Delta_2-4+J-2n}|z_{12}|^{-2\Delta_1}\left({ ( z- z_1)(  z- z_2)\over   z_{12}}\right)^{\ell}{1\over |z-z_2|^{2(\Delta+\ell)}}
\\
&\quad\times \int^\infty_0d\omega_2\,\sum_{p=0}^{J-2n-\ell}(-1)^{\ell+p}{J-2n-\ell\choose p}{\omega_2^{\Delta_2-\Delta_1-1+\Delta+\ell+2p}\left({ |z-z_1|^2\over |z_{12}|^2|z-z_2|^2}\right)^{J-2n-\ell-p}\over \left(\omega_2^2+{ |z-z_1|^2\over |z_{12}|^2|z-z_2|^2}\right)^{\Delta+J-2n}}
\\
&=2^{1-\Delta_1-\Delta_2-\Delta} m^{\Delta_1+\Delta_2-4+J-2n} \sum_{p=0}^{J-2n-\ell}(-1)^{\ell+p}{J-2n-\ell\choose p}
\\
&\quad\quad\times B\left({\Delta+\ell+2p-\Delta_1+\Delta_2\over 2},{\Delta-\ell+2J-4n-2p+\Delta_1-\Delta_2\over 2}\right)
\\
&\quad\times\left({ ( z- z_1)(  z- z_2)\over   z_{12}}\right)^{\ell}|z-z_1|^{\Delta_2-\Delta_1-\Delta-\ell}|z-z_2|^{\Delta_1-\Delta_2-\Delta-\ell}|z_{12}|^{\Delta+\ell-\Delta_1-\Delta_2}\,.
\fe
The three-point coefficient is 
\ie
C_{\Delta_1,\Delta_2;\Delta,\ell}^{(J)}
&={m^{\Delta_1+\Delta_2-4+J}\over 2^{\Delta_1+\Delta_2+\Delta+J-1}J!(\frac{1}{2})_J} \sum^{J\over 2}_{n=0} \sum_{p=0}^{J-2n-\ell} (-1)^{\ell+n+p}  {J\choose 2n}{J-2n-\ell\choose p}
\\
&\quad\quad\times {(J-|\ell|)!(J-2n)! (2J-1-2n)!! (2n-1)!!\over (J-|\ell|-2n)!}
\\
&\quad\quad\times B\left({\Delta+\ell+2p-\Delta_1+\Delta_2\over 2},{\Delta-\ell+2J-4n-2p+\Delta_1-\Delta_2\over 2}\right)\,.
\fe

\bibliographystyle{JHEP.bst}
\bibliography{refs.bib}

\end{document}